\documentstyle[titlepage,psfig]{article}
\begin{document}
 \large
 \newcommand{\qqq}{\textstyle}
 \newcommand{\qqt}{ }
 \newcommand{\qqw}{ }
 \newcommand{\qqm}{ }

 \title
 {Cosmic Rays: Studying the Origin}
 \author{Jacek Szabelski
 \vspace{0.5cm}\\
 The Andrzej So{\l }tan Institute for Nuclear Studies\\
 Cosmic Ray Laboratory\\
 ul. Uniwersytecka 5, 90-950 {\L }\'{o}d\'{z} 1, P.O.box 447\\
 js@zpk.u.lodz.pl}
 \date{2 July 1997}

 \maketitle
 \tableofcontents
 \listoffigures
 \listoftables
 
 \newpage
 \section{Introduction}
 Results from many experiments provide continuously 
 improving knowledge about cosmic rays (CR). 
 To interpret this knowledge 
 it is necessary to combine information from
 various areas of astronomy, particle physics,
 plasma physics, geophysics and others. 
 The phenomenon of CR relates to astronomy
 since CR come from outside the Earth,
 most of them from outside the Solar System.
 CR are detected due to their interaction
 with matter and the detectors are 
 the same as used in particle physics.
 CR are the phenomenon of Nature. All experiments
 are passive. This paper describes in some details
 author's contribution to interpretation of
 experimental results on the background of
 general trends of research in this area.
 Particular results are from experiments which sometimes 
 based on very different principles,
 from satellite detectors to deep underground detectors,
 all giving contribution to cosmic ray studies.\\

 \noindent
 Many years after the discovery of cosmic rays 
 their origin is still unknown. 
 There are problems with naming the sources (if they are),
 as well as with naming the mechanism of particle acceleration.
 It is accepted that acceleration is an ordinary (astro)physical
 process, 
 i.e. with the well known physical basic phenomena playing the
 crucial role,
 but placed in not understood, and therefore \lq \lq complicated",
 time--space scenario.
 To dramatize the problem a little, let us notice that
 the very sophisticated, man built accelerators
 can accelerate protons to the energy about 1000~GeV 
 (at the beginning of 1997), whereas 
 the measured flux of CR protons above energy 100,000~GeV is
 equal to about 2$\cdot$10$^{\qqt -5}$ particles per 
 (m$^{\qqm 2}\cdot {\rm sr}\cdot {\rm sec}$)
 (i.e. 3$\cdot$10$^{\qqt 10}$ 
 such protons are hitting the Earth's atmosphere every second)
 and more than 1000 of the most energetic CR of energies above 
 10,000,000,000~GeV (10$^{\qqt 10}$~GeV) have been registered.\\
 
 \noindent
 Let us accept following definition of primary cosmic rays
 as 
 {\it energetic particles of nonthermal origin,} 
 present in the space outside the Earth atmosphere.
 They are protons and other nuclei, electrons, 
 gamma photons, positrons, antiprotons and neutrinos.\\
 
 \noindent
 Different components have different abundances,
 different energy spectra, different ability of
 being measured.
 There is no evidence that all of them have common sources. 
 We can make relations between
 abundances of some of them (e.g. the bulk of antiprotons being
 secondary particles from CR protons and nuclei collisions
 with the interstellar medium
 \cite[{\em Szabelski et al., 1980}]{js:antypI}).
 The only common
 features are their relatively high energy and
 the problem with the explanation of their origin.\\
 There are many attempts to explain the mechanism of CR
 acceleration and I would like to mention 
 the shock wave acceleration mechanism 
 which has many achievements
 and is very fashionable in the recent years.
 Since there is no observational evidence pointing exclusively
 to this mechanism, as well as there are problems with
 acceleration efficiency and explanation of very high energy CR
 within this model, it looks reasonable to search for
 some other possible explanations of CR acceleration mechanisms
 as well.\\
 
 \noindent
 Principally, it is necessary to refer to the experimental
 studies of CR origin and acceleration.
 The most important are different measurements of CR
 and related events. We know the energy spectra of various
 components, absolute or relative abundances of
 components, anisotropy, time variation etc. 
 Thanks to them we have few models of CR ray propagation
 in our Galaxy, which provide some systematics to
 the bulk of information.\\
 
 \noindent
 Before describing some detailed studies of CR origin
 it might be worth summarizing the problems in the simplified manner.
 This would provide a reference level for the role
 and importance of CR studies.
 \begin{itemize}
 \setlength{\itemsep}{-2pt}
 
 \item Do observed CR originate within our Galaxy or
 are they of Universal origin~?
 \item Do they originate and are accelerated in the sources or
 are they of diffuse origin~?
 \item Is CR production uniform in time or intermittent~?
 \item Do all CR have the same origin or different
 components have principally different origin~?
 \item Are all CR accelerated by the same mechanism independent
 of their energy or by several different processes acting
 in limited energy regimes~?
 \item Is our position as \lq \lq observers" typical in the Galaxy and
 is it typical in time~?
 \item Does the intensity of CR vary from place to place in
 our Galaxy and vary with time~?
 \end{itemize}
 Almost each of above questions requires an extended comment to explain
 its meaning, since a lot of work has been devoted to
 such problems. We know a lot at present time and, although the
 above list is not complete, many past time problems have been
 solved already.\\
 However, the truly honest answer to each of above questions
 would be \mbox{\lq \lq YES or NO"}, at present.
 Sometimes YES, sometimes NO, or \mbox{\lq \lq YES, but"}, etc.
 Simply: there is no sufficient experimental evidence
 to provide the answer.\\
 
 \noindent
 This work does not provide answers to above listed problems.
 It only addresses them.\\
 The presented approach to study the CR origin shall not be
 treated as the unique, the best, nor even as the most promising.
 This is an approach. All other works are valuable, provided
 that their final results can be verified by experiment or
 observations.\\
 
 \noindent
 The most complete information about cosmic ray
 physics can be found in 
 the Proceedings of International Cosmic Ray Conferences;
 the last 24$^{\qqt th}$ was held in Rome, Italy in August 1995,
 the 23$^{\qqt rd}$ in Calgary, Canada in August 1993,
 the 22$^{\qqt nd}$ in Dublin, Ireland in August 1991,
 the 21$^{\qqt st}$ in Adelaide, Australia in January 1990,
 and the 20$^{\qqt th}$ in Moscow, the~USSR in August 1987.\\
 
 \noindent
 The studies of cosmic ray origin can be systematized in the following
 way:
 
 \begin{enumerate}

 \item {\bf Search for CR \lq \lq point sources"}.
 There are experiments which are 
 looking for enhancements of CR intensity
 around the well localized sources or enhancements
 of CR flux from such directions. 
 The idea is simple: if there are CR \lq \lq point sources"
 then the local intensity nearby the source is larger than
 average.\\
 The \lq \lq only" problems are with observations.
 Paths of CR charged particles are bent in interstellar magnetic
 fields and incoming direction of observed CR particle does
 not point to the source. Therefore such studies
 are limited to 
 \begin{enumerate}
 \setlength{\itemsep}{-2pt}
 \item gamma ray observations; $\gamma$--rays are CR secondaries from
 interactions of energetic particles with matter, 
 and the interstellar medium is relatively transparent for them,
 \item neutrino astronomy; as above, but they are difficult to
 detect,
 \item very high energy CR observations;
 at extremely high energy the paths of CR protons are only
 slightly bent in galactic magnetic field, provided that the 
 distance to the source is not too large, 
 \item examining the possibility of high energy 
 neutron flux from the nearby CR source,
 \item other astronomical methods; the very promising example
 is a comparison between the infra--red and radio emissivity
 in our Galaxy which allowed to point a number
 of non--thermal radio synchrotron emitting
 sources~\cite[{\em Broadbent et al., 1989}]{jlo:nontherm}.
 \end{enumerate}
 \item {\bf Studies of \lq \lq CR sources" chemical composition}.
 The direct measurements of CR mass composition were performed
 for CR energy below few hundred GeV.
 Results of local (at the Earth vicinity) 
 measurements of absolute or relative mass composition
 of CR and the dependence of CR mass composition on
 energy are related to CR source chemistry, mechanism of
 acceleration and propagation properties.
 The research here reminds of solving the puzzle problem:
 some pieces fit each other, some others fit
 in another corner, but we are not able to
 make a whole picture, and, possibly, we do not have
 all the pieces yet.
 \begin{enumerate}
 \setlength{\itemsep}{-2pt}
 \item 
 It is largely possible to measure the chemical composition
 of distant astronomical objects performing spectroscopic
 observation.
 Results of measurements of low energy CR mass composition
 are compared with observed star atmospheres
 to point out candidates for CR sources.
 The method has some \lq \lq achievements", e.g. 
 correlation between the abundances of CR isotopes, 
 the first ionization potential, and the chemistry of 
 Wolf--Rayet stars atmosphere 
 (\cite[{\em Blake and Dearborn, 1988}]{Erice:W-R}).
 \item 
 The accurate measurements of chemical CR composition
 and its variability with energy was used to study
 the possibility of CR acceleration (or deceleration)
 during propagation
 \cite[{\em Giler et al., 1989}]{bs:con_acc}.
 \item
 It is known that the CR intensity has a power law like energy
 spectrum, and its power index becomes smaller (steeper) by
 $\sim$~0.3 near few times 10$^{\qqt 15}$~eV.
 There is a question of CR mass composition change
 in a region of power index change and above.
 The energy dependence of this changing is
 very important for studies of propagation properties,
 acceleration mechanisms and CR source problem.
 \item
 Observation of \lq \lq secondary" CR particles energy
 spectra, i.e. particles originated in hadronic collisions
 of high energy CR with interstellar medium.
 These studies provide information on propagation
 properties of CR. The most important examples
 are studies of anti-proton flux, positron flux,
 $^{10}$Be, $^{54}$Mn and few other long--life radioactive 
 isotope fluxes, $^{3}$He, $^{2}$H and $^{3}$H fluxes.
 \end{enumerate}
 \item {\bf Diffuse CR anisotropy measurements}.
 Generally CR bombard uniformly the Earth's atmosphere.
 The observed anisotropies are on the level lower than
 10$^{\qqt -3}$. (Some anisotropies at low energy CR might
 be due to interplanetary propagation, and these are
 not considered here). The anisotropy studies might provide
 information on the nearby magnetic field configuration
 in the Galaxy.\\
 Anisotropy due to high energy photons from the Milky Way
 direction 
 (from interactions of nuclear CR in the Galaxy) 
 is predicted to be on the level of $\sim$5$\cdot$10$^{\qqt -5}$ 
 \cite[{\em Berezinsky et al., 1993}]{Gaisser:gamy}
 and experimentally observed the first harmonic
 amplitude of (12.7$\pm$1.2)$\times$10$^{\qqt -4}$ or
 excess within $\pm$10$^{\circ}$ in galactic latitude
 of 4$\cdot$10$^{\qqt -4}$ (1.5$\sigma$)
 \cite[{\em Alexeenko et al., 1993}]{BASA:Gal}.
 \item {\bf Time variability of diffuse CR intensity}.
 Generally diffuse CR flux seems to be very stable in time.
 Observations of \lq \lq cosmogenic nuclei"
 might provide information of CR variability over
 last $\sim$10$^{\qqt 5}$~years.
 Possible observation of large excess of CR flux
 about 3$\cdot$10$^{\qqt 4}$~years 
 ago~\cite[{\em Kocharov et al., 1990}]{kocz:1990}
 might be interpreted as being due to shock acceleration
 mechanism during the passage of a supernova remnant shock wave
 at that 
 time
 \cite[{\em Szabelska, Szabelski, Wolfendale, 1991}]{js:koczarov}.
 \end{enumerate}
 
 \noindent
 Studies of some areas of cosmic ray origin problem are presented
 in this work. The selection reflects the author's interest,
 and sometimes author's contribution (made in collaborations).
 \begin{itemize}
 \setlength{\itemsep}{-2pt}
 \item
 First experiments in gamma ray astrophysics are briefly presented. 
 Then the state of $\gamma$--ray point sources
 investigation is presented. 
 The next section
 is devoted to the studies of diffuse 
 emission of $\gamma$--rays, i.e. due to interaction of cosmic
 rays with the interstellar matter. Author's contribution
 is underlined.
 This is an area which develops very rapidly. 
 Some knowledge we had $\sim$10 years ago is compared
 with the present situation.
 \item
 Studies of CR with energies 10$^{\qqt 14}$~--~10$^{\qqt 17}$~eV
 are presented 
 in the Chapter~\ref{GRUPY-MIONOW}.
 This is the area of extensive air shower (EAS) physics.
 Interpretation of experimental results requires large Monte--Carlo
 computer simulations.
 The main target of these studies are high energy 
 nuclear interaction properties and CR mass composition end energy
 spectrum.
 Author's selected research in this area 
 relates to high energy muon groups
 as the promising method for studies of these problems.
 Results of computer simulations and experimental measurements
 are presented.
 \item
 Chapter~\ref{UHE-Chapter}
 is devoted to the highest energy cosmic ray
 studies. 
 The measurements of anisotropy are discussed. 
 The experimental search of the highest energy CR sources is presented. 
 Author's contribution is underlined.
 \end{itemize}
 
 \newpage
 
 \section{Gamma ray astrophysics}
 \subsection{Introduction}
 In this section some studies of cosmic photons with
 energies above 30~MeV will be presented.
 These $\gamma$--rays are almost unabsorbed in the interstellar
 matter. Their directions point to the places where they were produced.
 The known production mechanisms require energetic particle 
 (e$^{+}$, e$^{-}$ or nucleus). Therefore while studying
 these $\gamma$--rays one learns about cosmic rays at
 distant places.\\
 All observations of $\gamma$--rays made so far show
 approximately (in large energy scale) 
 power law decrease of $\gamma$--ray intensity
 with photon energy. 
 The detector should have large enough area
 to be able to register higher energy cosmic $\gamma$--ray photons. 
 The Earth atmosphere is not transparent to $\gamma$--rays.
 Observations outside the atmosphere are currently limited
 to $\gamma$--ray energy below $\sim$30~GeV.
 For satellite measurements a
 very efficient technique of discrimination of charged particle background
 has been developed, but
 energy measurement might require large and heavy devices.\\
 Experimental measurements of higher energy $\gamma$--rays
 are located on the ground at mountain altitude.
 Energetic photon interacts in the atmosphere 
 and produces electro--magnetic (E--M) cascade which consists
 of electrons, positrons and gammas (however, there is a chance
 of photo--production -- $\gamma$--nucleus interaction
 with hadron products, in the first or subsequent interaction).
 E--M cascades produced by cosmic $\gamma$--rays of energy
 below 30~TeV (= 3$\cdot$10$^{\qqt 4}$~GeV = 3$\cdot$10$^{\qqt 13}$~eV)
 have no chance of being detected on the ground. 
 However,
 some of e$^{+}$ and e$^{-}$ from these E--M cascades are
 faster than local phase velocity of light in the atmosphere
 and produce Cherenkov radiation which can be detected
 on the ground.
 Observation can be made at mountain altitudes during
 clear moonless nights.
 The lower $\gamma$--ray energy is limited by the strength  
 of the Cherenkov light signal 
 as compared
 with the dark--sky background
 (including stars in the field of view) and is equal
 to about 200~GeV at present.
 The upper limit is about 30~TeV due to statistics
 (falling intensity of $\gamma$--rays with energy).\\
 $\gamma$--rays of energy above 10$^{\qqt 14}$~eV can produce
 E--M cascades which reach the ground at mountain altitudes.
 Therefore they can be observed by 
 Extensive Air Shower (EAS) 
 arrays.
 
 \noindent
 Only satellite measurements 
 can clearly discriminate photon observation from
 nucleus induced one in a single event.
 Thus the satellite measurements of $\gamma$--rays
 allow to make observations of wide angle structures
 on the sky, and allow to place the upper limits
 for $\gamma$--ray fluxes from low intensity
 directions, like diffuse extragalactic flux.\\
 
 \begin{figure}[ht]
 \begin{center}
 \vspace{-0.6cm}
 \mbox{
 \psscalefirst
 \psfig{file=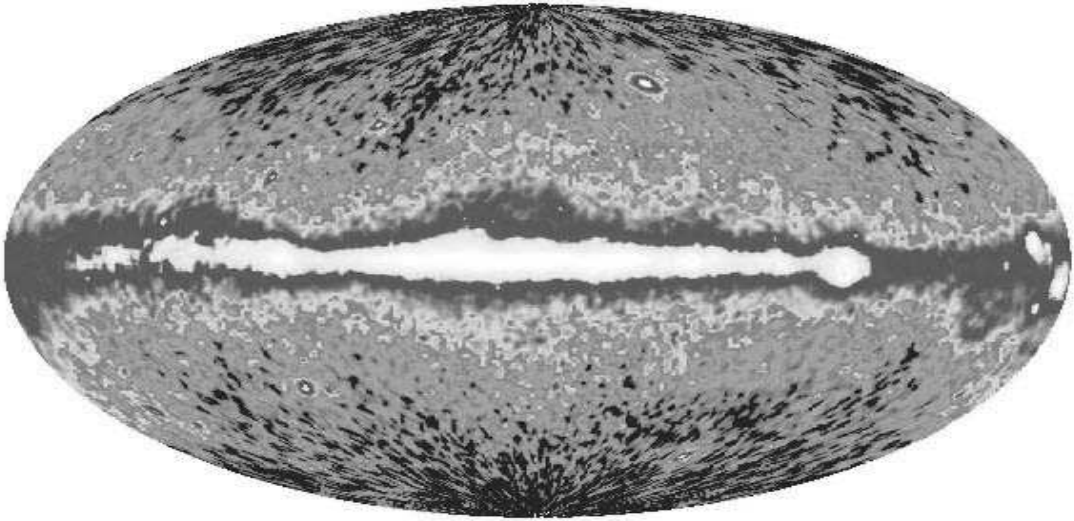,width=14.5cm,height=7.0cm}
 }\\
 \end{center}
 \vspace{-0.5cm}
 \caption[Compton Gamma Ray Observatory EGRET 
 detector intensity map (E$_{\qqt \gamma} >$ 100~MeV).]
 {The contour map
 of $\gamma$--ray intensity (E$_{\qqt \gamma} >$~100~MeV) observed 
 by the EGRET detector of Compton Gamma Ray Observatory.
 The map is in galactic coordinates, galactic centre in at the 
 centre, the galactic longitude goes left from the centre 
 (0$^{\circ} \rightarrow$~180$^{\circ}$);\\ 
 Crab 
 ({\em l$_{gal}$}~=~184.54$^{\circ}$, {\em b$_{gal}$}~=~--5.88$^{\circ}$)
 and
 Geminga
 ({\em l$_{gal}$}~=~195.12$^{\circ}$, {\em b$_{gal}$}~=~--4.27$^{\circ}$)
 are clearly seen at the right of the figure,
 and
 Vela
 ({\em l$_{gal}$}~=~263.52$^{\circ}$, {\em b$_{gal}$}~=~--2.78$^{\circ}$)
 is a white spot at a quarter right from the centre.
 The galactic plane
 is clearly seen. 
 (Map obtained 
 via computer network from NASA GSFC Science Support Center,
 April 1997)}
 \label{EGRET:Imap}
 \end{figure}

 \subsubsection{Direct $\gamma$--ray measurements.}
 $\gamma$--rays with energies $\sim$30~--~1000~MeV were measured
 in the past by two experiments: 
 \begin{itemize}
 \setlength{\itemsep}{-2pt}
 \item
 SAS--II (Nov.~19,~1972 -- June~8,~1973, due to \lq \lq a failure of
 a capacitor on the input portion of the low--voltage power supply") 
 \cite[{\em Fichtel et al., 1975}]{SAS-II},
 \cite[{\em Fichtel et al., 1978}]{SAS-II:tables}
 and 
 \item
 COS~B (1975 -- 1982)
 \cite[{\em Hermsen, 1980}]{Hermsen:PhD},
 \cite[{\em Strong et al., 1987}]{COS-B:database}.
 \end{itemize}
 At present 
 \begin{itemize}
 \setlength{\itemsep}{-2pt}
 \item
 the EGRET instrument (April 25, 1991 --~present time)
 \cite[{\em Kanbach et al., 1988}]{EGRET:project},
 \cite[{\em Thompson et al., 1992}]{EGRET:calib}
 on the NASA Compton Gamma Ray 
 Observatory (CGRO) measures $\gamma$--rays with energies
 $\sim$30~--~$\sim$40000~MeV
 (however, the detection of one photon above 100~GeV has been reported
 in the Compton Observatory Science Report no~164 
 of the 2$^{\qqt nd}$ of August 1994).
 \end{itemize}
 One can add to the list
 \begin{itemize}
 \setlength{\itemsep}{-2pt}
 \item
 the unsuccessful French--Soviet--Polish experiment
 GAMMA--1
 (\cite[{\em Agrinier et al., 1987}]{GAMMA-1:86}).
 \end{itemize}
 The principle of $\gamma$ detection is very similar in all these
 experiments. 
 The basic idea for this detector is
 to register the electron and positron tracks, the pair
 produced inside the detector by the energetic photon.
 The multilayer spark chambers are used for this.
 $\gamma$--ray energy is measured in calorimeter as
 a sum of energy deposited by e$^{+}$ and e$^{-}$.
 Thousands times larger
 proton flux is to be eliminated by the veto signal from
 shielding scintillation counters.\\
 The angular resolution plays important role
 in data interpretation.
 For the photon direction
 determination the most important is the estimation of 
 e$^{+}$ and e$^{-}$ tracks directions near to their
 place of origin. In about half of centimeter wide
 spark chambers of COS~B experiment the e$^{+}$e$^{-}$
 directions were estimated from spark position
 in subsequent chambers. There were problems with
 Coulomb scattering in spark chambers \lq horizontal' walls.\\
 The GAMMA--1 experiment had 3~cm wide spark chambers
 and it was possible (during calibration) to determine 
 e$^{+}$ and e$^{-}$ directions within one gap (there were
 3~scanning levels within each gap).
 Wide gaps of spark chambers required very high voltage
 supply and this created problems with electronic disturbance.
 Another problem is gas flushing of the spark chamber,
 and the gas storage limits the lifetime of the experiment.
 To improve angular resolution in the GAMMA--1 detector
 there were plans to use a special mask
 above the spark chambers to determine the entry point
 of $\gamma$--ray photon with still better precision
 (but this would decrease the effective area of detector nearly
 twice).
 The estimated
 accuracy of one $\gamma$--ray photon direction determination was 
 $\sim$2$^{\circ}$ at 100~MeV and
 $\sim$1$^{\circ}$ at 500~MeV
 \cite[{\em Akimov et al., 1987}]{GAMMA-1:Moskwa}
 end energy resolution 40\%~--~60\%
 \cite[{\em Akimov, Szabelski et al., 1990}]{js:GAMMA-1}
 (pre--flight calibration). 
 GAMMA--1 was specially designed to search and observe
 $\gamma$--ray point sources, but the failure of high voltage
 power supply for spark chambers eliminated this experiment.\\
 The EGRET detector has electronic read--out from spark chambers.
 The angular resolution of incident photon direction
 depends on its energy, position in the chamber and relative direction
 to detector's optical axis. The in--flight calibration
 \cite[{\em Thompson et al., 1992}]{EGRET:calib}
 (from Crab observation) give folowing values of
 HWHM (half width at half maximum): 
 4.4$^{\circ}$, 2.5$^{\circ}$, 1.1$^{\circ}$, 0.7$^{\circ}$ 
 and 0.4$^{\circ}$
 for energy ranges:
 30~--~70~MeV, 70~--~150~MeV, 150~--~500~MeV, 500~--~2000~MeV and
 2000~--~20000~MeV
 respectively.
 The corresponding values for the angular radius containing
 67\% of photons are:
 8.4$^{\circ}$, 5.6$^{\circ}$, 3.1$^{\circ}$, 1.6$^{\circ}$ 
 and 0.6$^{\circ}$.
 The energy resolution is 20\%~--~30\% (pre--flight calibration).\\

 \noindent
 Subsequent experiments were 
  
 \begin{itemize}
 \setlength{\itemsep}{-2pt}
 
 \item bigger in size (geometric factor) to register
 larger number of gammas,
 \item longer lasting (with exception of GAMMA--1, which was scheduled
 for 1--2~years, only),
 \item more efficient: SAS--II has registered 12500 gammas, COS~B -- 209537,
 and the EGRET of CGRO is registering photons with a rate
 $\sim$0.01~Hz of effective time.
 \end{itemize}
 Together with other features each of the steps 
 SAS--II $\rightarrow$ COS~B $\rightarrow$ CGRO EGRET
 was a milestone of the $\gamma$--ray astrophysics.\\
 Results of all--sky survey 
 of the CGRO mission obtained by the EGRET detector and
 reduced to the intensity Aitoff plot is presented
 in the Figure~\ref{EGRET:Imap}.\\

 \subsubsection{Cherenkov light detectors of TeV range $\gamma$--rays.
 \label{sect:Cherenkov}}
 $\gamma$--rays with energy around 1000~GeV 
 ($\equiv$~10$^{\qqt 12}$~eV $\equiv$~1~TeV)
 are measured by the ground based observation
 of Cherenkov light in the atmosphere. 
 Energetic photon produces an electro--magnetic
 cascade of gammas, electrons and positrons in the
 atmosphere. The cascade originates high in the atmosphere
 (6~--~50~km),
 develops in number of particles by subsequent cascading
 and dies out before reaching the ground.
 Energetic electrons and positrons
 run faster than light in the atmosphere
 and produce Cherenkov light. 
 This light
 undergoes various losses in the atmosphere.
 Part of it is lost due to
 mirror reflection
 and photomultiplier efficiency
 (e.g. for brief review see
 \cite[{\em Attallah, Szabelski et al., 1995}]{JS:ozone}).
 The signal
 can be observed on clean, moonless, dark nights
 using specially designed clusters of mirrors.
 The detection technique often involves very sophisticated
 and advanced technology.\\
 The lower $\gamma$--ray energy limit for these observations
 is about 200~GeV, i.e. $\sim$5~times above the
 CGRO limit (there are large efforts to reduce this gap from 
 both satellite and ground--Cherenkov sides, e.g. planned CELESTE
 Cherenkov experiments would have lower $\gamma$--ray energy
 limit about 20~GeV whereas the future satellite experiments
 $\gamma$--AMS and GLAST would have upper 
 $\gamma$--ray energy limit about 100~GeV
 \cite[{\em Dumora et al., 1996}]{Celeste}).\\
 
 \noindent
 There are two different types of Cherenkov light experiments:
 first is using \lq \lq imaging" method and
 the other 
 \lq \lq wave front sampling" method.
 \begin{itemize}
 \setlength{\itemsep}{-2pt}
 \item In the \lq \lq imaging" method the mirror (or set of submirrors)
 can reflect the light to the set of photomultipliers
 (or other light detectors). Light from one direction is reflected
 to one phototube and from another direction to another tube.
 Therefore one gets information about the angular distribution
 of Cherenkov photons (usually) at one place.
 The newest detector of this kind, the CAT imaging
 telescope placed at Themis site in the French Pyrenees,
 has 558 small phototubes spaced by 0.12$^{\circ}$~=~2.1~mrad
 in the central, physically most important part
 (\cite[{\em Punch et at., 1995}]{Punch:CAT}).\\
 This method is used in Durham group telescopes
 \cite[{\em Brazier et al., 1989}]{Durham:teles}, 
 Whipple telescopes
 \cite[{\em Weekes et al., 1989}]{Crab:Whipple_89}, 
 \cite[{\em Vacanti et al., 1991}]{Crab:Whipple_91},
 and many others.

 \item In the \lq \lq wave front sampling" method there is a number
 of mirrors (each with one phototube) spread over the
 field of size of the order of 10$^{\qqt 5}$~m$^{\qqm 2}$. All mirrors
 point to the same direction, and their angular acceptance
 is of the order of 10$^{\qqt -3}$~sr. 
 The arriving times of the front of Cherenkov light signal measured 
 at many detectors are used to determine the direction of an event.
 Amplitude can be used to estimate the primary energy.\\
 The THEMISTOCLE in French Pyrenees is the largest detector
 of this kind
 \cite[{\em Baillon et al., 1993}]{THEMISTOCLE}.
 \end{itemize}
 
 \noindent
 The main physical problem 
 in TeV $\gamma$--ray astronomy
 is that many times larger
 background is produced by similar cascades initiated
 by cosmic ray protons (and probably electrons)
 which are more numerous than CR photons.\\
 All these are \lq tracking' detectors not capable of
 measuring diffuse $\gamma$--ray radiation.
 In the search for $\gamma$--ray sources in the TeV energy region
 via atmospheric Cherenkov light observations
 the crucial parameter is the ability of suppression
 of showers initiated by nuclear component of CR in the bulk
 of observed events.
 This is largely achieved due to
 good angular resolution of event direction which is much better than
 in satellite experiments. 
 Most Cherenkov light $\gamma$--ray experiments can
 identify the shower direction with accuracy about 2~mrad
 ($\sim$0.1$^{\circ}$). Good angular resolution largely
 reduces the background from galactic CR in very narrow
 cone around the observed $\gamma$--ray source.\\
 
 \noindent
 Along with the angular resolution 
 special selection criteria of events are used
 in most Cherenkov experiments
 in order to increase the ratio 
 of events of electro--magnetic origin
 to the events of hadronic origin.
 In the \lq \lq imaging" method this is often a preference of the 
 events which gave a narrow angular image of Cherenkov photons
 observed (suppresses the number of hadronic events for which
 wide angular image is more likely). 
 In the \lq \lq wave front sampling" method this could be
 a preference of the events which gave a narrow spread
 in lateral distribution of the signal amplitudes or/and
 a narrow spread of the cone--like front of Cherenkov light.
 These methods are not very efficient.

 \subsubsection{$\gamma$--ray detections in air shower arrays.}
 Large effort has been made to observe
 photons of still higher energy. 
 Electro--magnetic (E--M) cascade produced by 
 $\gamma$--ray with energy more than $\sim$10$^{\qqt 14}$~eV
 can reach the ground.
 Great number of gammas, electrons and positrons would trigger 
 the EAS array.\\
 The EAS array is usually a set of charged particle detectors
 distributed
 on the ground. The detectors are separated by 
 $\sim$10~--~$\sim$300~m. A signal from a few detectors
 within a short time indicates the coherent event, EAS.
 Therefore, this technique allows to sample particles 
 from EAS.
 The method is quite old, but for the search for energetic 
 $\gamma$--rays, under the name \lq \lq search for high energy
 cosmic ray sources" 
 few new EAS arrays were specially
 designed and built recently:
 \begin{itemize}
 \setlength{\itemsep}{-2pt}
 \item
 HEGRA 
 \cite[{\em Aharonian et al. 1995}]{HEGRA:1995}, 
 \item
 CYGNUS 
 \cite[{\em Allen et al. 1995}]{CYGNUS:1995}, 
 \item
 CASA--MIA
 \cite[{\em Cronin et al., 1992}]{CASA-MIA},
 \item
 SPASE
 \cite[{\em van Stekelenborg et al., 1993}]{SPASE},
 \item
 TIBET 
 \cite[{\em Amenomori et al., 1995}]{Tibet-II})
 \end{itemize}
 with a very accurate EAS direction determination which allows
 to see \lq \lq the shadow of the Moon" (which is a half of
 degree in diameter).
 Most former EAS arrays were additionally equipped
 with precise clocks and other devices to search
 for \lq point sources'.

 \subsection{Gamma ray sources}
 There are three well known physical processes in which 
 $\gamma$--ray of energy above 10~MeV may be produced,
 all involve an energetic particle, 
 namely through $\pi^{\circ}$ decay 
 ($\pi^{\circ}$ as a product of nucleon--nucleon interaction),
 electron bremsstrahlung and inverse Compton scattering.
 Therefore one can expect that $\gamma$--ray sources are also
 cosmic ray sources (as a source of energetic particles necessary
 for $\gamma$--ray production).
 However, an increase of interstellar matter column density
 in particular direction would also be seen in $\gamma$--rays
 as a source in this direction, 
 since two of $\gamma$--ray production mechanisms
 give more gammas when there is more target material.\\ 
 The name \lq \lq gamma ray source" means that there is
 an excess of observed $\gamma$--ray flux from this
 particular direction. The excess could be indentified
 in DC signal (direct current) or due to periodicity
 analysis. The DC excess means that the number of 
 photons observed from this direction can be reasonably
 well seen above observed or predicted background level.
 For variable objects further information about time 
 variability of $\gamma$--ray flux helps to identify
 the source. For weak sources (very small excess
 above the background) or in a case where the background
 level is difficult to estimate (e.g. for many tracking
 detectors), the knowledge of the variability period 
 is crucial for source identification.

 \subsubsection{\label{gamma_sources}
 $\gamma$--ray energy range 30~--~1000~MeV}
 The following list contains results of satellite
 experiments in search for $\gamma$--ray point sources.
 (There were also balloon--borne observations
 which provided very interesting results, but
 they are not listed here.)
 
 \begin{figure}[t]
 \begin{center}
 \mbox{
 \psfig{file=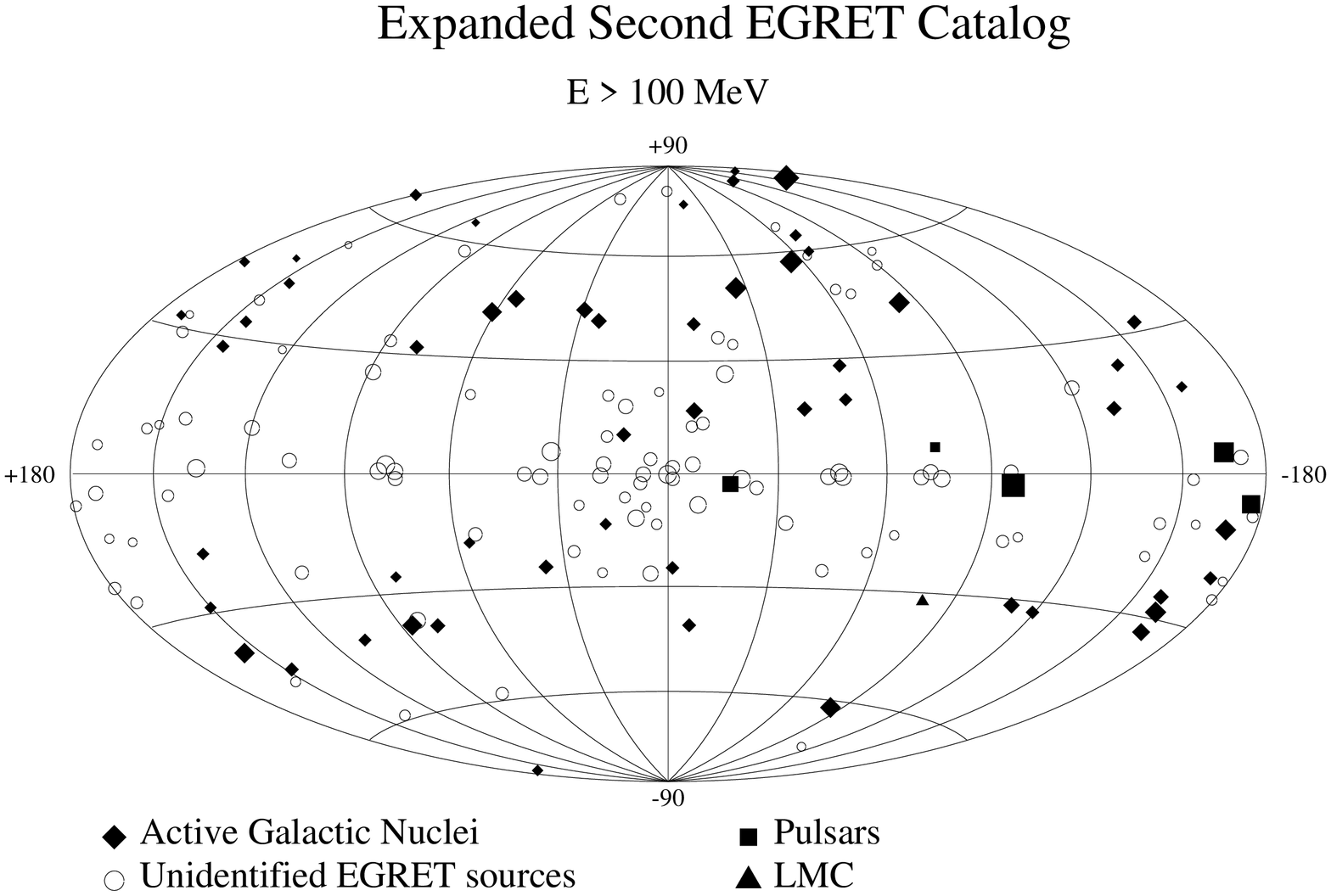,width=14.5cm,height=10.0cm,angle=90}
 }\\
 \end{center}
 \caption[Second EGRET catalogue of $\gamma$--ray sources.]
 {Second EGRET catalogue of $\gamma$--ray sources.
 (Map obtained 
 via computer network from NASA GSFC Science Support Center,
 1997).}
 \label{EGRET:2-nd}
 \end{figure}
 
 \begin{enumerate}
 \setlength{\itemsep}{-2pt}
 \item SAS--II. There were 3 clearly seen sources:
 Vela pulsar, Crab pulsar and Geminga. The large diffuse flux
 from the Milky Way was observed.
 \item COS~B.
 There were 3 clearly seen sources in the galactic plane:
 Vela pulsar, Crab pulsar and Geminga.\\
 The pulse identification of Vela (PSR0833$-$45)
 and Crab (PSR0531$+$21) were made during the flight.
 The periodicity of Geminga (2CG195$+$04)
 was found recently
 in agreement with X--ray data (ROSAT: \mbox{1E0630$+$178})
 (\cite[{\em Halpern and Holt, 1992}]{Rosat:Geminga})
 and CGRO data
 \cite[{\em Bignami and Caraveo, 1992}]{COSB:Geminga}.\\
 One COS~B source is extragalactic: 
 identified as 3C~273 in the COS~B catalog
 (but it probably contains two unresolved sources: 
 3C~279 and 3C~273, as follows from 
 the EGRET Source Catalog
 \cite[{\em Thompson et al., 1995}]{EGRET:IstCatalog}).\\
 The COS~B catalog of $\gamma$--ray sources 
 contained 25 directions of $\gamma$--ray flux excess and
 only 4 identifications with known astronomical objects.
 It is presented in the Appendix A: Table~\ref{tab:COSB} on 
 the page~\pageref{tab:COSB}
 (the 2CG Catalog \cite[{\em Swanenburg et al., 1981}]{COSB:2CGcat}).
 It is worth noticing that COS~B has not made a full sky
 survey.
 \item GAMMA--1, although with faulty spark chamber,
 has identified periodical signals from the Vela
 pulsar \cite[{\em Akimov et al., 1991}]{GAMMA-1:Vela}
 \item Compton Gamma Ray Observatory (CGRO) has found
 many new sources (see Figure~\ref{EGRET:2-nd} for map of EGRET
 sources).
 Vela and Crab were confirmed,
 the EGRET detector (30--5000~MeV) has found
 the periodic emission from Geminga (237~ms)
 in agreement with the ROSAT data
 \cite[{\em Bertsch et al., 1992}]{CGRO:Geminga}
 (however, the COMPTEL detector (0.7--30~MeV) on the CGRO has not
 seen this pulsation of Geminga
 (\cite[{\em Hermsen et al., 1993}]{COMPTEL:VelaGem}).
 New galactic $\gamma$--ray pulsars were found:
 \begin{itemize}
 \setlength{\itemsep}{-2pt}
 \item
 PSR1509$-$58 (below 2~MeV, i.e. only due to the COMPTEL detector) 
 \cite[{\em Bennett et al., 1993}]{COMPTEL:pulsary},
 \item
 PSR1706$-$44 \cite[{\em Thompson et al., 1992}]{CGRO:PSR1706}
 \item
 and PSR1055$-$52 
 \cite[{\em Thompson et al., 1995}]{EGRET:IstCatalog}.
 \end{itemize}
 Parameters of pulsars identified as $\gamma$--ray pulsars
 are presented in the Appendix A:
 Table~\ref{tab:PSR_EGRET} on the page~\pageref{tab:PSR_EGRET}.\\
 Many unidentified galactic $\gamma$--ray sources were observed.
 After Phase~3 of the mission 
 \lq \lq The Second EGRET Catalog of High--Energy Gamma--Ray Sources"
 \cite[{\em Thompson et al., 1997}]{EGRET:IIndCatalog}
 has 36 galactic ($b_{gal}~<~6^{\circ}$)
 sources with significance more than 5~$\sigma$,
 and 11 have significance bigger than 10~$\sigma$ at least in
 one observation period.\\
 Astrophysically the most interesting were discoveries 
 of extragalactic $\gamma$--ray sources, but
 we will concentrate on galactic sources for CR
 origin studies.
 \end{enumerate}
 
 \noindent
 The identified $\gamma$--ray galactic sources 
 (all but Geminga)
 are radio pulsars.
 They have the same periodicity in $\gamma$--rays as
 in other E--M frequences observed.
 Crab pulsar (period of 33~msec) 
 light curve has been observed in radio waves,
 optically, in X--rays and in $\gamma$--rays. 
 In all four modes Crab pulsar light curve
 has similar two peak structure at the same phase.\\
 However, in the other 4 identified EGRET $\gamma$--ray pulsars,
 $\gamma$--ray pulse has different phase than the pulse in radio waves.
 The shapes of light curves are also quite different for different
 wave frequences
 (but the periods are the same, of course).\\

 \noindent
 Geminga has not been observed as a radio pulsar and its variability
 was observed first in X--rays 
 \cite[{\em Halpern and Holt, 1992}]{Rosat:Geminga}
 and then confirmed in $\gamma$--rays.
 Geminga seemed to be a very near object.
 It was even suggested that it is as near as 30~pc away
 \cite[{\em Bailyn, 1992}]{Geminga:Nature},
 $\gamma$--ray data suggest an upper limit of the distance
 to be $\sim$380~pc 
 \cite[{\em Bertsch et al., 1992}]{CGRO:Geminga},
 and the recent estimations using the Hubble Space Telescope
 put Geminga at 
 157~(+57~--34)~pc (parallax observations by
 \cite[{\em Caraveo et al., 1996}]{Geminga:parallax}).

 \subsubsection{$\sim$1~TeV gamma ray search for sources}
 In the search for $\gamma$--ray sources in the TeV energy region
 via atmospheric Cherenkov light observations
 the crucial parameter is the ability of suppression
 of showers initiated by nuclear component of CR in the bulk
 of observed events.
 This is largely achieved due to
 good angular resolution of event direction which is much better than
 in satellite experiments which observe $\gamma$--rays
 in 30~MeV~--~30~GeV energy range.
 Most Cherenkov light $\gamma$--ray experiments can
 identify the shower direction with accuracy about 2~mrad
 ($\sim$0.1$^{\circ}$). Good angular resolution largely
 reduces the background from galactic CR in very narrow
 cone around the observed $\gamma$--ray source.\\
 Most experiments using \lq \lq imaging" method apply
 selection criteria. The idea is to reduce total bulk of events
 to a sample which is relatively enriched in
 $\gamma$ induced events as compared with hadronic events.
 Therefore the \lq \lq source signal" should be better seen.
 The \lq \lq imagers" register the number and directions
 of Cherenkov photons at one place at some distance from
 the Cherenkov shower centre. 
 The pattern 
 (in angular distribution of Cherenkov photon directions)
 has approximately
 eliptic shape with longer axis pointing to the shower centre.
 The principle of the selection relates to the angular spread
 of Cherenkov photon directions along the shorter axis of
 the eliptic shape. The $\gamma$ induced events are expected
 to have smaller spread of Cherenkov photon directions
 than hadron induced ones.\\

 \begin{figure}[h]
 \begin{center}
 \mbox{
 \psfig{file=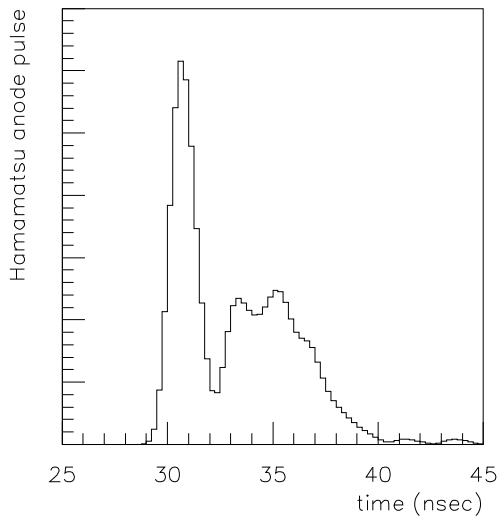,height=7.0cm,width=7.0cm}
 }\\
 \end{center}
 \caption[Time separation of muon signal in atmosheric 
 Cherenkov light detector.]
 {The simulated anode signal
 (arbitrary units)
 for photomultiplier Hamamatsu~H2083 showing
 peak from Cherenkov light generated by muon.
 The gap between \lq muon' signal and 
 \lq E--M' signal clearly separates the two.
 Such a pattern would indicate hadronic origin
 of detected event.
 See
 \cite[{\em Cabot, Szabelski et al., 1997}]{JS:muCher}
 for more details.}
 \label{fig:flashADC}
 \end{figure}
 
 \noindent
 The experiments using \lq \lq front sampling" method
 have not applied selection criteria to enrich
 the $\gamma$ induced events signal.\\
 The new technique of GHz sampling
 of the signal amplitude might provide a new
 method of determining hadronic origin of the event
 (\cite[{\em Cabot, Szabelski et al., 1997}]{JS:muCher}).
 The $\gamma$--ray produces E--M cascade which produces
 Cherenkov light. The Cherenkov light from the whole cascade
 has a characteristic wide--angle cone like front:
 at the plane perpendicular to the primary $\gamma$--ray
 direction Cherenkov photons at the centre are coming first,
 first photons at 50~m from the centre are delayed about 2~nsec,
 first photons at 100~m about 4~nsec etc. In each place
 all Cherenkov photons are coming within about 10~nsec
 after the first one. The times of the front observed
 at many places form a nice space/time cone.
 Most of these photons are produced at 1~--~10~km above 
 the detector.\\
 In case of primary CR proton the situation
 is similar: most of Cherenkov light is produced
 by E--M cascades from decays 
 of short lived hadrons (e.g. $\pi^{\circ}$).
 Also some muons are produced, which are charged
 particles going along straight lines in the air. When they have enough 
 energy they produce Cherenkov light in atmosphere
 (the total contribution is much smaller than from 
 e$^{+}$, e$^{-}$). If the muon fell in the vicinity of
 individual mirror of Cherenkov array,
 the light produced by this muon can be observed.
 This light is produced just above the detector (up to 100~m).\\
 The idea is to observe the time of signals with a nsec
 resolution. In the atmosphere the energetic muon 
 goes faster than light (muon goes with c, straight line,
 whereas e$^{+}$, e$^{-}$ in E--M cascade 
 undergo Coulomb scattering, Cherenkov light goes with
 v~=~c/n, where n is local refraction index).
 At 50~m from the centre the muon signal (Cherenkov light) can
 be about 2~nsec before the front of Cherenkov light
 from E--M cascade.\\
 The detailed Monte--Carlo simulation of cascade development
 in the atmosphere and with fast photomultiplier response
 gives the results presented in the Figure~\ref{fig:flashADC}.
 Observation of a signal, which does not fit to the
 space/time cone of Cherenkov front and has a muon peak
 would indicate the hadronic origin of the event.\\

 \noindent
 The observational atmospheric conditions required for
 Cherenkov light measurements in TeV astrophysics
 (i.e. dark, moonless, clear nights)
 significantly limit time available for observation
 of particular source. Most sources can be 
 visible at night only for few months a year, so
 the number of nights suited for observations are 
 limited.\\
 Below two examples of target of observations are presented.\\
 
 \noindent
 \underline{The Crab pulsar observation.}\\
 Results of the Crab observation 
 by the Whipple Observatory experiment 
 (\lq \lq imaging" method)
 were presented
 in \cite[{\em Vacanti et al., 1991}]{Crab:Whipple_91}.
 After 1808 minutes ($\sim$30~hours or $\sim$1.08$\cdot$10$^{\qqt 5}$~sec)
 of ON~source (and OFF~source) observation of the Crab 499798 (ON)
 and 494722 (OFF) events were observed. After application of
 proton shower suppression data processing the figure was:
 14622 (ON) and 11389 (OFF) with an excess of 3233 or
 20.0~$\sigma$ from the Crab direction.
 From Monte--Carlo simulations of $\gamma$--ray induced showers 
 the effective energy threshold was estimated to be 400~GeV
 and the effective collection area to be equal to
 4.2$\cdot$10$^{\qqt 8}$~cm$^{\qqm 2}$. So the flux from the
 Crab is 
 (7.0$\pm$0.4)$\cdot$10$^{\qqt -11}$~photons~cm$^{\qqm -2}$~sec$^{\qqm -1}$.
 For higher energies, above 4000~GeV, the ratio of
 ON source to OFF source events is higher, exceeding 2,
 although the statistics is much smaller and therefore 
 the observation of source, the Crab, is less significant.
 In the energy range 400~--~4000~GeV the differential
 $\gamma$--ray energy spectrum from the Crab was given by
 \[ 
 \frac{\qqq d~\rm N_{\qqw \gamma}(E)}{\qqq~d \rm E} 
 \, = \,
 2.5\cdot 10^{\qqt -13} \cdot 
 \left( \frac{\qqq \rm E}{\qqq \rm 400~GeV} \right) ^{\qqw -2.4\pm 0.3} 
 \,\,
 \frac{\qqq \rm photons}{\qqq \rm cm^{\qqm 2}\, s\, GeV}  
 \]
 No pulsed emission with the Crab pulsar
 period was observed.\\
 It might be important to notice that the Whipple
 observations were made during 5 months and the
 excess in the Crab ON direction was seen in each
 partial monthly data subset, although not with the same
 strength.\\
 
 \noindent
 \underline{
 Cyg~X--3.}\\ 
 $\gamma$--ray detections from
 the very distant X--ray and radio source 
 ($\sim$10~kpc away from the Solar System)
 with 4.8~h period -- Cyg~X--3 -- were frequently reported
 in the 80's.\\
 The search of periodic signal in COS~B data gave negative results 
 \cite[{\em Bennett et al., 1977}]{CygX-3:COSB}.\\
 The Durham University group 
 using \lq \lq imaging" method in their Cherenkov light experiments
 has reported 
 positive results of Cyg~X--3 observations, and
 they have found $\approx$12.5~msec periodicity in the
 four observations made in 1981, 1982, 1983, at the end of 1985,
 and 1988
 \cite[{\em Brazier et al., 1989}]{Durham:Cyg_rev}.
 However, positive detections of this source were not
 reported recently, nor the $\approx$12.5~msec periodicity
 was confirmed by another experiment.\\

 \subsubsection{Search for gamma ray sources 
 with E$_{\qqt \gamma}~>$~10$^{\qqt 14}$~eV.}
 $\gamma$--ray photon of energy above 10$^{\qqt 14}$~eV (=~100~TeV = 
 10$^{\qqt 5}$~GeV) entering the Earth atmosphere may produce
 an electro--magnetic (E--M) cascade which can reach the ground level.
 Therefore this search is performed using 
 extensive air shower (EAS) detectors --
 ground based array of charged particle detectors.
 There is no convincing experimental evidence for existence of 
 $\gamma$--ray sources at these energies.\\
 Several years ago the search for $\gamma$--ray 
 sources in EAS energy region
 was very popular and fashionable. It could be reasonable because
 such a discovery 
 (which would name the object) 
 would be a milestone in understanding the
 CR acceleration mechanism, efficiency of which exceeds any man--built
 accelerator by many orders of magnitude.
 In the eighties large number of CR point sources 
 \lq \lq had been found" in many EAS array data.
 The most \lq \lq famous" were probably the Crab pulsar
 and the variable radio and X--ray source Cyg~X--3.
 The Cyg~X-3 is a galactic source at the distance of $\sim$10--11~kpc
 from the Sun. It has been \lq \lq seen" in muon flux
 \cite[{\em Marshak et al., 1985}]{CygX3:muony},
 its 4.8~hour X--ray variability 
 \lq \lq was confirmed" by a number of EAS arrays together
 with another faster or slower periodicity. The search
 has been popularized in {\em the Scientific American} 
 \cite[{\em MacKeown and Weekes, 1985}]{CygX3:SciAm}
 and summarized by
 \cite[{\em Nagle et al., 1988}]{VHE:sources}.
 In such an atmosphere few new EAS arrays were specially
 designed and built for the very high energy $\gamma$--ray point
 sources search. 
 HEGRA 
 \cite[{\em Aharonian et al. 1995}]{HEGRA:1995}, 
 CYGNUS 
 \cite[{\em Allen et al. 1995}]{CYGNUS:1995}, 
 CASA--MIA
 \cite[{\em Cronin et al., 1992}]{CASA-MIA},
 SPASE
 \cite[{\em van Stekelenborg et al., 1993}]{SPASE},
 TIBET 
 \cite[{\em Amenomori et al., 1995}]{Tibet-II} 
 have very accurate EAS direction determination.
 None of these experiments confirmed previously \lq \lq observed"
 excess in DC (direct current) flux or pulsed emission
 from the Crab pulsar nor Cyg~X--3. The upper limits set by
 these observations contradict previous \lq \lq observations"
 (e.g. see
 \cite[{\em Borione et al., 1997}]{CasaMia-noCyg}).\\
 
 \begin{table}[t]
 \caption[Crab signal \lq \lq evidence" {\L }\'{o}d\'{z}
 EAS array]
 {\label{tab:Krab}
 The pixel map of the sky area in
 declination range 
 12$^{\circ}$~--~32$^{\circ}$ and 
 zenith angle 20$^{\circ}$~--~40$^{\circ}$ 
 in coordinates:
 the sidereal time vs. the azimuth angle.
 The passage of the Crab pulsar is indicated.
 The table from 
 \cite[{\em Dzikowski et al., 1984}]{Krab_Kosice}
 }
 \begin{center}
 \mbox{
 \psfig{file=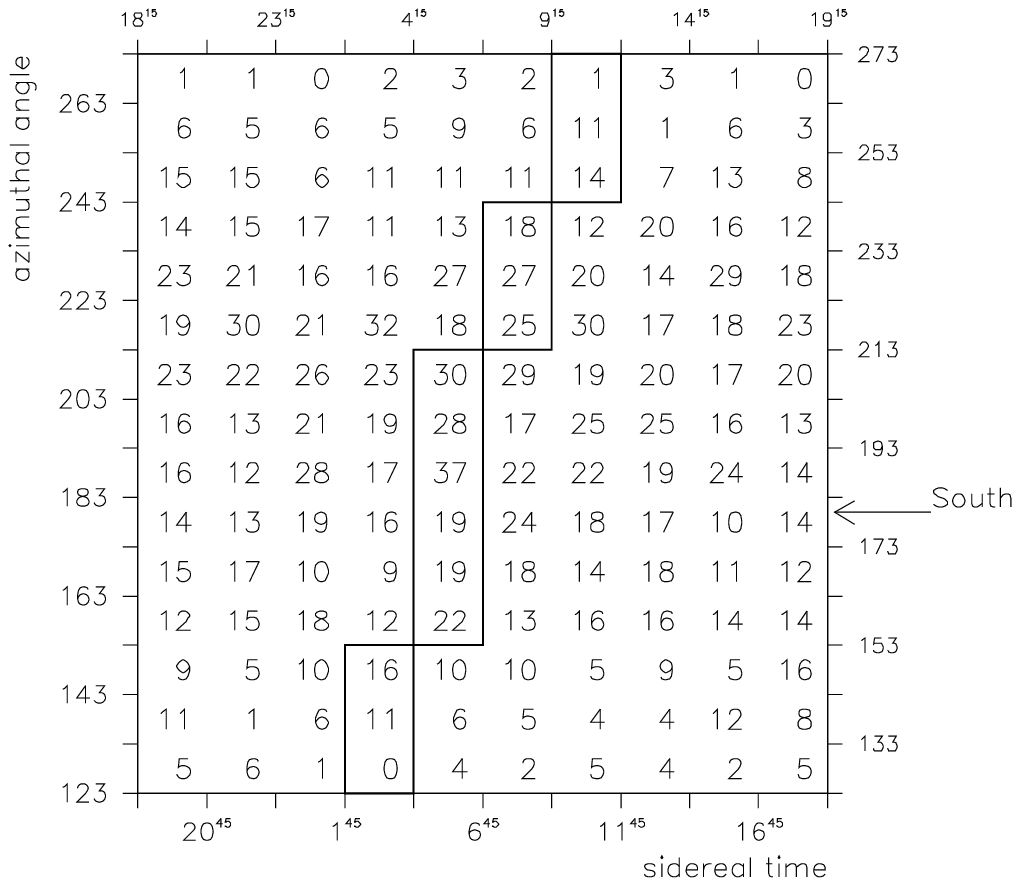,width=10.5cm,height=10.0cm}
 }\\
 \end{center}
 \end{table}
 
 \begin{figure}[t]
 \begin{center}
 \mbox{
 \psfig{file=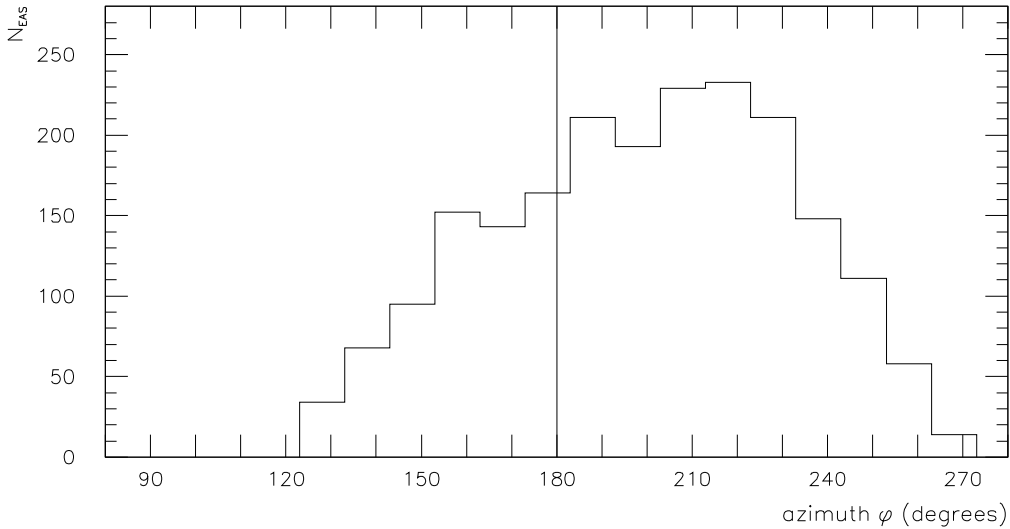,width=14.0cm,height=8.0cm}
 }\\
 \end{center}
 \caption[Erroneous azimuth angle distribution 
 related to the Crab signal search]
 {\label{Lodz:Crab_fi}
 The azimuth angle distribution of EAS presented
 in the Table~\ref{tab:Krab}. EAS observed from
 the sky area in declination range
 12$^{\circ} \leq \delta \leq$~32$^{\circ}$ and 
 zenith angle range 20$^{\circ} \leq \theta \leq $~40$^{\circ}$ 
 contribute.
 Since the EAS intensity is lower at bigger zenith angles
 one would expect the east--west symmetry 
 centered at the south direction.
 The distribution indicates an error in the
 data processing.}
 \end{figure}

 \noindent
 Also the {\L }\'{o}d\'{z} group reported the excess of EAS observed
 from direction of the Crab pulsar 
 \cite[{\em Dzikowski et al., 1981}]{Krab_RS},
 \cite[{\em Dzikowski et al., 1983}]{Krab_JP} and
 \cite[{\em Dzikowski et al., 1984}]{Krab_Kosice}.
 The measurements were made using old EAS array. The excess
 was not confirmed by the data analysis from the larger array
 with computerized data acquisition system. Instead some
 peculiarity was found in earlier results showing excess from the
 Crab pulsar.
 \begin{itemize}
 \setlength{\itemsep}{-2pt}
 \item In  
 \cite[{\em Dzikowski et al., 1984, the Table~1}]{Krab_Kosice}
 the detailed information of the measurement is
 presented (it is reproduced here as Table~\ref{tab:Krab}).
 The number of observed EAS from declination range 
 12$^{\circ}$~--~32$^{\circ}$ and zenith angle 
 20$^{\circ}$~--~40$^{\circ}$ was grouped
 in the pixel map of
 the sidereal time vs. the azimuth angle.
 The Crab pulsar has declination $\delta$~=~21$^{\circ}$59'
 and has the highest position on the sky 
 at sidereal time 5$^{\rm h}$31' (i.e. being at the south,
 azimuth angle $\phi$~=~180$^{\circ}$ in {\L }\'{o}d\'{z}
 at the local time corrected by the time difference to UT).
 The direction of the Crab pulsar falls into some pixels
 in the Table~\ref{tab:Krab} as indicated.\\
 The simplest cross check of that map is the azimuth
 angle distribution (the number of EAS summed over the sidereal
 time) presented in the Figure~\ref{Lodz:Crab_fi}.
 Because of the strong zenith angle dependence of EAS array
 acceptance, the highest position of this declination band 
 should point 
 to the south direction ($\phi$~=~0$^{\circ}$) and
 the azimuthal distribution should have
 east--west symmetry. 
 The asymmetry present in the Figure~\ref{Lodz:Crab_fi}
 indicates a serious error in the data processing.
 \item The distribution of the EAS arrival time
 difference along west--east timing arm was too wide.
 The EAS direction was calculated from two time differences
 (timing) between three scintillation detectors
 placed in the corners of a rectangular triangle.
 Timing distributions for both rectangular arms were presented
 in
 \cite[{\em Dzikowski, 1985, Figure~38}]{TDzik:PhD} and
 agree with later data processing results.
 The hypothetical maximum of the time difference
 corresponds to the distance between the scintillation
 counters divided by the speed of light and equal to 
 15~m~/~0.3~$\frac{\qqt \rm m}{\qqt \rm nsec}$~=~50~nsec and
 28~m~/~0.3~$\frac{\qqt \rm m}{\qqt \rm nsec}$~=~93.3~nsec
 for two timing arms in the case. Recorded distribution (W--E arm)
 in a smooth manner exceeds the limit in both
 (positive and negative) directions indicating
 very serious hardware failure.
 \end{itemize}
 
 \noindent
 Reanalysis of the old data and some new data from the 
 {\L }\'{o}d\'{z} old array
 did not confirm the previously reported excess (the
 above mentioned hardware problem is present 
 in the whole data set and it does not
 permit to use these data to EAS arrival direction analysis).\\
 Analysis of the data accumulated during several years 
 in the new {\L }\'{o}d\'{z} array
 did not confirm the previous positive detections.
 
 \subsection{Galactic diffuse emissivity of gamma rays.}
 Most $\gamma$--rays with energy above 30~MeV
 observed from the Milky Way
 direction originate due to cosmic ray (CR) interaction.
 Three processes can contribute here:
 \begin{itemize}
 \setlength{\itemsep}{-2pt}
 \item $\pi^{\circ}$ decay to 2 $\gamma$--photons;
 $\pi^{\circ}$'s are produced in  CR protons and other nuclei
 collisions with interstellar matter,
 \item CR electron bremsstrahlung in the interstellar matter,
 \item inverse Compton scattering of energetic CR electrons
 on galactic starlight photons and on cosmological microwave
 background photons.
 \end{itemize}
 The observed $\gamma$--ray intensity I$_{\qqt \gamma}$
 can have following components:
 \[
 \begin{array}{lcl}
 {\rm I}_{\qqw \gamma}({\rm E_{\qqw \gamma},l_{gal},b_{gal}}) 
 &  =  &
 \int~\left\{
 \frac{\qqq \rm q(E_{\qqw \gamma},R_{G})}{4 \pi}
 \, \cdot \,
 \left[
 \rm n_{\rm HI}(R_{G}) \, + \, 
 2 \, X'(E_{\qqw \gamma}) \, W_{CO}(R_{G})
 \right]
 \, + \right .\\
 & + & \left .
 \rm I_{IC}(E_{\qqw \gamma},R_{G},l_{gal},b_{gal})
 \right\}
 \, + \\
 & & \\
 &  +  & 
 \Sigma_{i} \, \, I_{i-source}({\rm E_{\qqw \gamma}},l_{gal},b_{gal})
 \, + \, 
 \rm I_{B}(E_{\qqw \gamma}),
 \end{array}
 \]
 where 
 the integral is along the line of sight,
 $\rm q(E_{\qqw \gamma},R_{G})$ is $\gamma$--ray 
 emissivity per hydrogen atom at the distance R$_{G}$ from galactic
 centre (assuming cylindrical symmetry) due to
 $\pi^{\circ}$ decay and electron bremsstrahlung,
 $\rm n_{\rm HI}(R_{G})$ is hydrogen atom density 
 (observed in 21~cm line),
 $\rm W_{CO}(R_{G})$ is CO (carbon monoxide) antenna temperature
 (at wavelength $\sim$2.6~mm related to CO rotational line
 \mbox{\it J = 1 $\rightarrow$ 0}),
 $\rm 2 \, X'(E_{\qqw \gamma}) \, W_{CO}(R_{G})$ is conversion
 CO temperature to H$_{\qqt 2}$ density (molecular hydrogen)
 and to equivalent HI,
 $\rm I_{IC}(E_{\qqw \gamma},R_{G},l_{gal},b_{gal})$
 is inverse Compton contribution,
 $\Sigma_{i} \, I_{i-source}({\rm E_{\qqw \gamma}},l_{gal},b_{gal})$
 is sum of discrete source contribution,
 and
 $\rm I_{B}(E_{\qqw \gamma})$ is an isotropic and experimental
 background
 (see
 \cite[{\em Bloemen, 1989}]{COSB:Bloemen}
 for discussion of these parameters
 and
 \cite[{\em Strong and Mattox, 1996}]{AWS:gradient}
 for currently best values).
 In the galactic plane emissivity $\rm q(E_{\qqw \gamma},R_{G})$
 dominates over
 $\rm I_{IC}(E_{\qqw \gamma},R_{G},l_{gal},b_{gal}=0^{\circ})$.
 In this sense
 diffuse $\gamma$--ray galactic emissivity is closely
 related to the CR distribution in our Galaxy.\\

 \noindent
 The studies of COS~B measurements of galactic 
 $\gamma$--ray emission were summarized in
 \cite[{\em Bloemen, 1989}]{COSB:Bloemen}
 and EGRET measurements in
 \cite[{\em Strong and Mattox, 1996}]{AWS:gradient}.
 In first approximation the diffuse $\gamma$--ray flux observed
 from the Milky Way direction is proportional to the
 column density of interstellar gas. Most of the gas is in the
 form of neutral hydrogen (HI) which is well observed
 in 21~cm radio waves. There is an important contribution
 to gas column density in galactic plane from the
 molecular hydrogen (H$_{\qqt 2}$). In low temperature regions of
 interstellar matter
 these molecules can not be observed. 
 Instead, rotation emission lines
 from carbon monoxide were measured, since CO molecules are
 excited due to collisions with H$_{\qqt 2}$ molecules. 
 The conversion factor from CO observations to H$_{\qqt 2}$
 column density 
 and further to HI equivalence
 depends on the gas distribution model, particular molecular cloud
 temperature, density etc. 
 Therefore the value
 $\rm X'(E_{\qqw \gamma})$ 
 could not be obtained from model calculations. 
 Its value \mbox{
 (1.9$\pm$0.2)~$\cdot$~10$^{20}~\frac{\qqq \rm mols}{\qqq \rm cm^{\qqm 2}}
 \left( \frac{\qqq \rm K~km}{\qqq \rm sec} \right)^{\qqm -1}$}
 was obtained from gamma ray data analysis
 \cite[{\em Strong and Mattox, 1996}]{AWS:gradient}.

 \noindent
 The ratio of diffuse $\gamma$--ray flux 
 to the interstellar gas column density
 is called the average $\gamma$--ray emissivity in this direction.
 In the galactic plane directions, 
 where the $\gamma$--rays are produced mostly
 in CR interactions with interstellar gas, the $\gamma$--ray emissivity
 is directly related to the CR intensity along the line of sight.
 This statement can be transformed to the form: 
 variation of $\gamma$--ray emissivity in the Galaxy would indicate
 variation of CR intensity 
 (studies of variability of CR electron intensity,
 which use radioastronomy observations of
 diffuse synchrotron radiation, are complicated, because
 the process of synchrotron emission depends on
 the galactic magnetic field).\\
 In the opposite option CR would be of universal origin,
 and the CR intensity would be everywhere the same 
 (i.e. the same in extragalactic space as inside the Galaxy).
 Still the $\gamma$--ray flux would be variable 
 (according to gas column density distribution)
 but the $\gamma$--ray emissivity is expected to be constant.\\
 Therefore variation of CR intensity in the Galaxy 
 would indicate the galactic origin of CR
 and the distribution
 of CR intensities could be correlated with distribution of
 many galactic objects 
 in a search of candidate for CR source.
 
 \subsubsection{Diffuse galactic $\gamma$--rays with energy above
 30~MeV.}
 Studies performed to find large variation of $\gamma$--ray
 emissivity failed.
 However, a small radial gradient of $\gamma$--ray emissivity
 with the galactic centre distance has
 been observed. Let the local $\gamma$--ray emissivity
 (Solar System at 8--10~kpc from the galactic centre)
 be the reference value 
 (1~kpc~= 3.08~10$^{\qqt 21}$~cm).
 \begin{enumerate}
 \setlength{\itemsep}{-2pt}
 \item
 At galactic radial distances 5~kpc 
   \begin{itemize}
   \setlength{\itemsep}{-2pt}
   \item Bloemen
      found 1.7--3~times
      more CR electron component intensity at 5~kpc than the local value,
      and practically no difference in CR proton component
      \cite[{\em Bloemen, 1985 p. 199}]{Bloemen:Phd},
   \item others argued 
      for 2 times higher CR intensity at 5~kpc than the local value
      \cite[{\em Bhat et al., 1985}]{BhatAWW:gradient},
  \item
      later, \cite[{\em Bloemen, 1989}]{COSB:Bloemen} presented
      50\% higher CR intensity
      and finally,
  \item
     from EGRET data
     it was shown that CR emissivity 
     (for $\gamma$--rays above 100~MeV -- mostly due to CR protons)
     at the distance of
     $\sim$5~kpc from the galactic centre is $\approx$~20\% 
     higher than the local value 
     \cite[{\em Strong and Mattox, 1996}]{AWS:gradient}.
  \end{itemize}
 \item
  Looking in the periphery of the Galaxy: 
  till 14--16~kpc from the galactic centre
   \begin{itemize}
   \setlength{\itemsep}{-2pt}
   \item Bloemen found no
     change in CR proton density and about 50\% less CR electron
     density than the local value
     \cite[{\em Bloemen, 1985 p. 199}]{Bloemen:Phd},
   \item we have
     found about 30\% less $\gamma$--ray emissivity above 300~MeV
     (due to CR protons) than local value,
     \cite[{\em Mayer, Szabelski et al., 1987}]{js:CRgrad},
    and 
   \item this is consistent with 
      final COS--B data analysis presented later in
      \cite[{\em Bloemen, 1989}]{COSB:Bloemen} and, 
   \item   
      EGRET results
      \cite[{\em Strong and Mattox, 1996}]{AWS:gradient}.
   \end{itemize}
 \item
   The galactic centre region at $\sim$3~kpc
   shows a similar or even a little bit smaller 
   emissivity as compared to the local value
   \cite[{\em Strong and Mattox, 1996}]{AWS:gradient}.
   This is not a very well understood result.
 \end{enumerate}

 No candidate for a CR source
 has been named from studies of radial density distribution
 since most galactic object distributions have much larger 
 radial gradient in the Galaxy.\\
 We have also performed studies to find $\gamma$--ray emissivity
 variation in directions of the outer Galaxy
 \cite[{\em Mayer, Szabelski et al., 1987}]{js:CRgrad}
 and in galactic spiral arm and interarm regions
 \cite[{\em Rogers, Szabelski et al., 1988}]{js:arm-interarm}.\\
 The $\gamma$--ray emissivity studies
 confirmed very weak radial gradient of CR intensity
 in the Galaxy. Comparison of energy spectra of emissivities
 in various directions also indicates changes 
 in CR energy spectra (at least relative ratio
 of proton to electron components might vary) from
 one direction to another.
 No candidate
 for galactic CR source was found 
 (all candidates have larger gradient of 
 the radial distribution in the Galaxy).\\

 \noindent
 In the studies of diffuse $\gamma$--ray emission
 it is necessary to subtract 
 contribution from unresolved
 $\gamma$--ray point sources
 present in the field of view. This is a difficult task
 since one can not know how many distant sources contribute
 to the flux when sources are weak and at too large distances
 to be identified. The contribution can be estimated by
 treating observed sources as typical and 
 extrapolating the knowledge of their properties to
 much larger space.\\
 The Crab pulsar, Vela pulsar and Geminga presented in the
 section~\ref{gamma_sources} 
 on the page~\pageref{gamma_sources}
 are very bright sources with observed flux well above
 the diffuse galactic $\gamma$--ray intensity (see
 Appendix A: Table~\ref{tab:COSB} on 
 the page~\pageref{tab:COSB}).
 These sources are not very distant on the galactic scale.
 Therefore one might expect that there are other not resolved
 sources of this kind somewhere in the galaxy.
 The above mentioned $\gamma$--ray sources are pulsars.
 However,
 there are quite a few radio pulsars
 within 2~kpc from the Sun, most of them are not observed in
 $\gamma$--ray observations, so the category \lq radio pulsar'
 should not be directly used as 
 a $\gamma$--ray source distribution pattern.
 (It might be worth mentioning here that it is likely that
 there are many more unobserved radio pulsars, since their
 radio emission directions missed the Earth).\\
 
 \noindent
 For the diffuse emission studies it is important to notice
 that if a source, like one of the EGRET 
 identified galactic pulsars,
 would be 3~times further away,
 it might not be noticed as a source in DC signal, 
 since its $\gamma$--ray
 flux would be $\sim$10 times weaker i.e. below the level of
 galactic diffuse intensity. 
 The contribution of such unresolved point sources
 to the galactic diffuse emission depends on 
 how numerous these sources are.
 It is very difficult to estimate the number
 of $\gamma$--ray sources in the galactic disc. 
 Very few were observed and identified. For these one might  
 know the distance and estimate the \lq \lq emissivity"
 (i.e. the flux multiplied by the distance squared).

 \begin{table}[t]
 \caption[The comparison of $\gamma$--ray source strength with
 the diffuse emissivity]
 {\label{tab:emiss}
 The comparison of $\gamma$--ray source strength with
 the diffuse emissivity -- see text on the
 page~\pageref{emiss:details} for details}
 
 \vspace{0.3cm}
 \begin{center}
 \begin{tabular}{lcrccrccrccrccrc}
 \hline
 &&&&&&&&&&&&&&&\\
 & \multicolumn{6}{c}{flux} &&&&
 \multicolumn{3}{c}{4$\pi$ source} &
 \multicolumn{3}{c}{related diffuse}  \\
 source name & \multicolumn{6}{c}{E$>$100~MeV} &
 \multicolumn{3}{c}{distance} &
 \multicolumn{3}{c}{emissivity} &
 \multicolumn{3}{c}{emissivity} \\
 &&&&&&&&&&&&&&& \\
 & \multicolumn{6}{c}{$\left( 
 {\rm 10}^{\qqw -6} 
 \frac{\qqq \rm photons}{\qqq \rm cm^{\qqm 2}~sec} \right)$} &
 \multicolumn{3}{c}{(pc)}
 & \multicolumn{3}{c}{$\left( 
 \frac{\qqq \rm photons}{\qqq \rm s} \right)$}
 & \multicolumn{3}{c}{$\left( 
 \frac{\qqq \rm photons}{\qqq \rm s} \right)$} \\
 &&&&&&&&&&&&&&& \\
 & \multicolumn{3}{c}{COS~B}
 & \multicolumn{3}{c}{EGRET} &&&&&&&&& \\
 &&&&&&&&&&&&&&& \\
 \hline
 &&&&&&&&&&&&&&& \\
 Geminga     &&  4.8 &&&  3.6 &&&  160 &&& 1.1$\cdot$10$^{\qqw 37}$ &
 && 5.9$\cdot$10$^{\qqw 37}$ &\\
 Vela pulsar && 13.2 &&&  9.2 &&&  500 &&& 2.7$\cdot$10$^{\qqw 38}$ &
 && 5.8$\cdot$10$^{\qqw 38}$ &\\
 Crab pulsar &&  3.7 &&&  2.3 &&& 2000 &&& 1.1$\cdot$10$^{\qqw 39}$ &
 && 9.2$\cdot$10$^{\qqw 39}$ &\\
 PSR1704-44  &&  -   &&&  1.2 &&& 3000 &&& 1.3$\cdot$10$^{\qqw 39}$ &
 && 2.1$\cdot$10$^{\qqw 40}$ &\\
 &&&&&&&&&&&&&&& \\
 \hline
 \end{tabular}
 \end{center}
 \end{table}

 \noindent
 Some estimation of contribution of unresolved point
 sources to the diffuse emissivity is presented in the
 Table~\ref{tab:emiss}.
 \label{emiss:details}
 The idea is to compare expected 4$\pi$ emissivities from the point sources
 with the diffuse emissivity due to CR interactions with
 the interstellar medium in the same volume. The volume is a cylinder
 of a radius of the distance to the source and a height of 100~pc
 (the local width of galactic disc).
 The presented in the Table~\ref{tab:emiss} diffuse
 emissivity was obtained by summing the emissivities
 within the cylinder volume. 
 The diffuse emissivity can be estimated assuming that:
 \begin{itemize}
 \setlength{\itemsep}{-2pt}
 \item CR intensity within the galaxy 
 is equal to the locally measured,
 \item the average interstellar matter density is equal to 1~hydrogen
 atom per cubic centimeter.
 \end{itemize}
 With the above assumptions it is possible to evaluate
 $\gamma$--ray emissivity due to CR electron bremsstrahlung and
 due to decay of $\pi^{\circ}$ from CR nuclear component
 interaction. It is equal locally to
 $\sim$2~$\cdot$~4~$\pi~\cdot$~10$^{\qqt -26}$~photons 
 per Hydrogen atom for $\gamma$--ray photons
 with energy above 100~MeV.\\
 The point source emissivity shown in the Table~\ref{tab:emiss}
 was obtained by assuming a 4$\pi$ $\gamma$--ray isotropic emission from
 the source. This assumption is not justified in the case
 of particular source and therefore the figure can not
 be interpreted as actual emissivity of listed sources.
 The $\gamma$--ray emission could be directional and it might have happened
 by chance that in case of these sources it is pointed towards
 the Sun. 
 If this is the case, then there are other $\gamma$--ray
 sources which are not observed because their emissions are
 pointed in other directions. In this case the above
 assumption of the point sources contribution is justified,
 but its interpretation is restricted to the
 comparison presented in the Table~\ref{tab:emiss}.\\
 Since we see one source of a given emissivity within a distance,
 we might estimate a corresponding volume per source as
 a cylinder of a radius equal to the distance and a height
 of 100~pc.
 If such cylinders would be repeated throughout the Galaxy
 it would represent a typical contribution to observed
 $\gamma$--ray flux from (mostly unresolved) sources and diffuse emission.
 A comparison between two last columns from the table
 suggests that the contribution to diffuse $\gamma$--ray galactic
 flux due to unresolved point sources is about 20\%.
 However, if 
 the interstellar matter density was smaller than assumed 
 or 
 source density was larger than observed locally,
 the source contribution to observed
 diffuse $\gamma$--ray flux would be more significant.

 \subsubsection{Diffuse galactic $\gamma$--rays with energy above
 10$^{\qqt 12}$~eV.}
 All detectors capable of observing cosmic $\gamma$--rays with
 energy around 10$^{\qqt 12}$~eV = 1~TeV are using Cherenkov light
 technique and they are tracking detectors, not suited to
 observe diffuse $\gamma$--ray flux.
 The main reason is large background of Cherenkov showers
 of hadronic (mainly proton) origin. To make this background
 compatible to expected signal from the point source
 the angular acceptance (or/and angular resolution) 
 has to be limited to a ring of radius of order of
 a few miliradians
 (1~mrad of arc is about 0.057$^{\circ} \approx$ 3.5 minutes of arc)
 and it is much smaller than dimension of diffuse emission
 in GeV range.
 This limits the statistics gained (most events are due to
 hadronic EAS) during one session of observation. 
 The atmospheric condition variablility influences the counting
 (trigger) rates from one observation session to another and therefore
 limits the validity of summing up observations from different
 sessions. Finally, the expected ratio of diffuse $\gamma$--ray
 event to hadronic background event would be of the order
 of 10$^{\qqt -5}$
 (\cite[{\em Berezinsky et al., 1993}]{Gaisser:gamy})
 which corresponds to one diffuse $\gamma$--ray event
 per a year of observation.\\
 Some improvements might be expected when an efficient method
 of discrimination between $\gamma$ and hadron Cherenkov showers
 will be found and successfully applied in experiments
 (e.g. see 
 section~\ref{sect:Cherenkov} on the page~\pageref{sect:Cherenkov}
 and
 \cite[{\em Cabot, Szabelski et al., 1997}]{JS:muCher}).
 It is necessary to notice that one can not
 expect any difference between Cherenkov showers 
 originated by the $\gamma$--ray and those
 originated by other electro--magnetic particles 
 (e$^{-}$ and e$^{+}$) which do not keep the direction
 to the place of their origin.\\

 \subsubsection{Diffuse galactic $\gamma$--rays with energy above
 10$^{\qqt 14}$~eV.}
 Air shower arrays capable of observing EAS with energy
 above 10$^{\qqt 14}$~eV can observe a diffuse $\gamma$--ray
 component looking for anisotropy or/and looking
 for \lq \lq muon poor showers", i.e. EAS which have
 abnormally small muon content. 
 Since muons are decay products of kaons and charged pions 
 of hadronic EAS, their presence in events generated
 by electromagnetic particles ($\gamma$--ray, e$^{+}$ and e$^{-}$)
 is suppressed. The main channel to produce hadrons in E--M
 cascades is the photo--production: hadron production in
 $\gamma$--ray interaction with nuclei. The expected average ratio
 of muon content in proton EAS to $\gamma$--ray induced shower
 depends on muon threshold, experiment altitude, distance to the EAS core 
 and primary particle energy. 
 From Monte--Carlo
 simulations using CORSIKA version 4.50,
 \cite[{\em J.~Knapp and D.~Heck, 1995}]{CORSIKAv450}
 for primary CR particle energy 10$^{\qqt 14}$~eV 
 the total number of muons with
 E$_{\qqt \mu} >$~1~GeV
 at sea level in proton EAS is about 50 times
 larger than in $\gamma$--ray induced showers 
 (i.e. $\approx$ 2.5 10$^{\qqt 4}$ muons in proton EAS 
 to $\approx$ 500 in $\gamma$--ray showers).
 At a higher altitude, where low energy muons have not decayed
 yet, the ratio is smaller, e.g. at 600~g/cm$^{\qqm 2}$ is
 about 35.
 The approximate formula is given in
 \cite[{\em Gaisser, 1990, p.~246}]{CR:Gaisser}:
 \[
 N_{\qqw \mu}^{\qqw (\gamma)} \, \sim \, R \, \, ln(E_{0}/1~GeV) 
 \, \cdot \, N_{\qqw \mu}^{\qqw (p)}
 \]
 \[
 R \, = \, \frac{\qqq \sigma_{\qqw \gamma \rightarrow hadrons}}
    {\qqq \sigma_{\qqw \gamma \rightarrow e^{+}e^{-}} }
    \approx 2.8 \, \cdot \, 10^{\qqw -3}
 \]
 This gives the ratio
 $N_{\qqt \mu}^{\qqt (\gamma)} / N_{\qqt \mu}^{\qqt (p)} \approx$~30
 for $E_{0}$~=~10$^{\qqt 14}$~eV.\\
 Since
 there is no clear
 evidence for observations of $\gamma$--rays in
 this energy range the muon ratios were not set experimentally.\\

 \noindent
 I should mention here old results of
 {\L }\'{o}d\'{z} group related to the detection of 
 \lq \lq muon poor showers". To verify this idea
 special muon detectors were constructed in the late 50's.
 The experimental search for \lq \lq muon poor showers" 
 gave positive results with a rate of
 0.6\%$\pm$0.2\% of ordinary showers
 \cite[{\em Gawin et al., 1963}]{Lodz:mupoorI}
 and about 0.7\% \cite[{\em Gawin et al., 1965}]{Lodz:mupoorII}.
 These results were not confirmed in the recent
 electronically collected data of {\L }\'{o}d\'{z} array.
 They would be also in contradiction to
 the theoretical prediction of
 \cite[{\em Berezinsky et al., 1993}]{Gaisser:gamy}
 on the rate of $\gamma$--ray showers to nuclear showers
 of the level 5$\cdot$10$^{\qqt -5}$ from the centre of
 Galaxy direction (i.e. direction of the highest 
 expected diffuse $\gamma$--ray flux, direction not seen from
 {\L }\'{o}d\'{z}).\\
 
 \noindent
 The diffuse $\gamma$--ray flux should be seen as an
 enhancement of events from direction of galactic disc,
 since this would reflect the column density distribution
 of the interstellar matter (target matter distribution
 for $\gamma$--ray creation processes). The observations
 are difficult: anticipated small anisotropy due to 
 $\gamma$--ray flux requires long time
 data acquisition with the very stable detection capability.
 The result is that no clear signal has been observed
 so far, although it is worth to mention a 1.5$\sigma$
 excess of EAS detection from the galactic plane
 direction by the Baksan group
 \cite[{\em Alexeenko et al., 1993~II}]{BASA:Gal}.
 
 \newpage
 \section{\label{GRUPY-MIONOW}
 Mass composition of high energy cosmic rays.}
 \subsection{Introduction.}
 CR mass composition is relatively well known for energies 
 below few hundred GeV. In this energy region fluxes 
 of different components of CR are measured directly,
 i.e. the detectors placed on satellites or high altitude 
 balloons are exposed directly to the primary CR particles.
 There are few techniques used for
 mass and charge determination (e.g. combinations of
 track curvature determination in magnetic field,
 charge determination through measurements of ionization
 losses, calorimetric energy measurements etc.).
 Review of low energy CR mass composition, its energy dependence and 
 interpretation for CR origin and propagation theories
 could be found in 
 \cite[{\em Engelmann et al., 1990}]{HEAO-3:nuclei}
 and
 \cite[{\em Swordy et al., 1990}]{Spacelab:nulc}.\\
 
 \noindent
 For higher energies the situation becomes more difficult,
 because fluxes of CR particles fall down with growing
 energy and the experiments need
 larger detector area and longer exposure
 to gain sufficient statistics. 
 The present upper energy limit of directly measured CR particles
 has been achieved by JACEE 
 (Japan--American Collaboration Emulsion Experiment).
 The exposure time and area 
 (both related to the statistics) 
 limit the energy of observed CR particles.
 CR proton flux was measured up to 10$^{\rm 6}$~GeV~(=10$^{\rm 15}$~eV).
 The results were presented in conference papers
 \cite[{\em Asakimori et al., 1993}]{CRspetr:JACEE1},
 \cite[{\em Asakimori et al., 1995}]{CRspetr:JACEE2},
 or published 
 \cite[{\em Burnet et al., 1990}]{JACEE:widmo}.
 JACEE results came from several balloon flights of 
 emulsion chambers.\\
 Another experiment of high energy primary CR measurement
 was placed on a satellite and can register CR protons of energy
 up to 10$^{\rm 5.5}$~GeV
 \cite[{\em Ivanenko et al., 1993}]{CRspetr:Sokol}.\\

 \noindent
 In this chapter the studies of 
 mass composition of primary cosmic rays (CR)
 of energy above $\sim$1~TeV~= 10$^{3}$~GeV~= 10$^{12}$~eV 
 will be presented.\\
 Ground based
 measurements of cosmic ray mass (or chemical) composition
 at energies above 1~TeV (per CR particle) 
 are very difficult. 
 For these energies long exposure and large area are required
 and {\em indirect} methods are used. The detectors are not
 exposed to the primary particle, but to products of its
 interactions in the atmosphere, or subsequent particle cascade.
 Various methods are in use. Classical extensive air shower (EAS)
 array registers electromagnetic component of EAS.
 There are calorimeters to register hadronic component of EAS.
 There are emulsion and X--ray films technics used at mountain
 altitudes. There are EAS muon detectors to measure
 penetrating component of EAS. 
 There are also detectors of Cherenkov light from EAS.
 In many cases some combinations
 of these methods are used to measure the same event.
 Most of these methods measure values related to
 primary CR energy per particle (not per nucleon).\\
 
 \noindent
 Interpretation of experimental data is very difficult.
 The characteristics of EAS 
 produced by a CR particle in the atmosphere
 depend on the particle mass and its energy, as well as
 properties of high energy interaction, which
 are not well known and subject to large fluctuations.\\
 One would like to know the energy spectra for each
 chemical component of primary CR. These would provide
 a lot of information about the acceleration sites
 and properties. Once the mass end energy spectra
 were known, the nuclear interaction properties
 could be studied at energies above these currently reached
 by accelerators, and for much, much lower cost
 of particle beam.

 \begin{figure}
 \begin{center}
 \mbox{
 \psfig{file=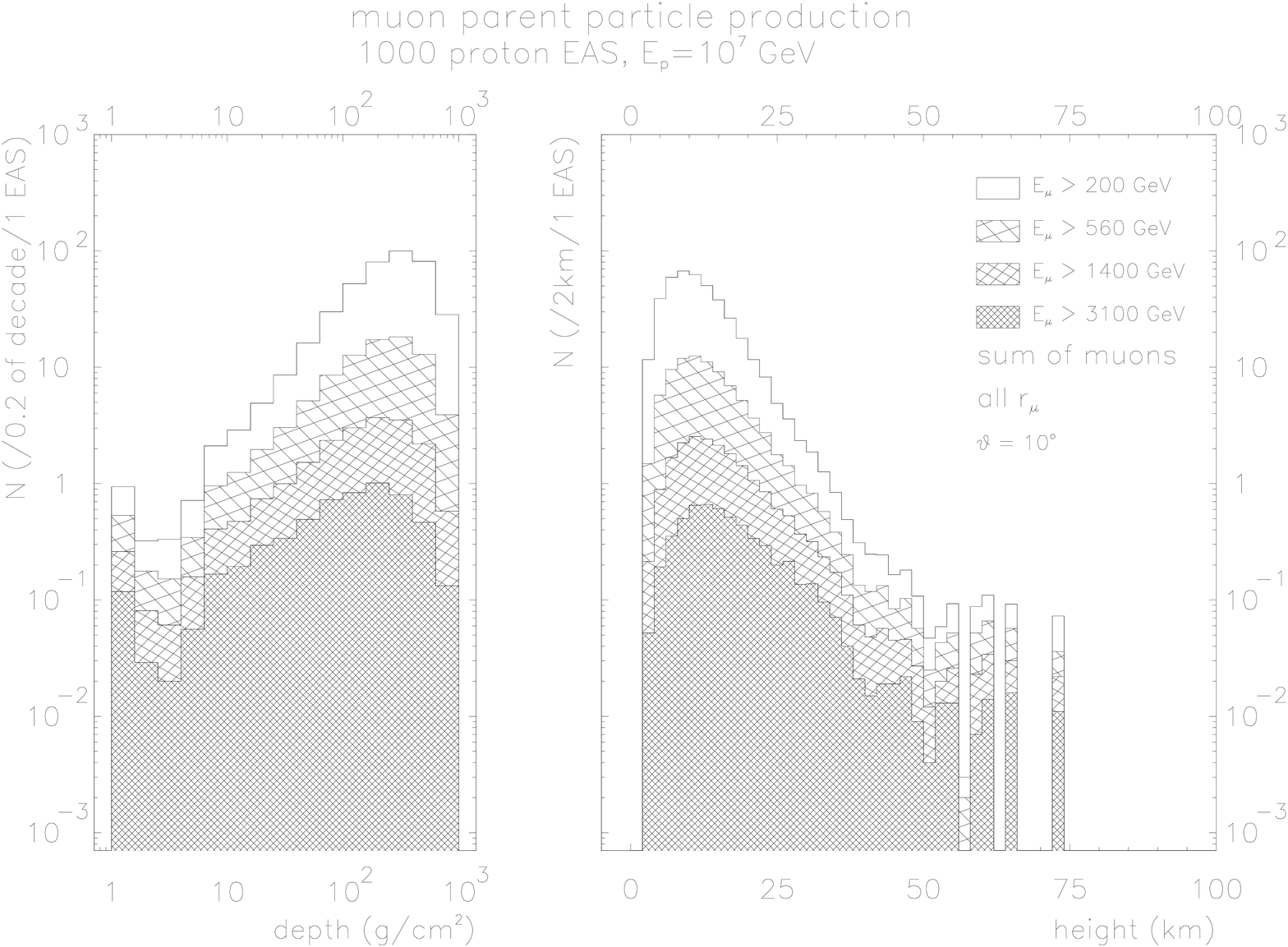,height=8.0cm,width=14.0cm}
 }\\
 \end{center}
 \caption[Muon origin levels for 10$^{\qqt 7}$~GeV: primary CR protons.]
 {Histograms of muon origin levels for 10$^{\qqt 7}$~GeV 
 protons
 (results of Monte-Carlo simulation)
 \cite[{\em Attallah, Szabelski et al., 1995}]{JS:muIPJ}).}
 \label{fig:pr700lev}

 \begin{center}
 \mbox{
 \psfig{file=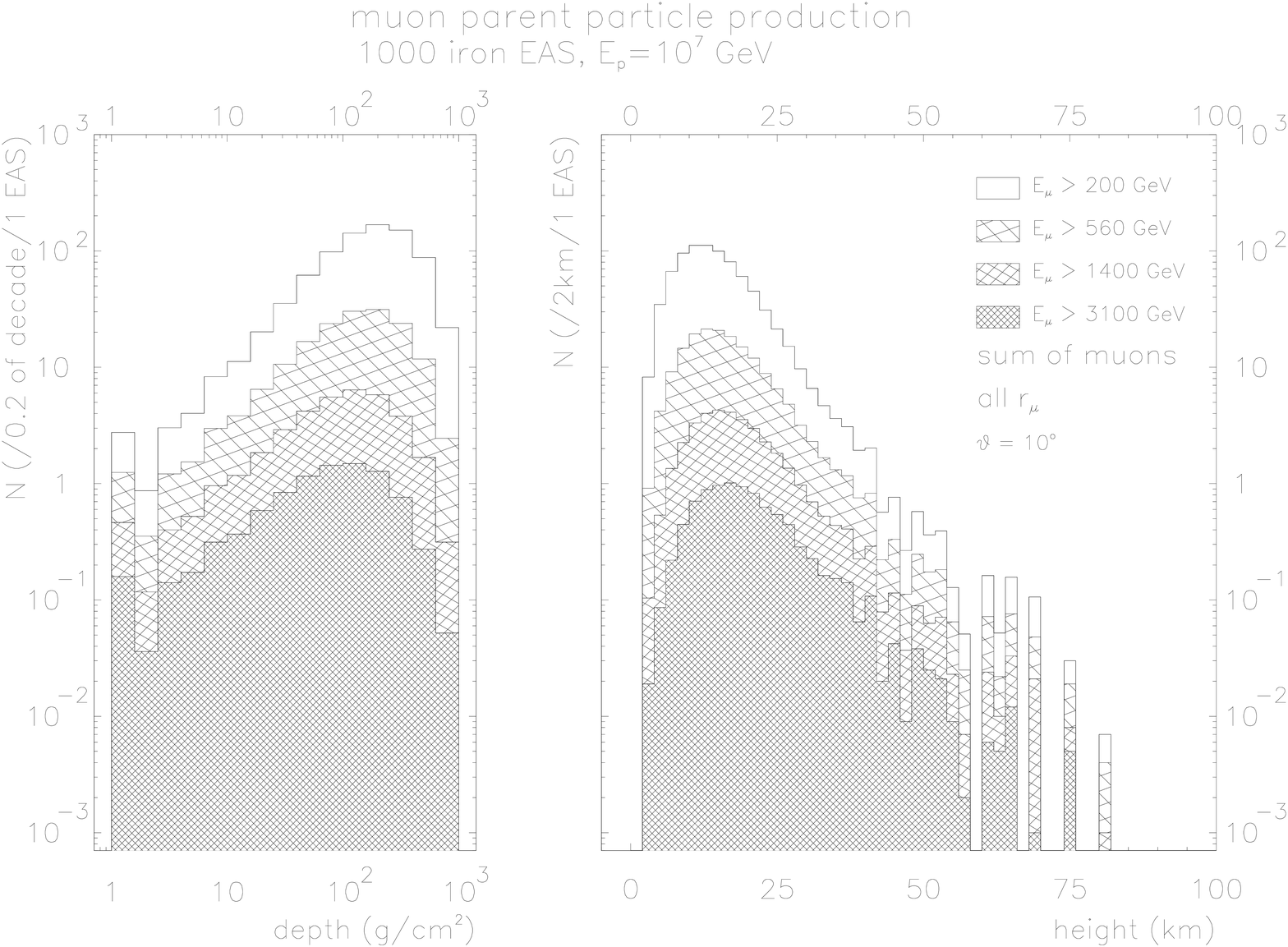,height=8.0cm,width=14.0cm}
 }\\
 \end{center}
 \caption[Muon origin levels for 10$^{\qqt 7}$~GeV: primary CR iron nuclei.]
 {Histograms of muon origin levels for 10$^{\qqt 7}$~GeV 
 iron nuclei
 (results of Monte-Carlo simulation)
 \cite[{\em Attallah, Szabelski et al., 1995}]{JS:muIPJ}).}
 \label{fig:fe700lev}
 \end{figure}
 
 \begin{figure}
 \begin{center}
 \mbox{
 \psfig{file=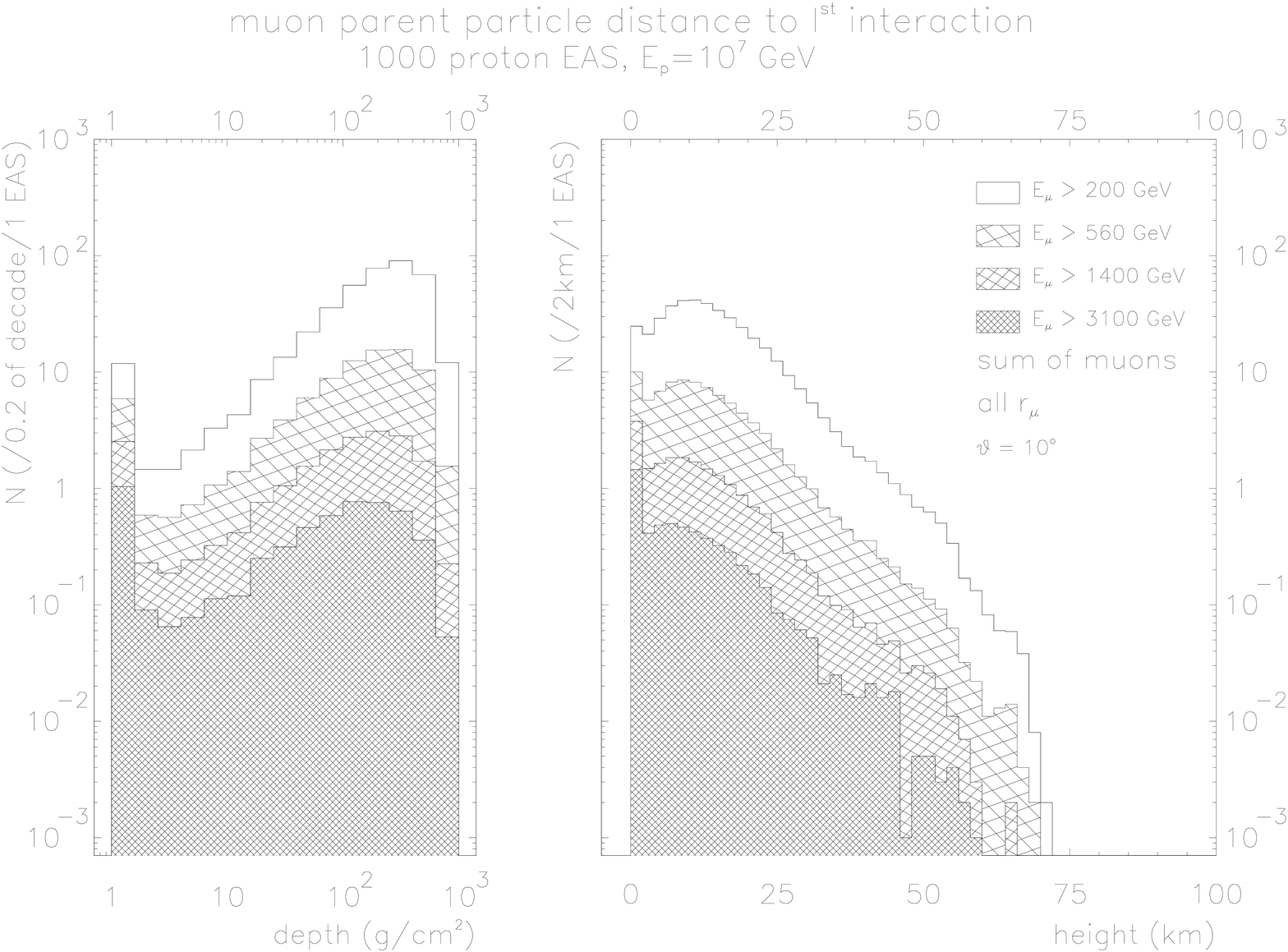,height=8.0cm,width=14.0cm}
 }\\
 \end{center}
 \caption[Muon origin levels relative to 1$^{\qqt st}$
 interaction for 10$^{\qqt 7}$~GeV: primary CR proton.]
 {Histograms of muon origin levels relative to
 the first interaction for 10$^{\qqt 7}$~GeV 
 protons
 (results of Monte-Carlo simulation
 \cite[{\em Attallah, Szabelski et al., 1995}]{JS:muIPJ}).}
 \label{fig:pr700rel}
 
 \begin{center}
 \mbox{
 \psfig{file=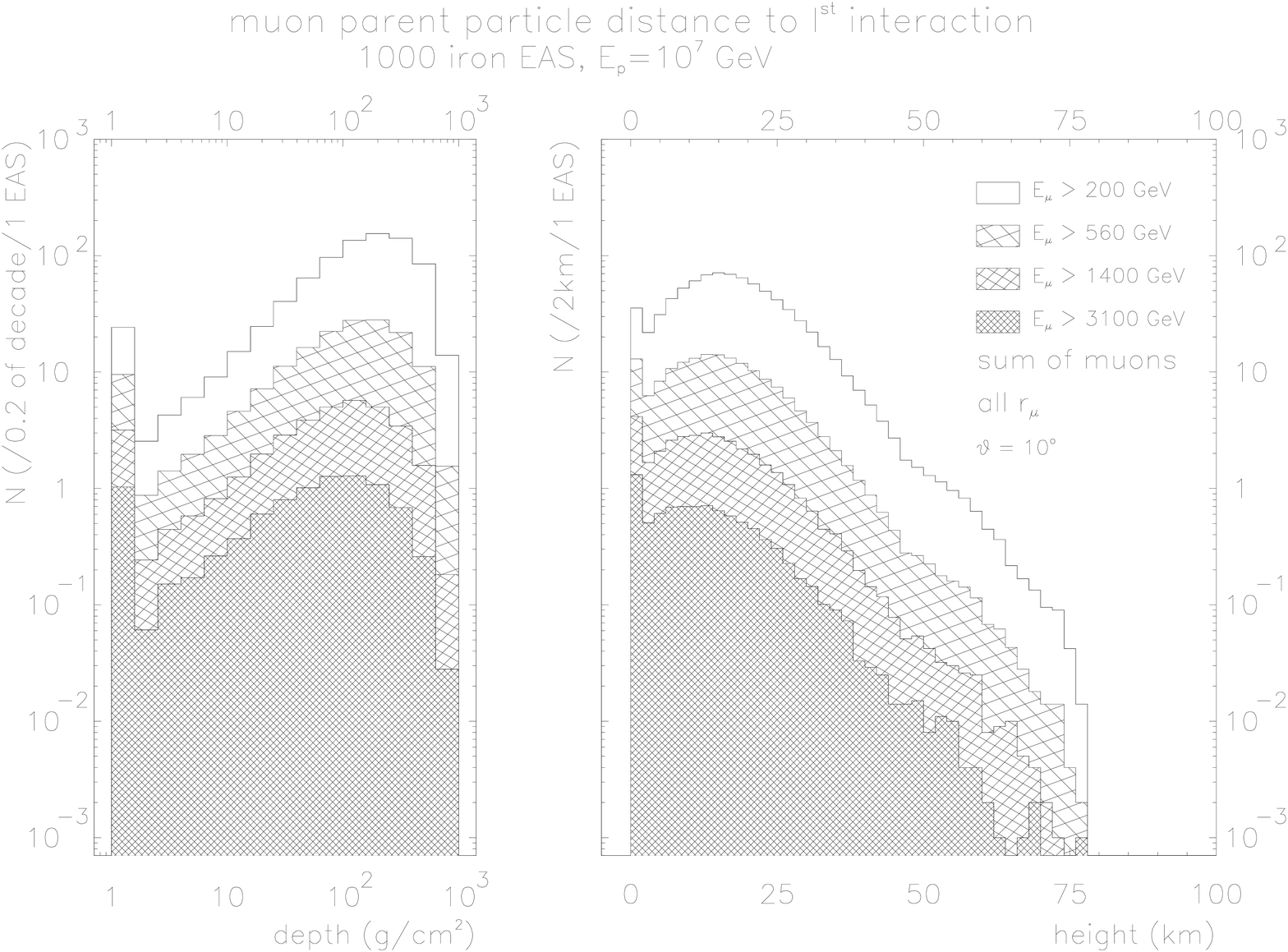,height=8.0cm,width=14.0cm}
 }\\
 \end{center}
 \caption[Muon origin levels relative to 1$^{\qqt st}$
 interaction for 10$^{\qqt 7}$~GeV: primary CR iron nuclei.]
 {Histograms of muon origin levels relative to
 the first interaction for 10$^{\qqt 7}$~GeV 
 iron nuclei
 (results of Monte-Carlo simulation)
 \cite[{\em Attallah, Szabelski et al., 1995}]{JS:muIPJ}).}
 \label{fig:fe700rel}
 \end{figure}
 
 \subsection{High energy cosmic rays mass composition and 
 underground measurements of muon groups.}
 Measurements of high energy CR muons can be used to study
 high energy nuclear interaction properties and 
 mass composition of the primary cosmic rays.
 These muons are decay products of high energy pions
 and kaons of extensive air shower~(EAS).
 Some pions and kaons can decay, other would interact.
 Therefore the number of muons depends on relation
 between probability of decay vs. probabilities of interaction or
 not--to--muon decay. It is clear that deeper in the atmosphere
 its density is larger and decays are relatively less probable than
 interactions as compared with the higher altitudes. 
 On another side the number of energetic pions and kaons in the EAS
 has a maximum at the altitude which depends on the primary
 particle energy, mass, first few interactions heights and 
 multiplicities of hadron production, etc. 
 So the parent muon particles can originate at relatively high
 altitude i.e. near to the first interaction, and bring 
 information about the EAS development 
 in the upper layer of the atmosphere.\\
 
 \begin{figure}[ht]
 \begin{center}
 \mbox{
 \psfig{file=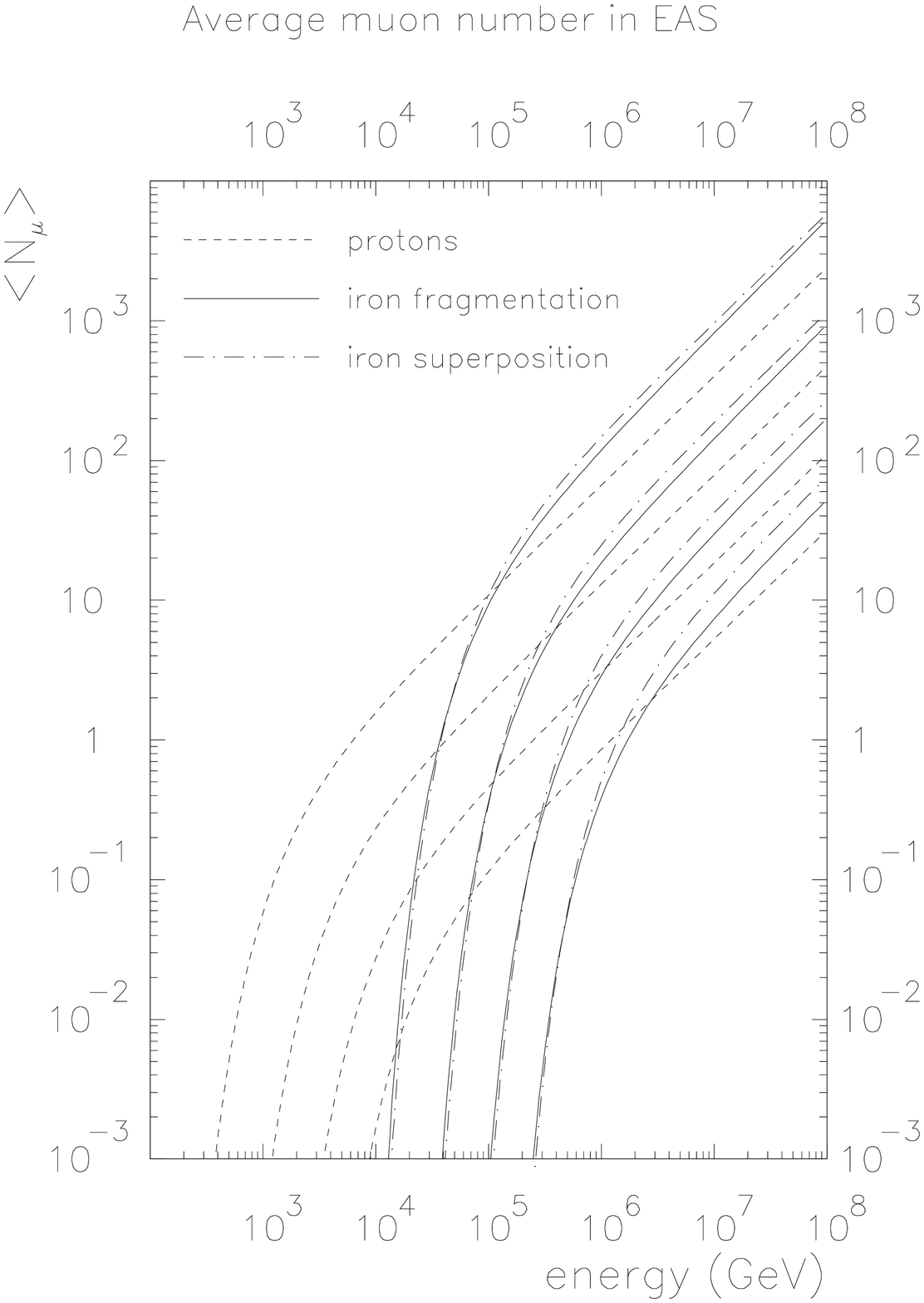,height=12.0cm,width=8.4cm}
 }\\
 \end{center}
 \caption[High energy muon production]
 {Average number of muons generated in EAS vs.
 energy of the primary particle as results
 of M-C calculations. Different
 curve modes relate to primary proton
 and iron nuclei. Results of
 superposition approach and fragmentation
 model are shown for comparison.\\
 Four sets of different curves relate to
 four muon energy thresholds; from top
 to bottom: 200~GeV, 560~GeV, 1400~GeV and
 3160~GeV.
 (Figure from \cite[{\em Attallah, Szabelski et al., 1995}]{JS:muIPJ}).}
 \label{fig:nmusuper}
 \end{figure}
 
 \noindent
 Measurements of muon groups, 
 i.e. simultaneous registration of number of parallel muons,
 play a special role.
 The experiments for muon group studies 
 have areas 
 from $\sim$10~m$^{\qqm 2}$ to $\sim$300~m$^{\qqm 2}$.
 They are placed relatively deep underground to limit the minimum
 energy of muons at the ground level to above $\sim$40~GeV
 (which is different for different experiments).
 The ground above the experiment absorbs the electromagnetic
 component (electron/positron and photons) of associated EAS
 as well as low energy muons. These are relatively low energy
 particles which originate much lower than most of high
 energy muons. Therefore underground measurements of high energy
 muons are 
 in some sense 
 equivalent to EAS observations at high altitude.\\

 \noindent
 In the proton induced showers the average number of high energy muons 
 depends on their energy, proton energy and the zenith angle.
 For CR proton energy (E$_{\qqt \rm p}$) above approximately
 20 times the muon threshold energy
 (E$_{\qqt \mu \, \rm th} > {\rm 40~GeV}$)
 (to escape from the complicated near--to--threshold relation)
 the predicted total average number of muons with energy
 above E$_{\qqt \mu \, \rm th}$ is proportional to
 E$_{\qqt \rm p}^{\qqt 0.7-0.8}$. 
 The power index in this relation depends on the high energy
 interaction model used in the calculations. Higher multiplicity
 models predict higher power index.\\
 \noindent
 Since high energy muons originate high in the atmosphere,
 calculations of predicted number of muons in EAS are
 sensitive to the interaction model used in Monte--Carlo
 program (see Figures~\ref{fig:pr700rel} and \ref{fig:fe700rel},
 where one can notice that most of high energy muons
 orginate within 200~g/cm$^{\qqm 2}$ from the first
 interaction). Results of measurements of single muon spectra in CR
 are related to properties of fragmentation region
 of high energy interaction models 
 whereas
 muon group rates
 correspond to high multiplicity (central) region of
 high energy interation.\\ 
 For CR nuclei (with atomic mass A) 
 having the energy E$_{\qqt \rm A}$ = A$\cdot$E$_{\qqt \rm n}$
 the total predicted average number of muons with energy above
 E$_{\qqt \mu \, \rm th}$ is larger than for proton shower 
 (E$_{\qqt \rm p}$=E$_{\qqt \rm A}$), provided that
 E$_{\qqt \rm n}$ is also well above E$_{\qqt \mu \, \rm th}$.
 When the primary CR particle nucleon has energy not much
 higher than the required muon threshold energy then
 the corresponding relation between nucleon energy and the number 
 or muons has not a power law form with the power index of
 0.7--0.8. The relation in this threshold region is much 
 steeper.\\
 The simplest approach to evaluate the number of muons generated
 by the primary CR nucleus is the superposition model.
 In this approach it is assumed that the number of high energy muons 
 (N$_{\qqt \mu\, \rm A}$) generated in the shower originated by primary
 nucleus of energy E$_{\qqt \rm A}$ = A$\cdot$E$_{\qqt \rm n}$ is
 equal (on average) to the number of muons which would
 be produced in A showers originated by protons (or neutrons)
 with energy E$_{\qqt \rm n}$:
 N$_{\qqt \mu\, \rm A}($E$_{\qqt \rm A}$) = 
 N$_{\qqt \mu\, \rm A}$(A$\cdot$E$_{\qqt \rm n}$) =
 A$\cdot$N$_{\qqt \mu\, n}$(E$_{\qqt \rm n}$).\\
 The more realistic model of primary CR nucleus interactions in the
 atmosphere assumes its destruction in a number of subsequent
 interactions. The destruction level depends on the interaction
 parameter. Such a model, presented in
 \cite[{\em Capdevielle, 1993}]{Capd:AbrEvap}
 assumes the abrasion of incident and target nuclei (which
 is the interaction parameter dependent) as well as some evaporation
 of nucleons from excited fragments after the collision.\\
 Calculations of predicted number of muons associated
 with primary CR nucleus were performed using both models
 for comparison. The important differences were noticed
 which in the first approximation show smaller high energy
 muon production for the abrasion and evaporation model
 as compared with the superposition model 
 (see Figure~\ref{fig:nmusuper}).
 Some results of these calculations are presented
 in Section~\ref{NNcolis} on page~\pageref{NNcolis}.\\
 
 \subsection{Monte--Carlo simulations of high energy
 muon shower development in the atmosphere.}
 To interpret the experimental data on muon groups
 they were compared with the results of Monte--Carlo
 simulations of EAS development.

 \subsubsection{General information about the
 Monte--Carlo simulations of EAS development.}
 To compare results of models of
 high energy CR particle interactions the Monte--Carlo
 simulations of EAS development were performed.
 The first results were obtained
 at the University of Bordeaux using the
 large VM--System IBM computer, 
 and then the code
 has been adapted to NDP~Fortran with the UNIX--System on the PC
 computer in {\L }\'{o}d\'{z}, 
 and then to Digital FORTRAN on Alpha stations in Perpignan and
 in {\L }\'{o}d\'{z} and 
 finally to PCs using Fortran to C converters, \lq DJGPP' compiler,
 and {\em go32.exe} program.
 During the Monte--Carlo simulations usage of most 
 program arrays was monitored, to prevent over--writing.

 \noindent
 In the presented analysis the EAS development has been simulated
 using the program code written by J.~N.~Capdevielle
 with some modifications related to the high energy muon group
 studies:
 \begin{enumerate}
 \setlength{\itemsep}{-2pt}
 \item information about muons at the program output,
 \item trace of hadrons with energy above 200~GeV, only
 (i.e. above the muon threshold).
 \end{enumerate}
 The hadron interactions were treated according to
 dual parton model
 (\cite[{\em Capdevielle, 1989}]{Capd:Model})
 with some further modifications to adjust results
 to experimental data
 on particle production in high energy interaction.
 The air (the target) has been treated as containing
 atoms with A=14, only.
 The atmosphere pressure--height relation has a form:
 \[
 \rm
 h (km) = \ln \left( \frac{\qqq 1034}{\qqq x} \right) \cdot
 (0.002375 x + 6.7625) \, \frac{\qqq 1}{\qqq \cos \it \theta}
 \]
 where x is depth in the atmosphere in $g/cm^{\qqm 2}$.
 This relation agrees with the US Standard Atmosphere
 (see preprint \cite[{\em Capdevielle et al., 1992}]{KfK4998}).\\
 
 \noindent
 The nucleus--nucleus interactions were treated according
 to the abrasion--evaporation model
 (\cite[{\em Capdevielle, 1993}]{Capd:AbrEvap}).
 Results were compared with the superposition approach
 to description of  nucleus--nucleus interactions,
 and this problem will be discussed later.\\
 
 \subsubsection{Brief description of high energy
 muon group rate evaluation.}
 In the first step number of EAS were simulated and results
 were stored in memory.
 The procedure looks as follows:
 \begin{enumerate}
 \setlength{\itemsep}{-2pt}
 \item The primary particle atomic numbers A were
 1, 4, 14, 28 and 56,
 \item the zenith angle $\theta$ was set to 10$^{\circ}$,
 \item the primary cosmic ray particle energy $E_{\qqt CR}$
 varied from the threshold energy related to 200~GeV
 muon production in EAS to $10^{\qqt 7.5}$~GeV/nucleus
 with the 0.1 step in logarithmic scale,
 \item for each A and particle energy $E_{\qqt CR}$ the number of EAS
 simulations have been performed and information on 
 first interaction and
 high energy muons was stored for further analysis.
 The number of simulated EAS varied from 1000 near the
 threshold CR particle energy to 100 at the highest
 energies. For each muon above the muon threshold
 energy (for this simulation: 200~GeV) its
 x,~y~position at the observation level, energy,
 parent particle type (i.e. pion or kaon)
 and the parent particle production height were stored.
 \end{enumerate}
 
 \noindent
 Some simple properties like the average muon number, muon
 number distribution, correlation of muon number with the
 parameters of the first interaction,
 muon lateral distribution, etc. could be performed
 at this stage (see
 Figures~\ref{fig:pr700lev},
 \ref{fig:fe700lev},
 \ref{fig:pr700rel} and
 \ref{fig:fe700rel}).\\

 \noindent
 While performing our
 calculations we have neglected some related problems.
 These seem to play less important role in the large
 muon group analysis. Incorporating them into the scheme
 of calculations would significantly enlarge the time of computing.
 Namely, we have neglected:
 \begin{enumerate}
 \setlength{\itemsep}{-2pt}
 \item the influence of the Earth's magnetic field
 on $\mu^{+}$, $\mu^{-}$ group:
 \[ r_{G}/(1~m) = \frac{\qqq p/ (1~GeV/c) }{\qqq 0.3 \, \cdot \, 
 B_{\qqq \perp}/ (1~T)}, 
 \rule{1cm}{0cm}
 p \approx 200~GeV,
 \rule{1cm}{0cm}
 B_{\perp} \approx 20 \mu T \, = \, 0.2~G
 \]
 $r_{G}$~=~3.3~10$^{\qqt 7}$~m is muon curvature radius.
 For muon production at altitude 15~km
 the displacement is $\approx$3.4~m due to magnetic field deflection.
 This can be compared with the displacement due to transverse
 momenta in pion or kaon production processes: 
 $< p_{t} > \approx$~0.4~GeV/c, gives 30~m for production at
 15~km and 200~GeV/c muon, i.e. nearly 10~times more.
 \item muon energy losses were treated as average for all muons
 (as an $E_{\qqt \mu \, \rm th}$ -- muon energy threshold), whereas for large
 muon energies, energy losses are not uniform.
 However, since the analysis is related to large muon
 groups, this effect does not play an important role.
 \item elastic scattering of muons in the rock was neglected
 i.e. their trajectories were straight lines.
 \item we have performed simulations for one zenith angle
 $\theta$ of incident primary CR particle, and used these results
 for the whole nearby range of zenith angles
 (e.g. for $\theta<20^{\circ}$ we have used results
 obtained from calculations assuming $\theta=10^{\circ}$).
 \item we assumed a circular shape of detector with the same
 effective area.
 \end{enumerate}
 At the present stage of muon group events analysis the above
 listed approximations do not seem to produce 
 effects which would change the interpretation of data.\\

 \subsubsection{Total number of high energy muons produced
 in EAS.}
 The total number of muons with energy above a threshold 
 (although not measured in most experiments)
 is a convenient value to
 compare between different calculations. Performing a number of
 M--C simulations of EAS for a fixed primary CR particle
 type and energy, the average number of muons can be
 found together with its distribution.
 The average number of muons depends on primary CR particle energy.
 It grows fast with particle energy near the muon production
 threshold energy and then grows according to the power law.

 \begin{figure}[ht]
 \begin{center}
 \vspace{-1.0cm}
 \mbox{
 \psfig{file=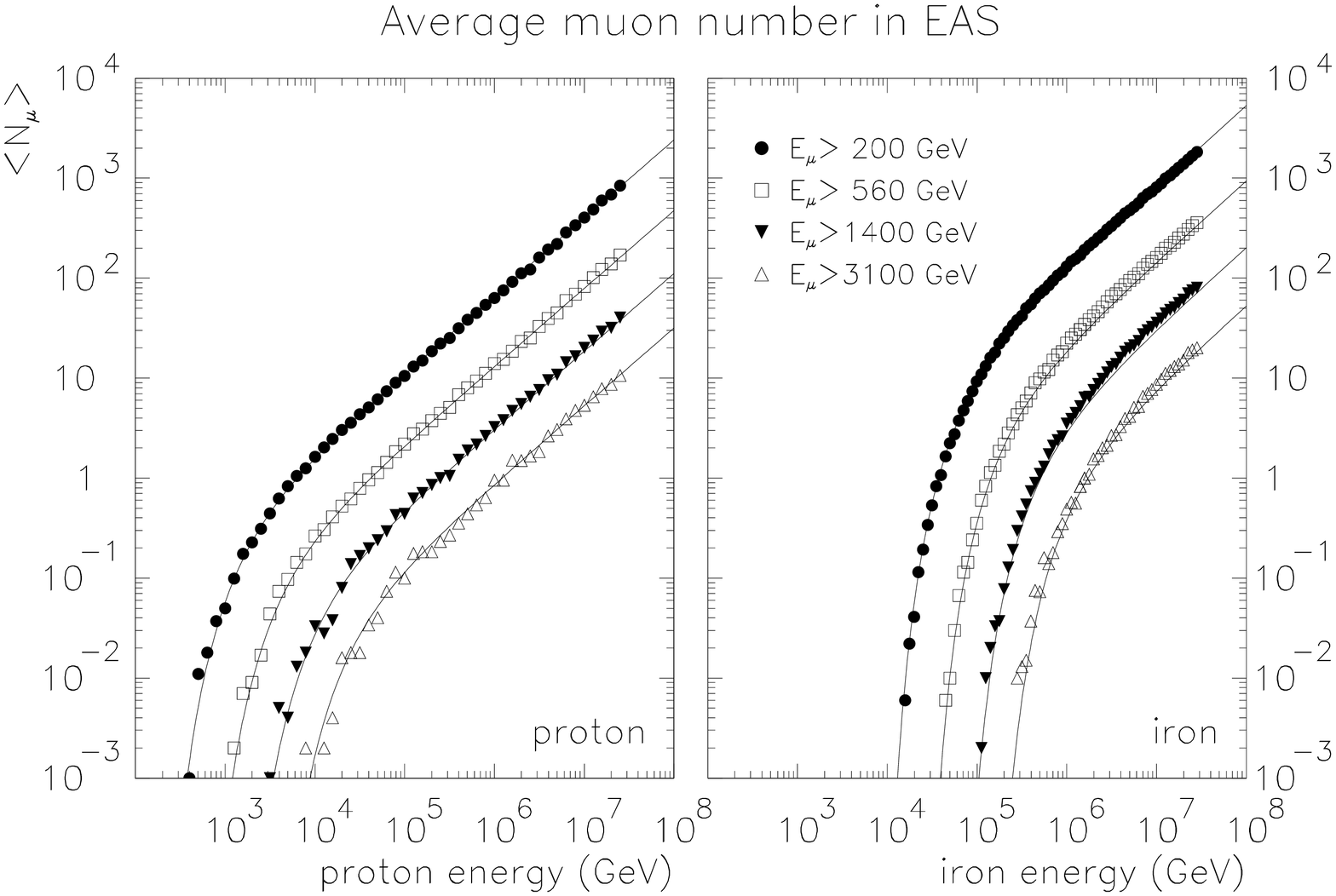,height=8.0cm,width=13.0cm}
 }\\
 \end{center}
 \caption[N$_{\qqt \mu}$ vs. E$_{\qqt \rm p}$ for simulated EAS]
 {Results of Monte-Carlo simulation of average
 number of high energy muons in EAS generated by protons
 and iron nuclei and approximate fit
 \cite[{\em Attallah, Szabelski et al., 1995}]{JS:muRoma}
 \cite[{\em Attallah, Szabelski et al., 1995}]{JS:muIPJ}.}
 \label{fig:avNmuvsE}
 \end{figure}
 
 \begin{table}
 \caption[Parameters for $<$N$_{\qqt \mu}>$ vs. energy.]
 {\label{tab:Nmuparam}
 Parameters for equation~\ref{eq:avnm}.}

 \vspace{0.3cm}
 \begin{center}
 \begin{tabular}{||c|cccccc||}
 \hline 
 &&&&&& \\
 & A & $\gamma$ & $C$ & $\alpha$ &  $B$ & $\beta$ \\
 &&&&&& \\
 \hline
 &&&&&& \\
 protons & 1 & 1.58 & 6.0 & 0.78 & 0.6 & 13.0 \\
 &&&&&& \\
 iron & 56 & 1.68 & 6.0 & 0.81 & 0.65 & 10.0 \\
 &&&&&& \\
 \hline
 \end{tabular} 
 \end{center}
 \end{table}

 \noindent
 In the Figure~\ref{fig:avNmuvsE}
 the average number of muons above
 four threshold energies is illustrated for primary
 protons and iron nuclei.
 Lines represent the parametrization:
 \begin{equation}
  <{\rm N}_{\qqw \mu\, \rm A}> = {\rm A} \cdot C \cdot
 ~\frac{\qqq E_{\qqw \rm n}^{\qqw \alpha}}
       {\qqq E_{\qqw \mu\, \rm th}^{\qqw \gamma}} 
 \, \cdot ~\,
 \left( 1.0 - B \cdot 
 \frac{\qqq E_{\qqw \mu\, \rm th}}{\qqq E_{\qqw \rm n}} 
 \right)^{\qqw \beta}
 \label{eq:avnm}
 \end{equation}
 
 \noindent
 where $E_{\qqt \rm n}$ = $E_{\qqt \rm CR}$/A, $E_{\qqt \mu\, \rm th}$ 
 is the muon energy threshold
 and other parameters are in the 
 Table~\ref{tab:Nmuparam}.\\
 For muon energy threshold $E_{\qqt \mu\, \rm th}$~=~200~GeV
 the distribution of N$_{\qqt \mu}$ is Gaussian
 around $<{\rm N}_{\qqt \mu}>$.
 For M--C simulations of 1000 EAS for each CR primary energy
 in the range (10$^{\qqt 5}$--10$^{\qqt 7.5}$~GeV) the number of  
 events with N$_{\qqt \mu}$
 can be parametrized as follows:
 \begin{equation}
 {\rm N}_{\qqw 1000}({\rm C},{\rm N}_{\qqw \mu},<{\rm N}_{\qqw \mu}>,\sigma ) =
  {\rm C} \cdot \frac{1}{\sigma \sqrt{2\pi}} \cdot
 \exp \left[ \frac{\qqq - ({\rm N}_{\qqw \mu} - 
 <{\rm N}_{\qqw \mu}>)^{\qqw 2}}{\qqq 2 \sigma^{\qqw 2}} \right] 
 \label{eq:n1000}
 \end{equation}
 \[
 \begin{array}{lrcccccl}
 \rm
 where~for~protons:
 & {\rm C} & = & 995 & - & 11.2 & \cdot & 
         {\rm log}_{10}(\rm E_{\qqw \rm p}~/{\rm 1~GeV}) \\
 & {\rm log}_{10}(\sigma) & = & -2.9 & + & 0.707 & 
       \cdot & {\rm log}_{10}(\rm E_{\qqw \rm p}~/{\rm 1~GeV}) \\
 \rm and~for~iron: 
 & {\rm C} & = & 1022 & - & 11.3 & \cdot & 
         {\rm log}_{10}(\rm E_{\qqw \rm Fe}/{\rm 1~GeV}) \\
 & {\rm log}_{10}(\sigma) & = & -2.8 & + & 0.683 & 
       \cdot & {\rm log}_{10}(\rm E_{\qqw \rm Fe}/{\rm 1~GeV}) \\
 \end{array} \]
 and
 $<{\rm N}_{\qqt \mu}>$ is given by the formula~\ref{eq:avnm}
 for E$_{\qqt \mu\, \rm th}$~=~200~GeV.\\

 \noindent
 The slope of $<{\rm N}_{\qqt \mu}>$~ vs. $E_{\qqt \rm p}$ dependence
 (at the power law part) is usually  related to
 the properties of the first interaction of CR particle,
 and particularly to the multiplicity of secondary particles
 produced then. For some calculations made in the past
 assuming the Feynman scaling 
 (\cite[{\em Feynman, 1969}]{Feynman:scal})
 the slope has a value of $\approx$0.7, and for extremely
 large scaling breaking of 
 \cite[{\em Wdowczyk, Wolfendale, 1979}]{WW:scalbr}
 model its value was equal to $\approx$0.85.
 The difference 0.15 in the power law dependence gives
 a factor of 1.4 in the difference in $<{\rm N}_{\qqt \mu}>$
 over the energy range of one order of
 magnitude and a factor of 2 over two orders
 of magnitude in the $E_{\qqt \rm p}$ energy range.
 The $<{\rm N}_{\qqt \mu}>({\rm E}_{\qqt \mu}\geq 
 {\rm E}_{\qqt \mu\, \rm th})$ vs. E$_{\qqt \rm p}$ relation
 can be normalized (and verified) in the near--to--threshold
 proton energy range by evaluation of predicted single
 muon flux:
 \[
 I_{\qqw \mu~m=1}({\rm E}_{\qqw \mu}\geq {\rm E}_{\qqw \mu\, \rm th}) =
 \int ~\, j_{\qqw \rm p}({\rm E}_{\qqw \rm p}) 
 ~\cdot <{\rm N}_{\qqw \mu}>({\rm E}_{\qqw \rm p}) 
 ~\, ~d{\rm E}_{\qqw \rm p},
 \]
 where $j_{\qqt \rm p}$(E$_{\qqt \rm p})$ 
 is a differential CR proton energy spectrum.
 The experimentally measured muon intensity
 for E$_{\qqt \mu} \geq$~200~GeV is about 3.2~$\cdot$~10$^{\qqt -2}$
 muons per (m$^{\qqm 2}$~sec~sr), and one gets the same value using
 CR energy spectrum and composition presented
 in the Appendix~B on page~\pageref{Appx:B}
 with about 
 30\%
 contribution from components heavier 
 than protons in CR flux.
 Using JACEE CR energy spectrum 
 \cite[{\em Asakimori et al., 1995}]{CRspetr:JACEE2}
 one gets
 2.5~$\cdot$~10$^{\qqt -2}$ muons per (m$^{\qqm 2}$~sec~sr)
 with about 
 40\% 
 contribution from heavier components.
 The single muon flux is relatively well measured
 (for E$_{\qqt \mu\, \rm th}>$200~GeV this means the accuracy 
 within a factor of 1.5, mostly due to the difference in
 muon energy estimation in different experiments;
 the project {\em \lq \lq L3+Cosmics"} gives opportunity
 to measure single muon flux with accuracy better than 1\%
 for E$_{\qqt \mu}< \sim$1000~GeV, using {\em \lq \lq L3"}
 detector at LEP in CERN).
 In these calculations the spectral index in the power law
 relation of $<{\rm N}_{\qqt \mu}>$~ vs. E$_{\qqt \rm p}$ 
 is equal to $\approx$0.8.
 This value is somewhat above presently accepted value 0.76
 which comes from simple consideration and relates to the 
 experimentally measured power index in the muon group
 multiplicity distribution, where in a large detector
 rate of the large muon group
 with m muons is proportional to m$^{\qqt -3.3}$.\\

 \noindent
 Some assumptions made in presented simplified 
 estimation of power index are not sufficiently exact.
 Since the CR energy spectrum has a power law energy dependence
 with the index of \mbox{$\approx$--2.7} (below 10$^{\qqt 6}$~GeV,
 at least) the average effective energy required to produce
 N$_{\qqt \mu}^{\qqt 0}$ in the shower is somewhat lower than the energy
 required to produce $<{\rm N}_{\qqt \mu}>={\rm N}_{\qqt \mu}^{\qqt 0}$.
 The ratio between these energies
 is not a constant factor since the N$_{\qqt \mu}$
 distribution around $<{\rm N}_{\qqt \mu}>$ for the constant CR particle
 energy is wider for smaller energy than for
 higher energy (where it seems to be of Gaussian shape with the
 width of 0.15 in $\log ({\rm N}_{\qqt \mu})$ scale, 
 and not $<{\rm N}_{\qqt \mu}>$,
 nor E$_{\qqt \rm CR}$ dependent at higher energies).
 The above effect of changed width in N$_{\qqt \mu}$ distribution
 for constant E$_{\qqt \rm CR}$ requires a bit faster growing of
 $<{\rm N}_{\qqt \mu}>$~ vs. E$_{\qqt \rm CR}$ 
 than predicted by the formula~\ref{eq:avnm}
 to get agreement with observed muon group multiplicity rate.
 It is necessary to mention that
 the power law index of CR particles energy spectra
 is not known absolutely well, and different measurements
 gave very inconsistent results
 (\cite[{\em Asakimori et al., 1993}]{CRspetr:JACEE1}).
 
 \begin{figure}[ht]
 \begin{center}
 \vspace{-1.0cm}
 \mbox{
 \psfig{file=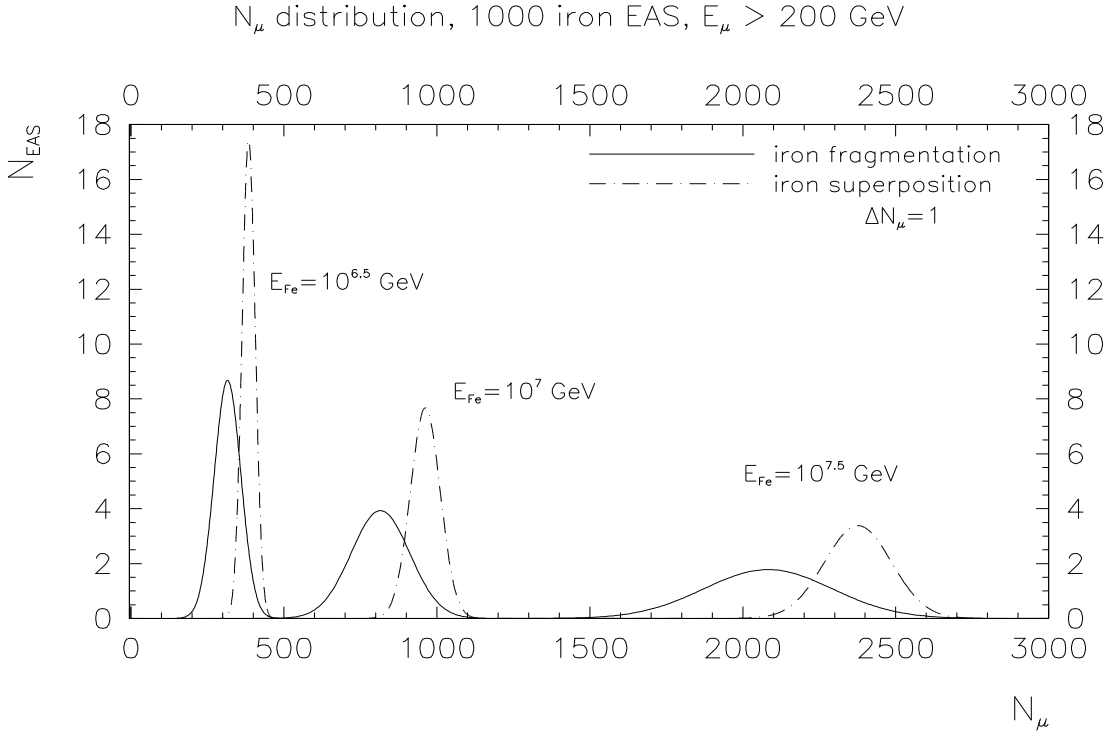,height=10.0cm,width=15.0cm}
 }\\
 \end{center}
 \caption[Compilation of Nmu distribution]
 {N$_{\qqt \mu}$ distributions (E$_{\qqt \mu} >$ 200~GeV) as result
 of M--C simulations for iron nuclei primary CR particle.
 Solid lines are results from fragmentation model for
 1000~EAS -- see formulae~\ref{eq:n1000} and \ref{eq:avnm}
 for iron -- for three primary energies
 E$_{\qqt \rm Fe}$ = 10$^{\qqt 6.5}$, 10$^{\qqt 7}$ and 10$^{\qqt 7.5}$.
 Dashed-dotted line represent predictions from the superposition
 approach -- rescaled from formulae~\ref{eq:avnm} and \ref{eq:n1000}
 for protons at 56~times smaller energy.}
 \label{fig:nmucomp}
 \end{figure}

 \subsubsection{High energy muon multiplicity 
 predicted in superposition model 
 and\\ abrasion/eva\-po\-ration model
 of nucleus--nucleus collision. \label{NNcolis}}
 Some comparison between two high energy models of 
 nucleus--nucleus interaction is presented
 (see Figure~\ref{fig:nmucomp}).
 \begin{enumerate}
 \setlength{\itemsep}{-2pt}
 \item The first model is the superposition model
 which is widely used in EAS physics.
 The result of interaction of a cosmic ray nucleus
 with nuclei of the atmosphere
 is replaced by superposition of each nucleon interaction
 with the atmosphere. If CR nucleus has the energy E$_{\qqt \rm A}$
 and the atomic number A then predicted results of the EAS development
 is obtained as a random superposition
 of A showers initiated by protons with the energy E$_{\qqt \rm A}$/A,
 i.e. nucleon energy. The results of nucleus interaction
 were replaced by the sum of A proton induced showers randomly
 sampled from the large storage or subsequently generated.
 \item The abrasion/evaporation model of nucleus--nucleus collision
 describes CR nucleus interaction in the atmosphere as
 subsequent fragmentation of incident particle together
 with evaporation of some nucleons from the excited
 fragment as presented in 
 \cite[{\em Capdevielle, 1993}]{Capd:AbrEvap}
 and 
 \cite[{\em Attallah, Szabelski et al., 1996}]{JS:AbrEvap}.
 \end{enumerate}
 In both cases nucleon and other hadron interactions
 with air--nuclei were described using
 developed version of \lq \lq dual parton model"
 presented in \cite[{\em Capdevielle, 1989}]{Capd:Model}.
 Details copncerning the atmosphere, particle decay parameters,
 cross sections parametrizations, multiplicities
 of secondary particles etc. were summarized in
 \cite[{\em Capdevielle et al., 1992}]{KfK4998}.

 \subsection{Experimental observation of muon groups.}
 \subsubsection{Muon and neutrino telescope of the
 Baksan Neutrino Laboratory of the Russian Academy of Sciences.}
 The Baksan Neutrino Observatory is placed in northern
 Caucassus in the valey of Baksan river,
 about 30~km from Mt.~Elbrus.
 The geographical coordinates are 
 N~43.42$^{\circ}$, E~42.67$^{\circ}$.
 The muon telescope is placed inside the mountain
 in a large cavity, which is made in the horizontal tunnel, 500~m
 from the entrance. The mountain slope is
 $\sim$30$^{\circ}$, so above the telescope there is
 $\sim$300~m of rock, i.e. $\sim$850~hg/cm$^{\qqm 2}$ 
 (rock's density is 2.70$\pm$0.03~g/cm$^{\qqm 3}$, Z/A=0.495,
 Z$^{\qqt 2}$/A=5.88, 
 \cite[{\em Gurencov, 1984}]{Baksan:tel1}).
 The rock above the telescope absorbs most of cosmic ray secondary
 particles; only neutrinos and high energy muons can penetrate
 to the detector. The muon threshold energy above the rock
 to penetrate it is $\sim$250~GeV for vertical direction,
 and $\sim$190~GeV for the direction of smallest rock's depth
 ($\theta~\approx$~30$^{\circ}$ inclined to the entrance).
 Muons coming from directions of large depths are also observed,
 even from direction pointing below the horizon. The latter
 are interpreted as being produced in energetic neutrino interactions
 in telescope vicinity.\\
 
 \noindent
 The Baksan muon and neutrino telescope has a size of
 \mbox{16.7~m$\times$16.7~m$\times$11.1~m}.
 It consists basically of 3150 liquid scintillation counters,
 each of size \mbox{0.7~m$\times$0.7~m$\times$0.3~m}.
 Scintillation counters make 4 horizontal layers (floors)
 separated by $\sim$3~m and 4 vertical side walls.
%(see figure~\ref{Baks:teleskope}).
 Every scintillation counter is seen by a single FEU--49
 photomultiplier. Two--fold information is collected:
 an impulse signal from the anode and an amplitude 
 signal from the 5$^{\qqt th}$ dynode. The signal from
 the anode gives the information that the energy released
 in the scintillation counter exeeded 12.5~MeV threshold
 (corresponding to 1/4 of energy released by relativistic,
 single charged vertical particle).
 The amplitude signal is used to determine energy released
 in the scintillation counter when this energy is
 in the range 0.5~GeV -- 500~GeV.
 \[
 {\rm E} \, = \, 0.5 \, {\rm GeV} \, \cdot \, 1.23^{\qqw \rm \, n - 1},
 \]
 where E is energy released and n is a channel number.
 The relative time of first impulse in each plane is registered and
 these times can be used to determine the direction 
 (e.g. up or down) of the event. 
 The absolute time of the event is also recorded.
 
 \noindent
 The telescope has many tasks:
 
 \begin{itemize}
 \setlength{\itemsep}{-2pt}
 
 \item measurements of single muon flux from different directions,
 \begin{itemize}
 \setlength{\itemsep}{-2pt}
 \item studies of muon attenuation in the rock,
 \item studies of cosmic ray anisotropy and time variation,
 \item search for cosmic ray point sources,
 \item studies of dependence of muon flux on atmospheric
   pressure and temperature of the atmospheric upper layers,
 \end{itemize}
 
 \item neutrino astrophysics,
 \begin{itemize}
 \setlength{\itemsep}{-2pt}
 \item search for signals from collapsing stars (the explosion
   of the Large Magellanic Cloud supernova in 1987 has been probably
   observed),
 \item search for neutrino sources,
 \end{itemize}
 
 \item search for proton decay,
 \item studies of high energy muon interactions,
 \item studies of muon groups.
 \end{itemize}
 
 \noindent
 The number of tasks requires large amount of information
 to be registered.
 
 \subsubsection[Data selection and muon group analysis.]
 {Data selection and muon group analysis. \footnotemark}
 \footnotetext{The studies presented in this and next section
 were performed in collaboration
 with V.~B.~Petkov and A.~A.~Voevodsky from the Russian Academy of
 Sciences and A.~Dudarewicz from the University of {\L }\'{o}d\'{z}}
 The muon groups are studied for information about
 high energy cosmic rays mass composition and
 about properties of high energy nuclear interactions.
 These studies require an extraction of rare muon group
 data from the bulk of all registered information stored
 on magnetic tapes. Muon group data are then stored
 in disk memories for further analysis. The data selection
 is not an easy task. 
 Small muon groups get some 
 \lq \lq muon group flags" already in the \lq \lq on line"
 analysis. Large muon groups data might be classified 
 incorrectly, or they might be too complicated for
 the \lq \lq on line" program and they might have
 no classification in that case.\\
 
 \noindent
 Originally we had used following criteria for
 large muon group selection:
 \begin{enumerate}
 \setlength{\itemsep}{-2pt}
 \item minimum 30 hit scintillation counters in each
 horizontal layer (floor) of the telescope,
 \item the number of hit scintillation counters
 in each horizontal layer should differ 
 from the average number of hit counters in
 horizontal layers for less than\\
 3~$\cdot~\sigma$~=~3~$\cdot~\sqrt{\rm average~number}$.
 \end{enumerate}
 
 \noindent
 The first criterion selects large events, and the second
 was applied to eliminate large electromagnetic or 
 hadronic cascades originated in high energy muon
 interaction in the telescope vicinity.\\
 After a couple of years we have learnt that electromagnetic cascades
 are more common in large muon groups than we had estimated
 originally. The second criterion has eliminated some part
 of muon groups (with $\sim$100 muons inside the telescope)
 in which electromagnetic cascades where also present.
 For the very large muon groups with more than 1/3 (above 1200)
 hit scintillation counters the second criterion was
 fulfilled despite the presence of large electromagnetic cascades.\\
 
 \noindent
 At present we use still another criterion: minimum number of hit
 scintillation counters in the whole telescope.
 For basic selection we use 100 as a minimum number,
 but we also use 300 and 600 for selection of very large
 muon group events.
 The basic criterion (100) is very often fulfilled by small
 muon groups (few muons) with electromagnetic cascades
 (few muon tracks are seen then). Of course most small muon group
 registrations do not fulfil the basic criterion, since
 they do not have electromagnetic cascades.\\
 For muon groups with $\approx$25 muons inside the telescope
 the basic criterion is of \lq \lq minimum bias" type, since
 the average muon hits about~4 scintillation counters
 while passing through the telescope.\\
 
 \subsubsection{Muon multiplicity distribution in Baksan telescope.}
 All muons in the group are parallel.
 The incoming direction of muon group can be found by searching
 for co--linear pattern of hit counters from different
 layers. The difference between time registration of the
 first hit counter in each plane can be used as a very
 crude information (for events with a large number of hit
 scintillation counters in a plane, the measurement of time
 of the first hit counter is uncertain for hardware
 reasons).
 Co--linearity is investigated by the computer program for 
 automatic data analysis. The program finds incoming
 directions for most events but large groups.
 Directions are verified \lq \lq by eye" using the program
 for visualisation of events. Corrected direction (when they are determined)
 are stored in a data base along with the automatic estimation.\\
 We have selected events with the zenith angle
 $\theta~<$~20$^{\circ}$ for the purpose of our studies,
 because it is easier to compare experimental
 results with predictions when there are events with similar
 threshold and the same zenith angle range.
 Let us notice that the rate of
 muon group events is related to the depth of rock,
 and the maximum rate is from
 direction to the entrance to the tunnel
 and $\theta~\approx$~30$^{\circ}$
 (i.e. outside our $\theta$ range).\\
 
 \noindent
 In most of registered muon group events it was possible to
 determine the gradient of muon tracks density in the
 plane perpendicular to event direction.
 The muon group core is near to extrapolated track of
 primary cosmic ray particle which has generated extensive
 air shower. 
 The average distance of a muon from the muon group core
 is about 25~m for muons with energy above $\sim$250~GeV.
 Large size of the telescope allows to measure the lateral
 distribution of high energy muons in the shower.
 In about a half of registered muon group events
 the muon group core is inside the telescope. Near the core
 the average muon energy is higher than far from the core
 because more energetic muons have narrower lateral
 distribution (for similar average lateral momentum
 they have larger longitudinal momentum).
 In this region the probability of muon interaction and
 creation of electromagnetic cascade is larger since
 muon interaction cross section depends on energy.\\

 \begin{figure}
 \begin{center}
 \vspace{-1.0cm}
 \mbox{
 \psfig{file=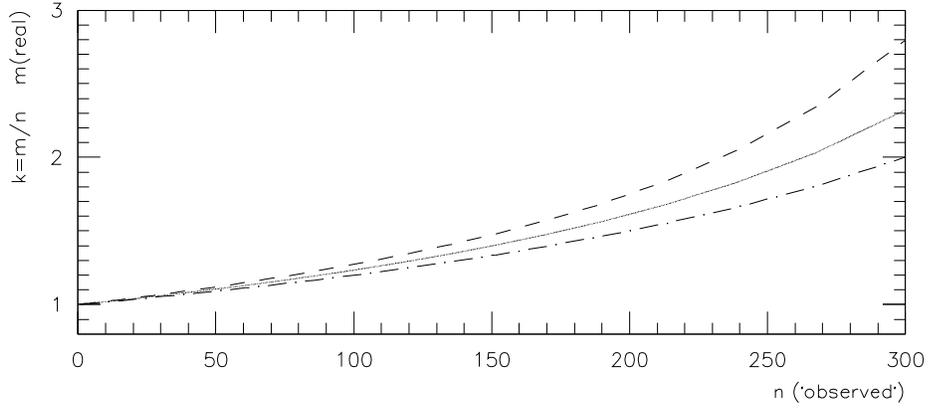,height=6.0cm,width=13.0cm}
 }\\
 \end{center}
 \caption[Muon multiplicity corrections in Baksan telescope.]
 {The figure presents correction factor from
 \lq \lq observed" number of muons to real muon number 
 which has been used in our muon group data analysis.
 The top line (dashed) was obtained by simulations
 when the EAS axis was inside the telescope,
 the bottom one (dashed--dotted): axis outside, and the central (solid)
 is the average taking into account acceptance;
 the central was used to produce the Figure~\ref{fig:mugr94}.}
 \label{fig:mugrcorr}
 \end{figure}
 
 \noindent
 Another very important problem is the spatial resolution of
 the telescope. It is limited to the single scintillation
 detector size:
 \mbox{0.7~m$\times$0.7~m$\times$0.3~m}.
 Very large muon showers could have muon densities high enough
 to allow two or more muons to pass through the same detectors.
 They would be interpreted as a single muon since the amplitude
 measurement starts at energy released above 500~MeV (i.e.
 above about 10 relativistic particles).\\
 This problem and \lq an opposite problem' of electromagnetic
 and hadronic cascades induced by muons in the detector
 was solved in detector simulations. We made a computer
 program to count the trajectories of muons taking into
 account the geometry of the telescope (the detectors layout).
 Then another program simulated the muon shower in the telescope
 taking into account muon lateral and energy distributions and 
 detector construction (i.e. muon interactions with telescope)
 for different muon groups with shower core inside and outside
 the telescope.
 This program gives results in the same format as real data output,
 so the program counting the tracks could be used.
 As a result we have worked out the correction factors
 (shown in the Figure~\ref{fig:mugrcorr})
 to transform the number of tracks identified by the
 program to number of muons in the telescope.\\
 Running the track counting program on the real data we 
 have obtained results shown in the table~\ref{tab:Moskwa94}
 \cite[{\em Voevodsky, Szabelski et al., 1994}]{JS:Moskwa94}.
 
 \begin{table}
 \caption[Counts of muon tracks in Baksan telescope]{\label{tab:Moskwa94}
 Counts of muon tracks (n) in Baksan telescope. Results were grouped in 
 number of track ranges. 
 The number of events (N) with
 the muon tracks (n) within the n range corresponds to 
 indicated time exposure. To produce muon multiplicity distribution
 shown in the Figure~\ref{fig:mugr94}, the number of tracks (n) 
 was corrected (see text).}

 \vspace{0.3cm}
 \begin{center}
 {\normalsize
 \begin{tabular}{||c|c|c|c|c|c|c|c|c||}
 \hline 
 &&&&&&&& \\
 n range    & 21--30            & 31--40            & 41--50            & 
       51--70        & 71--100            & 101--130           & 
       130--170      & $>$~170            \\
 &&&&&&&& \\
 \hline 
 & & & & & & & & \\
 $<$n$>$    & 24.5              & 34.6              & 45.0              &
       58.6          & 80.1               &  114.2             &
       143.8         &                    \\
 &&&&&&&& \\
 \hline 
 &&&&&&&& \\
 N          &  80               &  277              & 106               &
       98            & 39                 &  32                &
       13            & 4                  \\
 &&&&&&&& \\
 \hline 
 &&&&&&&& \\
 time (sec) & 1.7~10$^{\qqw 6}$ & 1.9~10$^{\qqw 7}$ & 1.9~10$^{\qqw 7}$ &
   1.9~10$^{\qqw 7}$ &  1.9~10$^{\qqw 7}$ &  4.2~10$^{\qqw 7}$ &
   4.2~10$^{\qqw 7}$ &  4.2~10$^{\qqw 7}$ \\
 &&&&&&&& \\
 \hline 
 \end{tabular}
 }
 \end{center}
 \end{table}

\begin{figure}[h]
 \begin{center}
 \vspace{-1.0cm}
 \mbox{
 \psfig{file=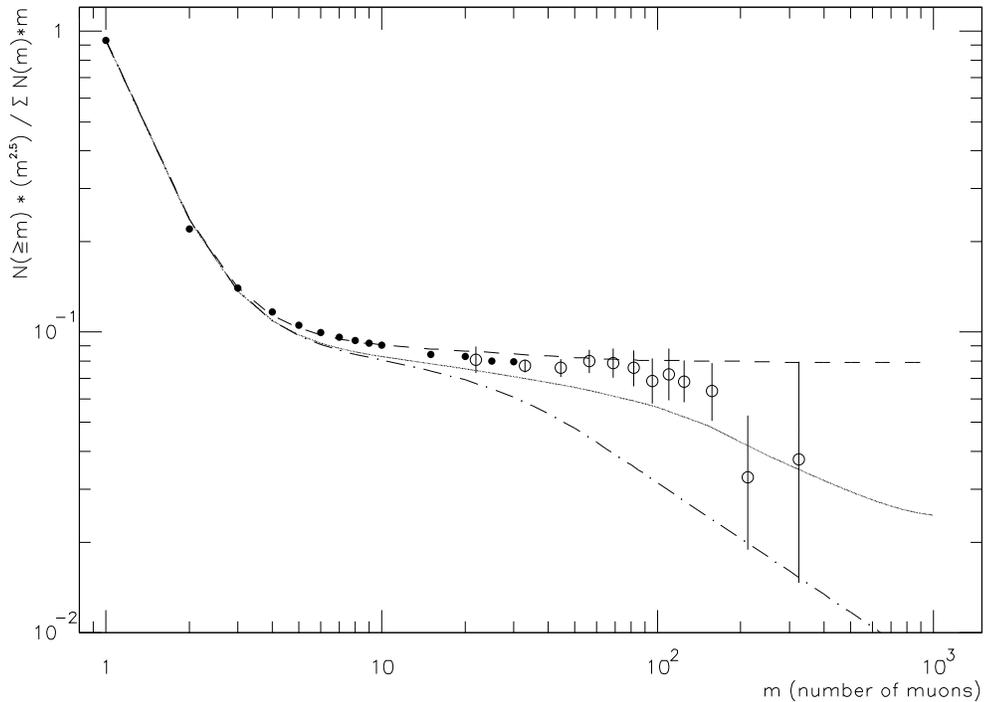,height=10.0cm,width=14.0cm}
 }\\
 \end{center}
 \caption[Muon multiplicity in Baksan telescope.]
 {The figure presents integral muon multiplicity distribution 
 (E$_{\qqt \mu\, \rm th}$~=~250~GeV) normalized to the total number
 of muons. The corresponding rates were multiplied by muon
 number in power 2.3. Black points were low multiplicity ones
 (no corrections),
 the open circles show high multiplicity muon rates
 \cite[{\em Voevodsky, Szabelski et al., 1994}]{JS:Moskwa94}.
 The curves correspond to calculated predictions
 using different primary energy spectra and mass compositions:
 dashed line -- no \lq \lq knee", 
 solid line -- \lq \lq knee" at 1.5~10$^{\qqt 15}$~eV/nucleon,
 dashed--dotted line -- \lq \lq knee" at 1.5~10$^{\qqt 15}$~eV/nucleus
 (see text for detailes).}
 \label{fig:mugr94}
 \end{figure}

 \noindent
 The number of tracks (n) obtained as a result of track counting 
 program needs to be related to the number of muons (m) in
 the telescope. We have used the average conversion factor
 presented in the Figure~\ref{fig:mugrcorr}. This
 takes into account the probability of a few muons going close together
 and mimiking one, cascades triggered by a muon in the telescope,
 and various positions of the EAS core relative to the detector area.\\
 Results are presented as a open circles in
 the Figure~\ref{fig:mugr94} (errors are statistical).
 Black points represent low muon multiplicity group rates,
 which were obtained from smaller exposure. For low multiplicity
 groups no correction was applied because the muon density was
 small enough.
 This figure is an integral muon multiplicity distribution
 for muons with E$_{\qqt \mu\, \rm th} >$~250~GeV, only muons within
 the telescope, group directions with zenith angle less than
 20~degrees. Different time exposures used for determining 
 number of events with different multiplicity were taken into
 account. The muon rate for the number of muons equal or greater
 than m
 was multiplied by m$^{\qqt 2.3}$ for presentation.
 The distribution was normalized to the total number of
 muons registered in the telescope
 (the method presented in
 \cite[{\em Chudakov, 1979}]{AECh:Kyoto}).\\

 \noindent
 The lines describe predictions obtained from calculations.
 Muon production for a given primary particle type
 and energy (i.e. EAS development, interaction
 parameters) were the same in all presented cases.
 The detailed calculations of high energy muon production
 using different interaction
 models show that the difference between models is 
 smaller than the difference which
 is due to primary particle mass (atomic number).
 Therefore it is reasonable to examine first the 
 prediction for various CR primary particles.
 The lines differ in the assumed cosmic ray mass composition
 and energy spectra. 
 The highest, 
 dashed line 
 represents the case where the
 integral CR energy spectra have constant power index (=~--1.7) 
 for all types of particles and in all related energy range
 (no \lq \lq knee"). The relative
 mass composition is the same as in low energy
 part of the next CR spectrum model. 
 This is probably not a realistic case, 
 since all EAS experiments show the \lq \lq knee" (change
 of power index of distribution) in shower size distribution
 (estimated total number of electrons in EAS at observation
 level). It can be noticed that it does not
 describe the muon group data above m~=~100.\\
 The next is a solid line which represents 
 the integral CR energy spectra which have
 a change of power index (the \lq \lq knee") 
 from --1.7 below 1.5~10$^{\qqt 15}$~eV/nucleon to
 --2.1 above this energy. The relative mass composition
 (in energy per nucleon) is constant.
 The \lq \lq knee" energy is given per nucleon, so 
 the mass composition as seen in energy per particle
 is heavier above the \lq \lq knee" than before,
 since the spectra of lighter components have the change 
 of power indexes for smaller energy per nucleon,
 than the spectra of heavier particles.
 This line lies generally below the experimental points.\\
 The bottom, dashed--dotted line 
 corresponds to
 integral CR energy spectra which have the
 change of power index (the \lq \lq knee")
 at the same energy per particle for all nuclei
 present in CR at these energies:
 from --1.7 below 1.5~10$^{\qqt 15}$~eV/nucleus
 to --2.1 above that energy.
 This spectrum has a constant relative mass composition 
 in energy per particle (i.e. per nucleus).
 The predictions are not consistent with muon group
 results presented in the Figure~\ref{fig:mugr94}.\\
 The spectra and mass compositions of cosmic rays
 are presented in the Appendix~B on the 
 page~\pageref{Appx:B}.\\
 It is interesting to compare the Figure~\ref{fig:mugr94}
 (muon multiplicity rates in Baksan telescope)
 with Figure~\ref{fig:nmusuper} (average number of
 muons produced in EAS).
 It gives an opportunity to estimate the primary CR particle
 energy related to observed muon multiplicity.
 Multiplicities 1 and 2 are mostly due to CR protons with
 energies 600~GeV -- 3000~GeV
 (6~$\cdot$~10$^{\qqt 11}$~--~3~$\cdot$~10$^{\qqt 12}$~eV). 
 The energies are lower than
 corresponding values for $<$N$_{\qqt \mu}>$ shown in the
 Figure~\ref{fig:nmusuper}, since the proton spectrum is very steep.
 When higher muon multiplicities are observed, the total
 number of muons above the threshold energy in EAS
 is about 4~--~5 times larger than the number of muons in the
 telescope, because of the lateral spread.
 Observed multiplicities around 10 correspond
 to primary energy about 
 1.5~$\cdot$~10$^{\qqt 14}$~--~4~$\cdot$~10$^{\qqt 14}$~eV.
 Observed multiplicities around 100
 require CR energies 3~$\cdot$~10$^{\qqt 16}$~--~10$^{\qqt 17}$~eV.
 The largest events with multiplicities above 300
 came from CR particles with energy above
 $\sim$1.5~$\cdot$~10$^{\qqt 17}$~eV.
 Therefore a single detector can register events triggered
 by CR with primary energy from $\sim$600~GeV to
 more than 10$^{\qqt 17}$~eV, i.e. from relatively known
 energy region (interaction properties, mass composition)
 to energies 10$^{\qqt 5}$ times higher.\\
 Results from two other large underground experiments
 have been published recently.
 \begin{itemize}
 \setlength{\itemsep}{-2pt}
 \item
 The Soudan~2 detector
 (\cite[{\em Kasahara et al., 1997}]{Soudan2})
 is at the depth of 710~m 
 (muon energy threshold $\sim$~700~GeV)
 and has dimensions 8~m~$\times$~14~m and
 5.4~m in height.
 The results of muon groups measurements 
 (multiplicity range from 1 to 12 in the detector) 
 correspond to primary particle energy range
 10$^{\qqt 12}$~--~1.3~$\cdot$~10$^{\qqt 16}$~eV.
 \item
 The MACRO detector has six \lq supermodules',
 each is 
 12~m~$\times$~12~m~$\times$~9~m in size
 \cite[{\em Ambrosio et al. 1996}]{MACRO-I}.
 The muon energy threshold is $\sim$~1400~GeV.
 The results of muon groups measurements 
 (multiplicity range from 1 to 39 in the detector) 
 correspond to primary particle energy range
 3~$\cdot$~10$^{\qqt 12}$~--~10$^{\qqt 17}$~eV
 \cite[{\em Ambrosio et al. 1996}]{MACRO-II}.
 \end{itemize}

 \subsection{Future prospects of cosmic ray
 mass composition measurements.}
 The measurement of mass composition of cosmic ray energy
 spectra in energy range 10$^{\qqt 14}$~--~10$^{\qqt 16}$~eV
 is a target of a few large currently working or being constructed
 extensive air shower (EAS) experiments. They combine
 measurements of a few components of EAS for each event 
 to more efficiently measure the primary CR
 particle energy and distinguish between different primary
 particle masses.
 Only in Europe we have:
 \begin{itemize}
 \setlength{\itemsep}{-2pt}
 \item
 Experiment HEGRA placed at mountain altitude on
 Canary Islands, which registers electro--magnetic (E--M)
 component of EAS, low energy muons and Cherenkov radiation.
 It has been working already for a few years already.
 \item
 The KASCADE experiment placed near the sea level,
 beginning to register E--M particles, low energy muons
 and hadrons (large hadron calorimeter).
 \item
 The EAS--Top array, measuring at mountain altitude 
 E--M component, low energy muons,
 Cherenkov light in coincidence with high energy muon
 (E$_{\qqt \mu\, \rm th)} >$~1400~GeV) large detector MACRO.
 \item
 The Andyrchi array, measuring E--M particles
 in coincidence with high energy muons 
 (E$_{\qqt \mu\, \rm th)} >$~250~GeV) 
 registered in the Baksan telescope (mountain altitude)
 (\cite[{\em Alexeev et al., 1993}]{Baksan:EAS},
 \cite[{\em Alexeev et al., 1994}]{Andyrchii}). 
 \end{itemize}
 The general idea in most of these experiments is to
 measure components related to EAS energy (e.g.
 Cherenkov light, low energy muons or E--M particles)
 and relate them to
 components finally determined by the early stage of EAS
 development (i.e. high energy muons or Cherenkov light)
 which almost do not interact later, and therefore
 provide information closely related to primary particle.\\

 \noindent
 The combination of Andyrchi EAS array with large 
 underground muon telescope at Baksan looks very
 promising for CR mass composition measurements.
 The EAS array detectors lie on the area of 
 5~$\cdot$~10$^{\qqt 4}$~m$^{\qqm 2}$,
 it is on average 350~m above the muon telescope and
 the array covers 0.35~sr as seen from the telescope.
 This provides about one event per 10~sec in coincidence
 with the muon inside the underground telescope.
 This rate is more than 100 times higher than in the
 case of EAS--Top/MACRO experiments, since EAS--Top
 array is much higher above the muon detector (smaller
 solid angle) and the muon threshold is about 5 times bigger.\\
 The results from Andyrchi/muon telescope
 EAS measurements should allow to narrow energy ranges
 discussed at the end of last section.
 This should provide some information about the
 CR mass composition in important high energy range.\\

 \noindent
 The interpretation of EAS data becomes much reliable
 recently due to fast development of computing power.
 It is very important that computers are fast,
 have large memories and large disk space, which allow
 to run the EAS shower programs tracing millions of
 particles through the atmosphere.
 Even more important is the existence of reliable,
 publicly available and \lq \lq friendly"
 large simulation programs for EAS studies.
 The most famous in our area is 
 the CORSIKA code 
 \cite[{\em Capdevielle et al., 1992}]{KfK4998}
 developed in Forschungszentrum Karlsruhe.
 It enables theoretical studies of different components of EAS. 
 The comparison of its results with experimental data provides
 certain confidence about the model of hadronic interactions
 used in the program, as well as the overall modeling of
 tens of physical processes, which EAS particles undergo.

 \newpage
 \section{\label{UHE-Chapter}
 The highest energy cosmic rays}
 The highest energy cosmic ray (CR) events exceed
 10$^{\qqt 20}$~eV per particle.
 In this section some results of studies of CR with energy above
 10$^{\qqt 19}$~eV are presented.\\
 The main reasons for studying the highest energy cosmic 
 rays are:
 \begin{itemize}
 \setlength{\itemsep}{-2pt}
 \item nuclear interaction properties,
 \item limits on cosmic ray energy,
 \item acceleration mechanism,
 \item source problem.
 \end{itemize}

 \noindent
 However, it should be clear that the present stage of
 research in this direction might be called: the beginning.
 There are no scientifically justified answers to above
 listed interesting problems.

 \subsection{Detectors of the highest energy cosmic rays}
 CR particles with energy above 10$^{\qqt 19}$~eV are very rare --
 about 1~event per 1--2 years on 1~km$^{\qqm 2}$~sr 
 \cite[{\em Cunningham et al., 1980}]{UHE:HP_I}.
 Each event produces huge cascade of secondary particles 
 in the atmosphere. For these two reasons the arrays for
 observing such events should have large area 
 (to increase the number of observed events),
 but they do not 
 need to have detectors very close to each other (since
 the size of extensive air shower region with detectable
 particle density is large). The observations last long and 
 detectors undergo modernizations and changes during the
 operation. Usually the array sensitivity depends on EAS
 energy, zenith angle, position. 
 For running experiments the exposure grows in time.
 Therefore given below sizes or exposures are
 approximate and valid for the time of reference.\\
 The largest arrays for detecting CR of the highest energies
 are:
 \begin{itemize}
 \setlength{\itemsep}{-2pt}
 
 \item Haverah Park (England, $\sim$12~km$^{\qqm 2}$
 \cite[{\em Bower et al., 1983}]{UHE:HP_2})
 -- 45~km$^{\qqm 2}$~yr 
 \cite[{\em Bell et al., 1974}]{UHE:SUGAR_2},
 \item SUGAR - Sydney (Australia, $\sim$110~km$^{\qqm 2}$
 \cite[{\em Winn et al., 1986}]{UHE:SUGAR_1})
 -- 175~km$^{\qqm 2}$~yr
 \cite[{\em Bell et al., 1974}]{UHE:SUGAR_2},
 \item Volcano Ranch (USA),
 -- 30~km$^{\qqm 2}$~yr
 \cite[{\em Bell et al., 1974}]{UHE:SUGAR_2},
 \item Yakutsk (Syberia)
 -- 33~km$^{\qqm 2}$~yr
 \cite[{\em Bell et al., 1974}]{UHE:SUGAR_2},
 -- 170~km$^{\qqm 2}$~yr~sr 
 \cite[{\em Bower et al., 1983}]{UHE:HP_2},
 \item Akeno Giant Air Shower Array (AGASA) (Japan -- 
      $\sim$100~km$^{\qqm 2}$)
 \cite[{\em Chiba et al., 1991}]{Akeno_1},
 \item Fly's Eye (Utah, USA, another method 
 of detecting high energy CR via detection of light from
 fluorescence of atmospheric molecular nitrogen due to ionization
 and excitation caused by relativistic particles of the extensive air
 shower
 (\cite[{\em Baltrusaitis et al., 1985}]{UHE:Fly's_eyeI}
 and \cite[{\em Linsley, 1978}]{UHE:Fly's_eyeIII}),
 Fly's Eye acceptance varies from
 $\sim$1~km$^{\qqm 2}$~sr at 10$^{\qqt 17}$~eV to
 $\sim$1000~km$^{\qqm 2}$~sr at 10$^{\qqt 20}$~eV,
 although the observation time efficiency is $\sim$6\%
 \cite[{\em Cassiday, 1985}]{UHE:Fly's_eyeII},
 (the highest energy observed event of 3$\cdot$10$^{\qqt 20}$~eV
 was reported recently, 
 \cite[{\em Bird et al., 1993}]{UHE:Fly's_eyeIV}).
 
 \end{itemize}
 Only Akeno, Yakutsk and Fly's Eye detectors are working now.
 
 \subsection{Data availability}
\begin{figure}[h]
 \begin{center}
 \vspace{-1.0cm}
 \mbox{
 \psfig{file=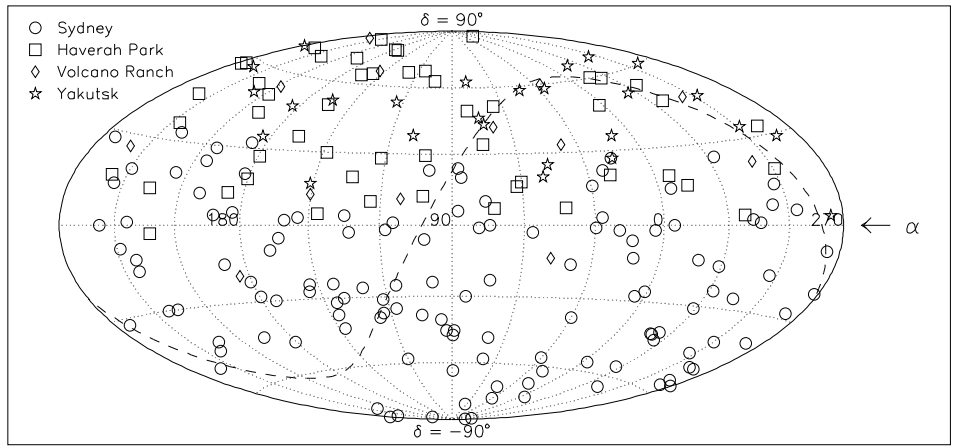,width=13.5cm,height=6.8cm}
 }\\
 \end{center}
 \caption[The sky map of UHE~CR (E$_{\qqt CR} >$~3$\cdot$10$^{\qqt 19}$~eV)]
 {\label{fig:AitoffUHECR}
 Hammer--Aitoff plot (celestial coordinates) 
 with highest energy data compilation 
 (E$_{\qqt CR} >$~3$\cdot$10$^{\qqt 19}$~eV).}
 \end{figure}

 The data on the highest energy events and
 descriptions of detectors
 from Volcano Ranch, Haverah Park, SUGAR and Yakutsk
 were printed 
 \cite[{\em Linsley, 1980}]{UHE:VRAcat},
 \cite[{\em Cunningham et al., 1980}]{UHE:HPAcat},
 \cite[{\em Winn et al., 1986}]{UHE:SUGcat} and
 \cite[{\em Efimov et al., 1988}]{UHE:YKScat}. 
 The estimated energy,
 zenith and azimuth angles, direction in equatorial
 coordinate system, time of the event
 and position of the shower core are available.
 For some events the responses of all detectors
 of the array are also given.
 Figure~\ref{fig:AitoffUHECR} shows
 directions of showers with energy above 3$\cdot$10$^{\qqt 19}$~eV.
 
 \subsection{Astrophysical information}
 The classical discussion of astrophysical problems
 related to the highest energy CR can be found in
 \cite[{\em Hillas, 1984}]{UHE:Hillas}.\\
 The energy spectrum as measured by Haverah Park,
 Volcano Ranch (the Northern hemisphere) and SUGAR
 (the Southern hemisphere) is discussed. It has a
 power law energy dependence below 10$^{\qqt 19}$~eV
 (with a power index $\sim$--3 for the differential 
 energy spectrum)
 and shows some flattening for the higher energies.
 Despite the difficulty
 of energy estimation and different methods used in different
 experiments for this purpose the energy spectra show
 similar energy dependence. However, to get the same intensities
 for these 3 experiments it was necessary to renormalize
 the SUGAR events energy by 15\% (which is within the
 possible systematic error since SUGAR detected
 only low energy muon component of EAS)
 \cite[{\em Szabelski et al., 1986}]{js:UHEold1}.
 The intensity difference could be more important as it might
 reflect the difference between the Northern and Southern
 hemispheres, which could relate to mass composition
 of highest energy CR and the galactic magnetic fields
 configuration in the Solar System vicinity 
 ({\em M.~Giller, private communication, 1993}).
 Mass composition of CR at these energies is not known
 experimentally, of course.\\

 \noindent
 The incoming direction of very high energy CR is measured
 quite well. This information and time of the event
 are much more reliable than any other measured value 
 (e.g. energy) for this energy CR.
 No important time variation nor correlation
 were found during the data analysis. 
 Also the directional distribution seems to be 
 very uniform. 
 These observed features are very difficult to
 interpret \lq \lq constructively" for studies of
 origin and source problem of CR in this energy range.\\
 Charged particles have bent paths due to galactic magnetic
 fields. 
 \label{Larmor_rad}
 The CR particle Larmor radius 
 r$_{L}$~= 1.08~E$_{\qqt 15}$/(Z~B$_{\qqt \mu\rm G}$)~pc,
 where E$_{\qqt 15}$ is equal to particle energy in 10$^{\qqt 15}$~eV unit,
 Z{\em e} is its charge, B$_{\qqt \mu\rm G}$ is the magnetic field normal
 to the particle momentum in $\mu$G and 
 1~pc~$\approx$~3.08$\cdot$10$^{\qqt 18}$~cm.
 For typical average values of B$_{\qqt \mu\rm G}\approx$~3 and particle
 energy of 10$^{\qqt 19}$~eV,
 r$_{L}$~$\approx$~3~kpc for protons, 
 and r$_{L}$~$\approx$~50~pc for iron nuclei.
 The \lq \lq local" thickness of the galactic disc
 defined as a length of column containing half of neutral hydrogen
 is equal to $\sim$100~pc. 
 For these reasons one might expect to observe anisotropy of
 arrival direction of the highest energy CR if their sources
 were galactic or relatively local.
 
 \subsubsection{\label{n-PCF} {\em n}--point correlation function}
 Since there are no obvious directions on the sky which the 
 highest energy CR events would \lq prefer',
 it is good to have a well defined method of searching for
 possible local anisotropies in experimental data.
 This method should meet following requirements:
 \begin{itemize}
 \setlength{\itemsep}{-2pt}
 \item work for small and large statistics,
 \item allow for different angular size of clustering,
 \item allow for statistical comparison between observations
  and predictions. 
 \end{itemize}
 The last point is particularly complicated because experiments
 are located at different positions on the earth, they have different
 acceptance (which is also zenith angle dependent). So the expected
 sky coverage is not uniform.
 We found that the {\em n}--point correlation function
 meets above requirements.\\
 
 \begin{figure}[ht]
 \begin{center}
 \vspace{-1.0cm}
 \mbox{
 \psfig{file=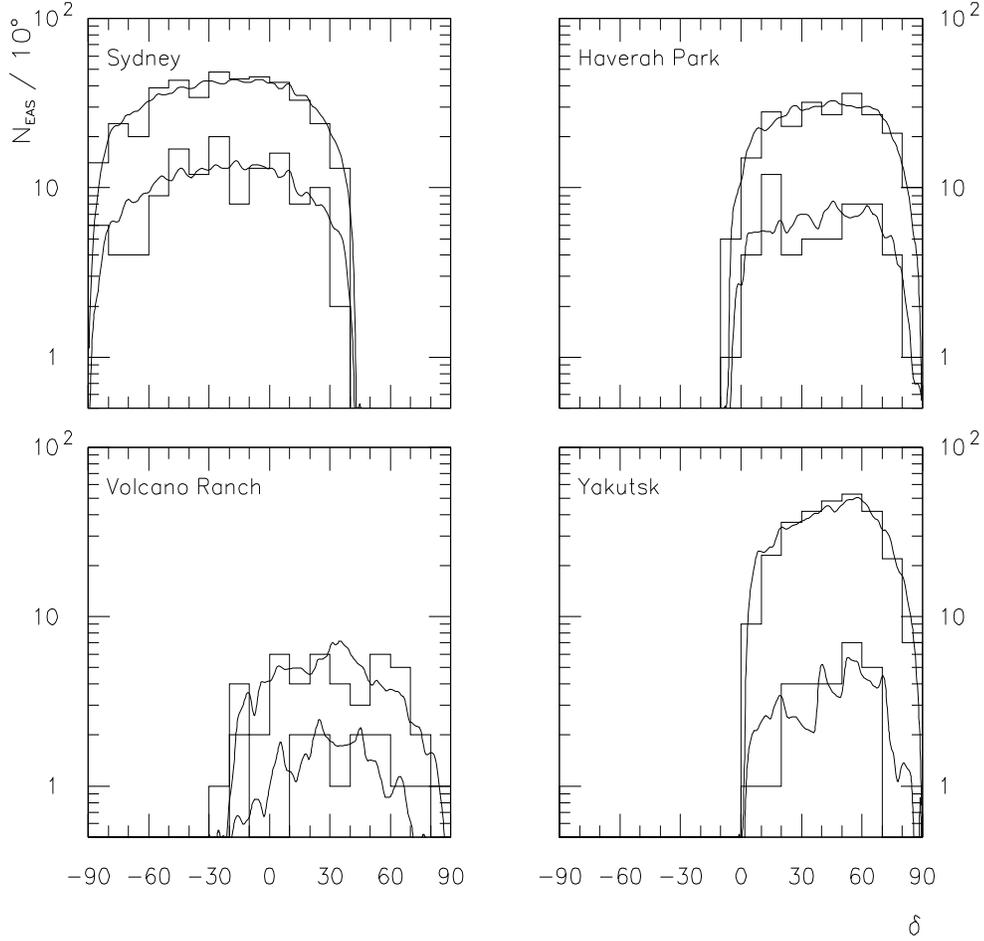,width=13.5cm,height=13.5cm}
 }\\
 \end{center}
 \caption[UHE~CR declination distribution]
 {\label{fig:UHEdecl}
 The declination distribution of arrival direction of CR from the 
 zenith angle distribution (lines) and those observed (histograms).
 The compilation of data from experiments at Sydney, Haverah Park,
 Volcano Ranch, and Yakutsk.
 Upper histogram and line represent CR events
 with energy E$_{\qqt CR} >$~10$^{\qqt 19}$~eV whereas
 lower histogram and line correspond to 
 E$_{\qqt CR} >$~3$\cdot$10$^{\qqt 19}$~eV.}
 \end{figure}

 \noindent
 We define the {\em n}--point correlation function for a sample containing
 {\em M} events as follows 
 \cite[{\em Chi, Szabelski et al., 1991}]{js:uheJap}:

 \[ {\rm w}_{n}(\theta )~=~\sum_{\qqq i=1}^{\qqq M} \, \vartheta 
 \left( \sum_{\qqq j=1,j \neq i}^{\qqq M} R(\theta_{ij})~-~n \right) \]
 
 where 
 \hspace{0.4cm}
 $\vartheta (x) = \left\{ \begin{array}{rrc}
 1 & {\rm if} & x \geq 0 \\
 0 & {\rm if} & x < 0
 \end{array} \right. $
 \hspace{0.6cm}
 and
 \hspace{0.4cm}
 $ R(\theta_{ij}) = \left\{ \begin{array}{rrc}
 1 & {\rm if} & \theta_{ij} \leq \theta \\
 0 & {\rm if} & \theta_{ij} > \theta
 \end{array} \right. $\\
 
 \noindent
 $\theta_{ij}$ is the angular separation of two events on the sky.\\
 It is possible to calculate the predicted values of w$_{n}(\theta)$
 for each declination (declination band). Let $\varrho$ be the density
 of events in a declination range (i.e. number of events per steradian)
 and let there be {\em M} events in this range. Then, in an area limited
 by the circle of radius angle $\theta$, equal to 
 S~=~2~$\pi$~(1--$\cos \theta$),
 the average number of particles {\em m}~=~S~$\varrho$ is expected.
 The predicted value of w$_{n}(\theta)$ can be taken from the Poisson
 distribution:
 \[ {\rm w}_{n}(\theta) = M \cdot 
 \sum_{\qqq i=n}^{\qqq \infty} 
 \left( \exp(-m) \frac{\qqq m^{\qqw i}}{\qqq i\, !} \right). \]
 
 \noindent
 The evaluation of the predicted values of w$_{n}(\theta)$ was performed
 for declination bands (but not for right ascension nor galactic latitude
 bands etc.), because the exposure coverage is approximately the same
 along the whole band due to the daily rotation of the Earth. Then it is
 possible to search for any enhancement within one declination band
 (e.g. crossing with galactic plane) or even for larger structures
 appropriately evaluating predicted average number~{\em m}.
 
 \subsubsection{Search for enhancements of 
 ultra high energy cosmic ray arrival directions}
 In the paper 
 \cite[{\em Chi, Szabelski et al., 1991}]{js:uheJap}
 arrival directions of ultra high energy (UHE) CR were examined
 and here these results are presented. The CR events
 with energy above 10$^{\qqt 19}$~eV were used for the analysis.
 For these energies the Larmor radius is larger then
 3~kpc for protons and 50~pc for iron nuclei
 (see page~\pageref{Larmor_rad}). Therefore one might expect
 some directional enhancement in the case of galactic orgin
 of these CRs. The bulk of observed events do not show any
 anisotropy, so the galactic disc is not a favourite place
 of UHE~CR sources. We had searched for
 some grouping of UHE~CR events, since it might indicate
 the direction of a (weak) source within several hundreds of
 parsecs from the Sun.\\

 \noindent
 About 700 events from 
 Volcano Ranch \cite[{\em Linsley, 1980}]{UHE:VRAcat},
 Haverah Park \cite[{\em Cunningham et al., 1980}]{UHE:HPAcat},
 SUGAR \cite[{\em Winn et al., 1986}]{UHE:SUGcat} and
 Yakutsk \cite[{\em Efimov et al., 1988}]{UHE:YKScat}
 were summed up for the direction analysis of CR with energy
 above 10$^{\qqt 19}$~eV.
 Special attention was paid to the analysis method to be
 independent of any astrophysical model of UHE~CR origin or
 propagation, i.e. instead of verifying some ideas we
 used the correlation analysis presented 
 in the section~\ref{n-PCF}.\\
 First the directions of UHE~CR were grouped according
 to their zenith angles and according to their declination
 angles. Assuming the azimuth angle symmetry and
 the equal coverage of the sidereal day by the working time
 and taking the
 actual zenith angle distribution it was possible to evaluate
 the predicted declination distribution
 for each experiment separately. The sum of declination
 distribution from all four experiments together with
 predictions is shown in the Figure~\ref{fig:UHEdecl}.

 The predicted and actual distributions agree well. 
 They were used to evaluate the average \lq \lq predicted"
 density of UHE~CR events in declination bands for
 the {\em n}--point correlation function analysis.\\
 
 \begin{figure}[t]
 \begin{center}
 \vspace{-1.0cm}
 \mbox{
 \psfig{file=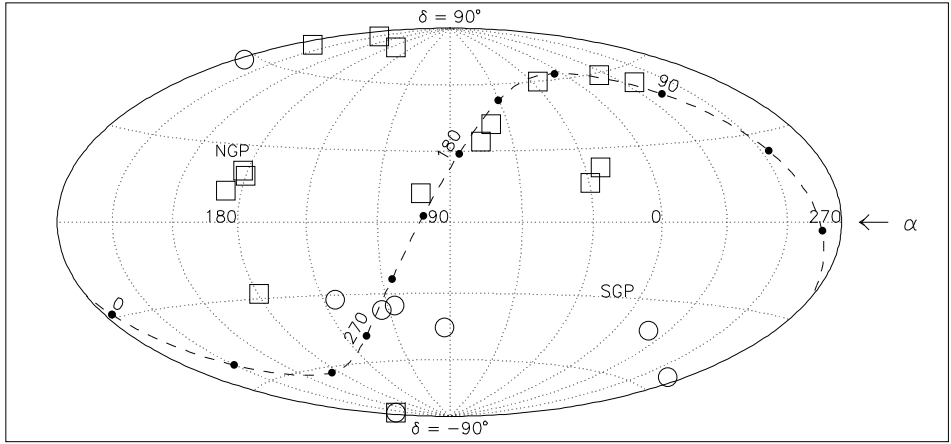,width=13.5cm,height=6.8cm}
 }\\
 \end{center}
 \caption[The sky map of UHE~CR groups]
 {\label{fig:AitoffGR}
 Hammer--Aitoff plot (celestial coordinates) 
 of \lq \lq groups" of more than 5 CR events
 (E$>$10$^{\qqt 19}$~eV) within 6$^{\circ}$ of an event
 with E$>$3$\cdot$10$^{\qqt 19}$~eV (plotted as squares)
 and groups of more than 3 CR events within
 (E$>$3$\cdot$10$^{\qqt 19}$~eV) within 6$^{\circ}$
 listed in the table~\ref{tab:UHEgroups}
 (plotted as circles)
 (those statistically significant are 
 listed in the table~\ref{tab:UHEgroups}
 and discussed in the text).
 The galactic plane and galactic poles are indicated. 
 }
 \end{figure}
 
 The {\em n}--point correlation function analysis
 (see section~\ref{n-PCF}) 
 was performed for {\em n} up to 5 and $\theta$ up to 40$^{\circ}$.
 Different event energy requirements were examined:
 \begin{enumerate}
 \setlength{\itemsep}{-2pt}
 \item all events with energy above 10$^{\qqt 19}$~eV,
 \item all events with energy above 3$\cdot$10$^{\qqt 19}$~eV and
 \item the central particle with energy above 3$\cdot$10$^{\qqt 19}$~eV
  but other with energy above 10$^{\qqt 19}$~eV.
 \end{enumerate}
 We have found a general agreement between observed and predicted
 values of w$_{n}(\theta)$ in different declination regions
 and energy bands, with two exceptions, both when all
 events have energy above 3$\cdot$10$^{\qqt 19}$~eV
 (see the table~\ref{tab:UHEgroups} for more details):
 \begin{itemize}
 \setlength{\itemsep}{-2pt}
 \item for $-$90$^{\circ}<\delta<-$60$^{\circ}$ there is 1 event with 
 4 other events within 6$^{\circ}$ of it, whereas
 0.01 is expected (14 events in the region and a chance probability
 6.6$\cdot$10$^{\qqt -4}$),
 \item for $-$60$^{\circ}<\delta<-$30$^{\circ}$ there are
 3 events with 3 other events associated within 6$^{\circ}$
 whereas 0.067 is expected. There are also 3 other events in this
 region with 2 events associated within 6$^{\circ}$, whereas
 0.57 is expected.
 \end{itemize}
 
 \begin{table}[t]
 \caption[Coordinates of the UHE~CR groups]{\label{tab:UHEgroups}
 Positions of the central event of the UHE~CR groups on the sky.
 All events in listed groups have E$_{CR} >$~3$\cdot$10$^{\qqt 19}$~eV.}

 \vspace{0.3cm}
 \begin{center}
 \begin{tabular}{||crc|crc|r|r|r|r||}
 \hline
 &&&&&&&&&\\
 \multicolumn{3}{||c|}{central CR} &
 \multicolumn{3}{c|}{number of} &&&& \\
 \multicolumn{3}{||c|}{event energy} &
 \multicolumn{3}{c|}{events} &
 \multicolumn{1}{c|}{$\alpha$} & 
 \multicolumn{1}{c|}{$\delta$} & 
 \multicolumn{1}{c|}{{\em l}$_{gal}$} &
 \multicolumn{1}{c||}{{\em b}$_{gal}$} \\
 \multicolumn{3}{||c|}{(10$^{\qqw 19}$~eV)} &
 \multicolumn{3}{c|}{within 6$^{\circ}$} &&&& \\
 &&&&&&&&&\\
 \hline
 &&&&&&&&&\\
 & 4.06 &&& 5 && 
 255$^{\circ}$ & $-$82$^{\circ}$ & 311$^{\circ}$ & $-$23$^{\circ}$ \\
 & 7.20 &&& 4 && 
 340$^{\circ}$ & $-$42$^{\circ}$ & 355$^{\circ}$ & $-$60$^{\circ}$ \\
 & 6.61 &&& 4 && 
 281$^{\circ}$ & $-$55$^{\circ}$ & 341$^{\circ}$ & $-$21$^{\circ}$ \\
 & 4.33 &&& 4 && 
 93$^{\circ}$ & $-$45$^{\circ}$ & 253$^{\circ}$ & $-$25$^{\circ}$ \\
 & 5.44 &&& 3 && 
 123$^{\circ}$ & $-$37$^{\circ}$ & 255$^{\circ}$ &  $-$1$^{\circ}$ \\
 & 4.45 &&& 3 && 
 116$^{\circ}$ & $-$35$^{\circ}$ & 250$^{\circ}$ &  $-$5$^{\circ}$ \\
 & 4.65 &&& 3 && 
 144$^{\circ}$ & $-$32$^{\circ}$ & 262$^{\circ}$ & $+$15$^{\circ}$ \\
 &&&&&&&&&\\
 \hline
 \end{tabular}
 \end{center}
 \end{table}
 
\begin{figure}[hh]
 \begin{center}
 \vspace{-0.5cm}
 \mbox{
 \psfig{file=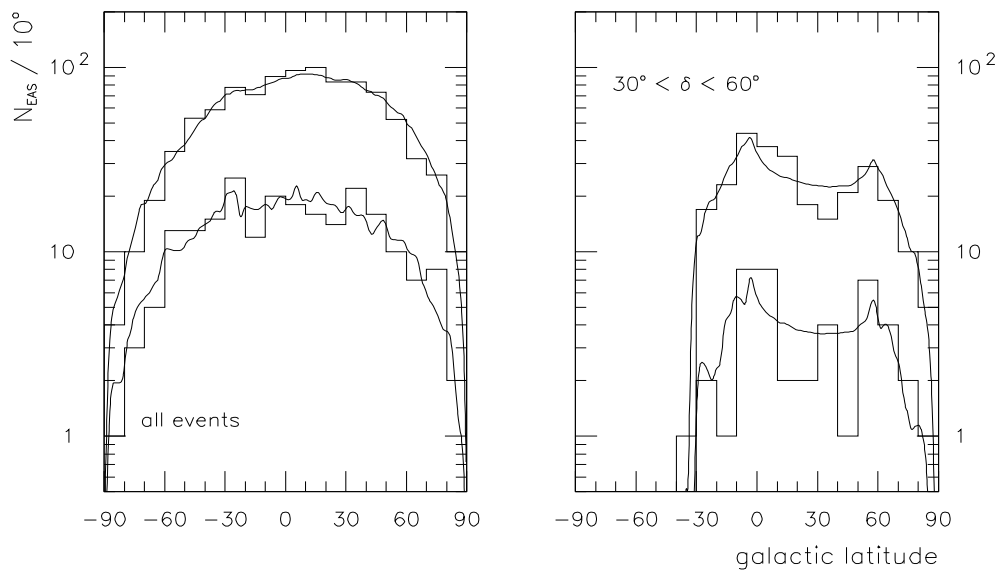,width=13.5cm,height=8.2cm}
 }\\
 \end{center}
 \vspace{-0.5cm}
 \caption[UHE~CR Galactic plane enhancement in the northern
 hemisphere]
 {\label{fig:UHE30_60}
 Left figure shows the observed and predicted (from declination
 distribution) galactic latitude distribution of CR events
 with energy E$_{CR} >$~10$^{\qqt 19}$~eV (upper histogram and line)
 and E$_{CR} >$~3$\cdot$10$^{\qqt 19}$~eV (lower histogram and line).
 Data from Sydney, Haverah Park, Volcano Ranch and Yakutsk were 
 summed.\\
 Right figure shows similar distributions
 of events in declination range
 30$^{\circ}$--60$^{\circ}$ shown in the Figure~\ref{fig:UHEdecl}.
 The \lq grouping' near the galactic plane 
 (see Figure~\ref{fig:AitoffGR})
 for E$_{CR} >$~3$\cdot$10$^{\qqt 19}$~eV
 manifests here as 
 16 events observed and 10.8 predicted
 in -10$^{\circ} <$~b~$<$10$^{\circ}$ range.}
 \end{figure}

 \noindent
 All these events are in the south celestial hemisphere,
 where only the SUGAR data were available. In all cases the
 excess was seen for $\theta \leq$~6$^{\circ}$. Therefore
 we performed a search of grouping of EAS directions
 within 6$^{\circ}$ around the events 
 with E$_{CR} >$~3$\cdot$10$^{\qqt 19}$~eV
 separately for other events with energies
 E$_{CR} >$~10$^{\qqt 19}$~eV 
 and E$_{CR} >$~3$\cdot$10$^{\qqt 19}$~eV.
 Results are presented in the Figure~\ref{fig:AitoffGR}
 on the Aitoff map of the sky in
 celestial coordinates. 
 It is possible to notice some \lq \lq by eye correlation"
 of positions of the 6~CR event groups with the galactic
 plane in the northern celestial hemisphere
 (energy requirement: 
 the central particle with energy above 3$\cdot$10$^{\qqt 19}$~eV
  but other with energy above 10$^{\qqt 19}$~eV,
  squares in the Figure~\ref{fig:AitoffGR}).
 There are 4 such groups (6 events within each group)
 in part
 of the sky limited declination range 30$^{\circ}$--60$^{\circ}$.
 This part
 has very good exposure coverage (see figure~\ref{fig:UHEdecl})
 and the statistical significance of each group is small.
 In this declination region there are 212 events with 
 E$>$10$^{\qqt 19}$~eV and 33 among them with E$>$3$\cdot$10$^{\qqt 19}$~eV
 and 207.3 and 35, respectively, are expected from the zenith
 angle distribution, so there is no excess in DC signal
 in this declination range.
 Figure~\ref{fig:UHE30_60} presents the expected and observed
 galactic latitude distribution of these events. One can 
 notice that an excess of events close to the galactic plane
 ({\em b}$_{gal}$=0$^{\circ}$)
 seen in the Figure~\ref{fig:AitoffGR} in
 declination range 30$^{\circ}$~--~60$^{\circ}$ 
 does not exceed significantly the predicted values.
 
 \subsubsection{Conclusions and future development.}
 We have shown that observed CR events with
 energy E$>$10$^{\qqt 19}$~eV have isotropic distribution
 of incoming directions. There are some weak
 indications for possible grouping of ultra high energy
 CR events which have at present 
 too low statistic  
 to be interpreted as genuine source signals.
 However, there are plans to make very large CR arrays
 to measure CR events at these energies. There were
 attempts to build 1000~km$^{\qqm 2}$ array named
 EAS--1000 in the USSR (then in Kazakhstan and then in Russia).
 Now there are plans to build arrays ($\sim$1000~km$^{\qqm 2}$ each)
 in international collaboration named the Auger Project.
 The main target of these projects is to measure as high
 CR energy as possible and to examine their directions.
 Since they would have an order of magnitude better 
 acceptance than existing detectors, much better statistical
 information might be expected.

 \newpage
 \section{Summary}
 Some directions of research 
 in the past $\sim$10 years 
 in the field of cosmic rays (CR) have been presented. 
 There are also other important areas 
 of CR studies which are not addressed properly here. 
 The presented selection was based mainly on the author's contribution.
 The situation in other areas of CR studies is
 similar to the ones presented here in the sense
 that they are also in a state of rapid development.
 \begin{enumerate}
 \item
 In gamma ray astrophysics the development
 was achieved mainly due to the progress in experiments.
 \begin{itemize}
 \setlength{\itemsep}{-2pt}
 \item
 In the energy range below 30 GeV results 
 provided by the Compton Gamma Ray Observatory experiments
 (EGRET and COMPTEL) extended our knowledge
 about $\gamma$--ray sources, CR diffuse emission
 and abundance of some material 
 (i.e. $^{26}$Al, H$_{2}$)
 in our Galaxy.
 The EGRET experiment reached a regular state of art
 in measurement based on observation of 
 e$^{+}$,e$^{-}$ pair production by $\gamma$--ray:
 \begin{itemize}
 \item
 the nuclear CR background was successfully eliminated,
 \item
 the event scanning was automatic,
 \item
 the whole sky has been viewed, 
 \item
 and the statistics is reasonably good.
 \end{itemize}
 The COMPTEL experiment based on the Compton scattering
 effect covered 1~--~30~MeV $\gamma$--ray energy region.
 This was the first large and long--lasting experiment
 based on this principle and it has opened a new
 important window in experimental $\gamma$--ray astrophysics.\\
 The main discovery of EGRET was the observation and identification
 of extragalactic sources. 
 Earlier experiments did not view the whole sky.
 They seldom looked outside the galactic plane,
 mainly for background calibration reasons. 
 It was believed that
 observable $\gamma$--rays are emitted within our Galaxy.\\
 During CGRO mission
 \begin{itemize}
 \item
 half of COS~B galactic $\gamma$--ray sources were confirmed 
 (but another half were not),
 \item
 a small gradient of $\gamma$--ray emissivity
 in the outer Galaxy 
 ($\approx$~--0.3~/~4~kpc~$\times$ local emissivity)
 which we have found in COS~B data
 was confirmed. This indicates the galactic origin
 of CR in GeV range. \\
 The observations of the outer Galaxy give much more
 clear picture, since there is little 
 molecular H$_{\qqt 2}$ in this direction and 
 one might expect few unresolved sources.\\
 The $\gamma$--ray emissivity in the inner Galaxy
 shows an \lq uncomfortable' picture. The gradient
 observed in 5~--~10~kpc galactic radius range is
 very small ($\approx$~--0.2~/~5~kpc~$\times$ local emissivity). 
 Distribution of any other galactic objects 
 (e.g. pulsars, supernova remnants, young stars etc.
 as candidates for cosmic ray sources) 
 has much steeper gradient.
 The central part of Galaxy (galactic radius less than 4~kpc)
 seems to have $\gamma$--ray emissivity on the level
 observed at 10~kpc distance (i.e. smaller than at 5~kpc).
 However, unknown molecular hydrogen distribution
 and possible contribution from unresolved sources
 make this conclusion \lq model dependent'.
 \end{itemize}
 \item
 In energy range 500~GeV~--~30~TeV several Cherenkov light 
 cosmic ray detectors have been running. The Crab is probably
 the best observed source. However no pulsed emission
 related to the radio pulsar has been found.\\
 The methods of Cherenkov light measurements are still developing.
 The most important is the problem of elimination of background
 due to hadronic particles.\\
 Many efforts are made to lower the $\gamma$--ray energy threshold
 to minimalize the energy gap between satellite and
 ground based gamma ray measurements.
 \item
 Above 100~TeV a few new extensive air shower (EAS) arrays were
 specially constructed to discover CR point sources.
 No statistically significant positive observation was reported,
 which contradicts earlier rumour. It seems that the quality 
 of previous measurements, mostly due to the techniques used,
 did not provide legitimacy for drawing conclusions about
 point sources signals at the level claimed.
 \end{itemize}
 \item
 Investigations of mass composition of cosmic rays 
 at energies above 10$^{\qqt 14}$~eV/particle using 
 indirect ground based methods 
 require large experiments and a long time of data collection.
 The main reason for that is low intensity of CR particles
 with these energies.\\
 For studying the origin of CR the important information
 would be the energy spectrum of each CR component
 (i.e. separately for each nuclei or isotopes) or
 at least the energy spectrum separately for
 each group of CR nuclei (i.e. protons, helium,
 CNO group, heavy group around A$\sim$28 and iron group).
 The grouping of atomic masses in CR aboundance 
 (i.e. presence of elements with atomic masses within
 some limits and relative absence of others)
 is known from direct measurements at lower energies
 and
 might not be present at higher energies
 (however, the grouping at lower energies has nuclear and atomic 
 physics justification).\\
 The important role of the knowledge of CR mass composition
 at energies above 10$^{\qqt 14}$~eV/particle is known
 for at least 40~years. 
 Many different techniques and methods were used:
 \begin{itemize}
 \setlength{\itemsep}{-2pt}
 \item
 extensive air shower (EAS) arrays to detect electro--magnetic
 ($\gamma$--ray, e$^{+}$ and e$^{-}$) or/and low energy (less than 20~GeV)
 muon component of EAS,
 \item
 large area X--ray films exposed at mountain altitudes,
 \item
 atmospheric Cherenkov light detectors,
 \item
 EAS hadron detectors at ground level,
 \item
 large deep underground detectors for high energy muons 
 (single muons and muon groups),
 \item
 \lq \lq Fly's eye" detector to see atmospheric scintillations
 due to EAS,
 \end{itemize}
 and combinations of these techniques at the same place and time.\\
 However, the EAS seems still to be very complicated event
 especially to draw physically important conclusions from its
 observation.
 The main progress made during last ten years in this area
 is in Monte--Carlo simulations of the event and in 
 precision of measurements of particular EAS components.
 \begin{itemize}
 \setlength{\itemsep}{-2pt}
 \item
 The new, large computer programs to simulate EAS development 
 were made; CORSIKA is probably the best example of them.
 The programs were publicly available. 
 So far there was no standard method of comparing different results. 
 Now they can be related to Monte--Carlo results 
 obtained by the same computer code. 
 This is very important since there is no standard EAS array. 
 Also different computer programs can be related to
 one standard. 
 The \lq \lq standard" is not perfect, it has some approximation 
 and some extrapolation to physically unknown areas. 
 It is interesting to see whether changing models of unknown
 physics would change predicted observations. 
 It can be shown that 
 many model parameters of physical processes in EAS 
 have no correlation with observed
 results in some EAS components.
 \item
 The EAS experiments are more \lq \lq carefully" designed.
 For studies of many physical and astrophysical properties 
 the altitude of experiment plays very important role.
 \lq Higher' means not only nearer to the first interaction
 (smaller number of interactions in EAS development) but
 also places experiment in the better position 
 in EAS size development. Cherenkov light is strongly
 absorbed in the air below altitude of $\sim$1.5~km.\\
 The individual detectors of EAS array provide detailed
 information about the time and particle densities
 at the place. The quality of this information has
 improved (one has to keep in mind that these
 experiments are made outside laboratories and run for years).\\
 All detectors are continuously monitored for their proper
 work and detailed data are stored in computer memories
 for further processing.
 \end{itemize}
 For the question of mass composition of CR from EAS measurements
 the most zing experiments can be selected.
 Here it was shown that high energy muons observations
 can provide results sensitive to primary mass composition.
 The predictions of muon multiplicity rates can be only a
 little affected by the interaction 
 models at high energy whereas they are sensitive to
 primary CR mass composition.\\
 However, the interpretation of experimental results is still difficult
 due to effects produced by the steepness of
 primary CR energy spectrum, fluctuations in EAS development 
 and the extreme acceptance, near to sensitivity limits, of currently
 operating detectors.
 Two large underground muon detectors have \lq \lq top" EAS
 arrays above them on the mountain surface: 
 Baksan muon telescope (Russia)
 and
 Gran Sasso MACRO (Italy).
 Probably more spectacular results from these experiments will
 appear when enough data from coherent measurements 
 will be collected and processed.
 \item
 Studies of the highest energy cosmic rays are going \lq slower' 
 because it takes many years to accumulate required number of events. 
 The problem is important since there are plans to
 make one or two very large detectors to study CR with 
 these energies. Situation presented here shows that our knowledge   
 about CR with energy above $\sim$10$^{\qqt 19}$~eV is very
 limited. It is possible to measure the direction of such event and
 its time, but other characteristics might have large 
 errors. Especially conversion from the measured size 
 (or strength) of the event to primary CR particle energy
 is unclear. There are only very approximate simulations of
 the highest energy CR EAS development, and there are clear
 disagreements in energy spectra obtained by different experiments.\\
 Therefore we have concentrated on directional properties
 of these CR, which is very important in the studies of their
 origin. We adopted special mathematical
 methods of searching for possible clustering of the 
 directions of events. Our conclusion is that the 
 assumption of isotropy is valid, despite some 
 interesting grouping observed \lq by eye'.\\
 We are far from scientific explanation of origin of
 these extremely energetic particles.
 \end{enumerate}

 \noindent
 While studying the origin of cosmic rays one can find
 a message from the {\em Nature}: \lq \lq not yet".

 \newpage
 \section{Appendix A: Gamma ray sources}
 \begin{table}[h]
 \caption[The 2CB Catalog of Gamma--Ray Sources (COS~B)]
 {\label{tab:COSB} 
 (Appendix A)
 The 2CB Catalog of Gamma--Ray Sources (COS~B) from 
 \cite[\em Swanenburg et al., 1981]{COSB:2CGcat}.
 COS~B sources which are confirmed in 
 \lq \lq The Second EGRET Catalog of High--Energy Gamma--Ray Sources" 
 \cite[{\em Thompson et al., 1997}]{EGRET:IIndCatalog}
 are indicated.}

 \vspace{0.3cm}
 \begin{center}
 {\normalsize
 \begin{tabular}{lrrrll}
 \hline
 &&&&&\\
 & \multicolumn{2}{c}{position} & \multicolumn{1}{c}{flux} && \\
 \cline{2-3}
 &&& \multicolumn{1}{c}{E$>$100~MeV} &
 \multicolumn{1}{c}{2-nd EGRET}
 & \\
 \multicolumn{1}{c}{COS~B} &
 \multicolumn{1}{c}{\em l$_{gal}$} & \multicolumn{1}{c}{\em b$_{gal}$} 
 & \multicolumn{1}{c}{${\rm 10}^{\qqw -6} \times$} &
 \multicolumn{1}{c}{Catalog}
 & \\
 \multicolumn{1}{c}{source name} & \multicolumn{2}{c}{(degrees)} & 
 \multicolumn{1}{c}{$\left( 
 \frac{\qqq \rm photons}{\qqq \rm cm^{\qqm 2}~sec} 
 \right)$} &
 \multicolumn{1}{c}{source name}&
 \multicolumn{1}{c}{identification} \\
 &&&&& \\
 \hline                                              
 &&&&& \\                                           
 2CG~006$-$00 &   6.7 &  $-$0.5 &  2.4 & 2EG~J1801$-$2312 & \\
 2CG~010$-$31 &  10.5 & $-$31.5 &  1.2 && \\
 2CG~013$+$00 &  13.7 &  $+$0.6 &  1.0 && \\
 2CG~036$+$01 &  36.5 &  $+$1.5 &  1.9 && \\
 2CG~054$+$01 &  54.2 &  $+$1.7 &  1.3 && \\
 &&&&& \\
 2CG~065$+$00 &  65.7 &     0.0 &  1.2 && \\
 2CG~075$+$00 &  75.0 &     0.0 &  1.3 & 2EG~J2019$+$3719 & \\
 2CG~078$+$01 &  78.0 &  $+$1.5 &  2.5 & 2EG~J2020$+$4026 & \\
 2CG~095$+$04 &  95.5 &  $+$4.2 &  1.1 && \\
 2CG~121$+$04 & 121.0 &  $+$4.0 &  1.0 && \\
 &&&&& \\
 2CG~135$+$01 & 135.0 &  $+$1.5 &  1.0 & 2EG~J0241$+$6119 & \\
 2CG~184$-$05 & 184.5 &  $-$5.8 &  3.7 & 2EG~J0534$+$2158 & PSR~0531$+$21 - Crab pulsar \\
 2CG~195$+$04 & 195.1 &  $+$4.5 &  4.8 & 2EG~J0633$+$1745 & Geminga \\
 2CG~218$-$00 & 218.5 &  $-$0.5 &  1.0 && \\
 2CG~235$-$01 & 235.5 &  $-$1.0 &  1.0 && \\
 &&&&& \\
 2CG~263$-$02 & 263.6 &  $-$2.5 & 13.2 & 2EG~J0835$-$4513 & PSR~0833$-$45 - Vela pulsar\\
 2CG~284$-$00 & 284.3 &  $-$0.5 &  2.7 & 2EG~J1021$-$5835 & \\
 2CG~288$-$00 & 288.3 &  $-$0.7 &  1.6 & 2EG~J1049$-$5847 & \\
 2CG~289$+$64 & 289.3 &  $+$64.6 & 0.6 & 2EG~J1229$+$0206 & 3C~273 \\
 2CG~311$-$01 & 311.5 &  $-$1.3 &  2.1 & 2EG~J1412$-$6211 & \\
 &&&&& \\
 2CG~333$+$01 & 333.5 &  $+$1.0 &  3.8 && \\
 2CG~342$-$02 & 342.9 &  $-$2.5 &  2.0 & 2EG~J1710$-$4432 & PSR~1706$-$44 (only in EGRET) \\
 2CG~353$+$16 & 353.3 &  $+$16.0 &  1.1 && $\rho$~Oph (molecular cloud)\\
 2CG~356$+$00 & 356.5 &  $+$0.3 &  2.6 && \\
 2CG~359$-$00 & 359.5 &  $-$0.7 &  1.8 & 2EG~J1747$-$3039 & \\
 &&&&& \\
 \hline
 \end{tabular}
 }
 \end{center}
 \end{table}

 \begin{table}
 \caption[Pulsars of CGRO EGRET detector]
 {\label{tab:PSR_EGRET}
 (Appendix A)
 Parameters of gamma ray pulsars observed by EGRET detector at
 the Compton Gamma Ray Observatory from 
 {\em The First EGRET Source Catalog} 
 \cite[{\em Thompson et al., 1995}]{EGRET:IstCatalog}}
 
 \vspace{0.3cm}
 \begin{center}
 {\normalsize
 \begin{tabular}{rlrrrrrc}
 \hline
 &&&&&&&\\
 Pulsar & 
 name & 
 \multicolumn{1}{c}{\em l$_{gal}$} &
 \multicolumn{1}{c}{\em b$_{gal}$} &
 \multicolumn{1}{c}{period P} & 
 \multicolumn{1}{c}{\em \.{P}} & 
 \multicolumn{1}{c}{\em \"{P}} &
 Pulsed Flux \\ 
 &&&&&&&
 ($E>$100~MeV) \\
 &&&&&&& \\
 &&&&
 \multicolumn{1}{c}{(ms)} &&
 \multicolumn{1}{c}{(sec$^{\qqw -1}$)}&
  10$^{\qqw -6} \, \frac{\qqq \rm photons}{\qqq \rm cm^{\qqm 2} \, sec}$\\
 &&&&&&& \\
 \hline
 &&&&&&& \\
 B0531$+$21 & Crab & 184.54 & $-$5.88
 & 33.39 & 4.214E$-$13 & $-$1.18E$-$24 &
 1.8 $\pm$ 0.1
 \\
 J0630+178 & Geminga & 195.12 & 4.27
 & 236.97 & 1.095E$-$14 & $-$2.14E$-$26 &
 2.9 $\pm$ 0.1
 \\
 B0833$-$45 & Vela & 263.52 & $-$2.78
  & 89.286 & 1.244E$-$13 & 3.01E$-$25 &
 7.8 $\pm$ 1.0
 \\
 B1706$-$44 & & 343.2 & $-$3.0
  & 102.46 & 9.301E$-$14 & &
 1.0 $\pm$ 0.2
 \\
 B1055$-$52 & & 286.1 & 6.4
  & 197.2 & 5.840E$-$15 & &
 0.24 $\pm$ 0.04
 \\
 &&&&&&& \\
 \hline
 \end{tabular}
 }
 \end{center}
 \end{table}
 
 \noindent
 In the Table~\ref{tab:COSB} the list of COS~B sources is presented.
 These sources are mostly galactic, since COS~B spent most of its time
 observing the galactic plane. 13 COS~B sources (out of 25) are
 also present as $\gamma$--ray sources in
 \lq \lq The Second EGRET Catalog of High--Energy Gamma--Ray Sources" 
 \cite[{\em Thompson et al., 1997}]{EGRET:IIndCatalog}.
 More precisely:
 \begin{itemize}
 \setlength{\itemsep}{-2pt}
 \item
 2CG~342$-$02 identification with 2EG~J1710$-$4432 -- PSR~1706$-$44
 is not indicated in
 \cite[{\em Thompson et al., 1997}]{EGRET:IIndCatalog},
 however the positions are very nearby,
 \item
 identification of 2CG~288$-$00 with 2EG~J1049$-$5847 has
 a question mark (?),
 \item
 the 2CG~289$+$64 
 ({\it l$_{gal}$}~=~289.3$^{\circ}$, {\it b$_{gal}$}~=~64.6$^{\circ}$)
 (named as 3C~273 in COS~B Catalog, weak source)
 is not identified with
 2EG~J1229$+$0206 
 ({\it l$_{gal}$}~=~289.87$^{\circ}$, {\it b$_{gal}$}~=~64.40$^{\circ}$)
 (3C~273 in
 \cite[{\em Thompson et al., 1997}]{EGRET:IIndCatalog},
 relatively weak source);\\ 
 instead there were suggestions that
 2CG~289$+$64 could be mainly related to
 EGRET source 2EG~J1256$-$0546
 ({\it l$_{gal}$}~=~305.10$^{\circ}$, {\it b$_{gal}$}~=~57.06$^{\circ}$)
 (3C~279 in EGRET Catalog, more than 10 times stronger source
 than 3C~273).
 \end{itemize}

 \noindent
 In about 15 years (between COS~B results and 
 2$^{\qqw nd}$ EGRET Catalog)
 approximately half of the sources were confirmed,
 whereas the rest of excesses in $\gamma$--ray flux seem to be 
 due to diffuse emission. Progress has been achieved mainly
 due to larger statistics and better angular resolution
 of the new experiment.\\

 \noindent
 \lq \lq The First EGRET Source Catalog"
 \cite[{\em Thompson et al., 1995}]{EGRET:IstCatalog}
 lists 10 other 
 {\em high confidence detections} (above 6~$\sigma$) of galactic sources
 with $|b| \leq 10^{\circ}$, 
 8~sources (above 5~$\sigma$) with $|b| > 10^{\circ}$ and 
 25~positive detections of radio loud quasars and BL~Lac objects.
 (The catalog contains also few lists of less confident detections,
 and upper limits for selected pulsars, 
 radio loud quasars, BL~Lac objects, radio galaxies
 and Seyfert galaxies and radio quiet quasars).
 From normal galaxies
 only Large Magellanic Cloud was detected
 (indicated as LMC in the Figure~\ref{EGRET:2-nd} 
 on page~\pageref{EGRET:2-nd}).\\
 In \lq \lq The Second EGRET Catalog of High--Energy Gamma--Ray Sources" 
 \cite[{\em Thompson et al., 1997}]{EGRET:IIndCatalog}
 the list of 129 sources (+28 in the Supplement) is presented.
 There are 5 pulsars, 40 identifications with
 active galactic nuclei (AGN), 
 11 possible identifications with AGN
 and 71 sources unidentified so far.\\
 Some parameters of $\gamma$--ray pulsars
 are presented in the Table~\ref{tab:PSR_EGRET}.

 \newpage
 \section{\label{Appx:B} Appendix B: Cosmic ray energy spectra.}
 There are different presentations of cosmic ray (CR)
 energy spectra, especially when mass composition
 is presented. The most frequently used are
   (see 
 \cite[{\em Gaisser, 1990, p.~10}]{CR:Gaisser} for
 comparison and comments):
 \begin{itemize}
 \setlength{\itemsep}{-2pt}
 \item[--] number of particles per energy per nucleon, 
 \item[--] number of particles per energy per nucleus,
 \item[--] number of nucleons per energy per nucleon,
 \item[--] number of particles per rigidity 
 R $\equiv \frac{\qqq \rm p~c}{\qqq \rm Z~e}$.
 \end{itemize}

 \subsection{Low energy cosmic ray proton energy spectra.}
 In low energy range ($<$100~GeV per nucleon) it might be useful
 to have a description of the CR proton energy spectrum.
 However, it is necessary to notice the difference
 between interstellar spectrum and the spectrum
 at the top of atmosphere below about 10~GeV
 due to solar modulation. The interstellar proton
 spectrum can be approximately described as:
 \[
 \frac{\qqq d {\em j_{\qqw \rm p}}}{\qqq d {\rm T_{\qqw \rm p}}} \, = \,
 \frac{\qqq 2 \, {\rm T_{\qqw \rm p}}^{\qqw -2.75}}
 {4.26 \, {\rm T_{\qqw \rm p}}^{\qqw -0.876} \, + \, 1}
 \, \, \,
 \frac{\qqq \rm protons}{\qqq \rm cm^{\qqm 2}~sec~sr~GeV} ,
 \]
 where T$_{\qqt \rm p}$ is proton kinetic energy in GeV.\\
 The balloon  measurements made in 1976 and 1979
 \cite[{\em Webber et al., 1987}]{pr:Webber}
 provide the proton spectrum which can be described as:
 \[
 \frac{\qqq d {\em j_{\qqw \rm p}}}{\qqq d {\rm T_{\qqw \rm p}}} \, = \,
 1.6 \, ( {\rm T_{\qqw \rm p}} + {\rm m_{\qqw \rm p}} )^{\qqw -2.7}
 \, \, \,
 \frac{\qqq \rm protons}{\qqq \rm cm^{\qqm 2}~sec~sr~GeV} ,
 \]
 where m$_{\qqw \rm p}$ is proton mass in GeV/c$^{\qqt 2}$.\\

 \subsection{Cosmic ray energy spectra at high energies.}
 We present here two models of CR energy spectra and
 mass composition at high energies (above $\sim$1000~GeV). 
 We present formulae of integral spectra of CR,
 i.e. spectra with energy above the threshold energy.
 These are not measured spectra, since energy of CR particles
 is measured directly only in the lower part of presented energy range.
 These spectra were used to calculate predictions
 shown in the Figure~\ref{fig:nmusuper}.\\

 \begin{figure}[t]
 \begin{center}
 \vspace{-1.0cm}
 \mbox{
 \psfig{file=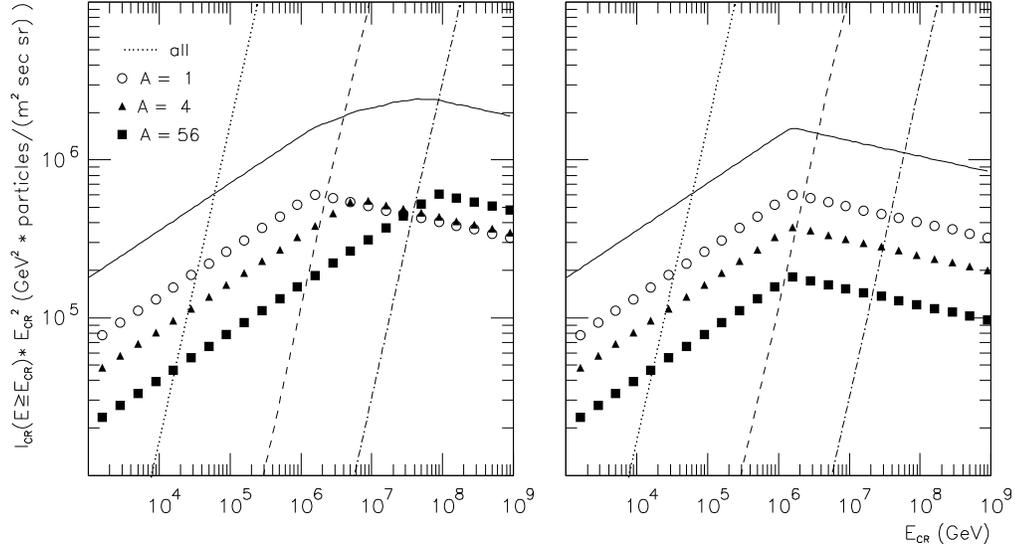,height=8.0cm,width=14.0cm}
 }\\
 \end{center}
 \caption[Models of cosmic ray energy spectra.]
 {Left figure presents a model of cosmic ray energy spectrum
 with a change of power index at constant energy per nucleon.
 The right figure presents a similar model but with 
 the change of power index at the same particle energy.
 The corresponding formulae are in the text.
 The figures present integral energy spectra multiplied by
 the particle energy squared. Only 3 components are shown,
 but solid lines represent sum of all components.\\
 The straight lines represent constant fluxes:
 dashed--dotted lines -- 
 one particle per 100~m$^{\qqm 2}$ per steradian per year, 
 dashed lines -- 
 one particle per 100~m$^{\qqm 2}$ per steradian per day,
 and dotted lines --
 one particle per 100~m$^{\qqm 2}$ per steradian per minute.}
 \label{fig:wbak}
 \end{figure}

 \subsubsection{\label{ApB:nucleon}
 Cosmic ray spectrum with a \lq \lq knee" at the constant
 energy per nucleon.
 }
 (See left Figure~\ref{fig:wbak}).
 \\

 \[ 
 I_{\qqw \rm A}(>{\rm E})\, 
    = \,C_{\qqw i}\, \cdot \,\varrho_{\qqw \rm A} \, \cdot \,
 ({\rm E/A})^{\qqw \alpha_{\qqw i}} 
 \, \, \,
 \frac{\qqq \rm particles}{\qqq \rm m^{\qqm 2} \, s \,sr}, 
 \]
 \noindent
 where CR nucleus energy - E is in GeV, 
 A is nucleus atomic number,\\
 {\em i} \,= \,1\, \, for\, \, E/A\, $\leq$\, 1.5 $\cdot$ 10$^{\qqt 6}$~GeV, 
 otherwise {\em i} \,= \,2,\\
 $\alpha_{\qqt 1}$ \,= \, --1.7, $\alpha_{\qqt 2}$ \,= \,--2.1,\\
 C$_{\qqt 1}$ \,= \,9.14 $\cdot$ 10$^{\qqt 3}$, 
 \, C$_{\qqt 2}$ \,= \,2.70 $\cdot$ 10$^{\qqt 6}$,\\

 \begin{tabular}{||c|cccccc||} 
 \hline           
  & \rule{1.2cm}{0cm} & \rule{1.2cm}{0cm} & \rule{1.2cm}{0cm} & 
  \rule{1.2cm}{0cm} & \rule{1.2cm}{0cm} & \rule{1.2cm}{0cm} \\
  A  &           
       1    &  4    & 9      & 14     & 28      & 56 \\
  &&&&&& \\
  $\varrho_{\qqw A}$ &
      0.939 & 0.055 & 0.0009 & 0.0035 &  0.0011 & 0.0003 \\ 
  &&&&&& \\
  \hline
 \end{tabular}

 \vspace{0.3cm}
 \noindent
 For E = 10$^{\qqt 5}$~GeV   
 \hspace{0.5cm}
 $I_{\qqt \rm total}$($>$E) = 7.13~$\cdot$~10$^{\qqt -5}$ 
 (m$^{\qqm 2}$\,s\,sr)$^{\qqt -1}$ and\\
 for E = 10$^{\qqt 7}$~GeV   
 \hspace{0.5cm}
 $I_{\qqt \rm total}$($>$E) = 2.14~$\cdot$~10$^{\qqt -8}$ 
 (m$^{\qqm 2}$\,s\,sr)$^{\qqt -1}$.\\

 \subsubsection{\label{ApB:jadro}
 Cosmic ray spectrum with a \lq \lq knee" at the constant
 energy per particle (nucleus).
 }
 (See right Figure~\ref{fig:wbak}).
 \\

 \[ 
 I_{\qqw \rm A}(>{\rm E})\, 
   = \,C_{\qqw i}\, \cdot \,\varrho_{\qqw \rm A} \, \cdot \,
 ({\rm E/A})^{\qqw \alpha_{\qqw i}} 
 \, \, \,
 \frac{\qqw \rm particles}{\qqw \rm m^{\qqm 2} \, sec \,sr}, 
 \]
 \noindent
 where CR nucleus energy - E is in GeV, 
 A is nucleus atomic number,\\
 {\em i} \,= \,1\, \, for\, \, E \, $\leq$\, 1.5 $\cdot$ 10$^{\qqt 6}$~GeV, 
 otherwise {\em i} \,= \,2,\\
 $\alpha_{\qqt 1}$ \,= \, --1.7, $\alpha_{\qqt 2}$ \,= \,--2.1,\\
 C$_{\qqt 1}$ \,= \,9.14 $\cdot$ 10$^{\qqt 3}$, 
 \, C$_{\qqt 2}$ \,= \,2.70 $\cdot$ 10$^{\qqt 6} \cdot$ A$^{\qqt -0.4}$,
 and other parameters are the same as before.\\
 \noindent
 For E = 10$^{\qqt 5}$~GeV   
 \hspace{0.5cm}
 $I_{\qqt \rm total}$($>$E) = 7.13~$\cdot$~10$^{\qqt -5}$ 
   (m$^{\qqm 2}$\,s\,sr)$^{\qqt -1}$ and\\
 for E = 10$^{\qqt 7}$~GeV   
 \hspace{0.5cm}
 $I_{\qqt \rm total}$($>$E) = 1.33~$\cdot$~10$^{\qqt -8}$ 
 (m$^{\qqm 2}$\,s\,sr)$^{\qqt -1}$.

 \begin{figure}[t]
 \begin{center}
 \vspace{-1.0cm}
 \mbox{
 \psfig{file=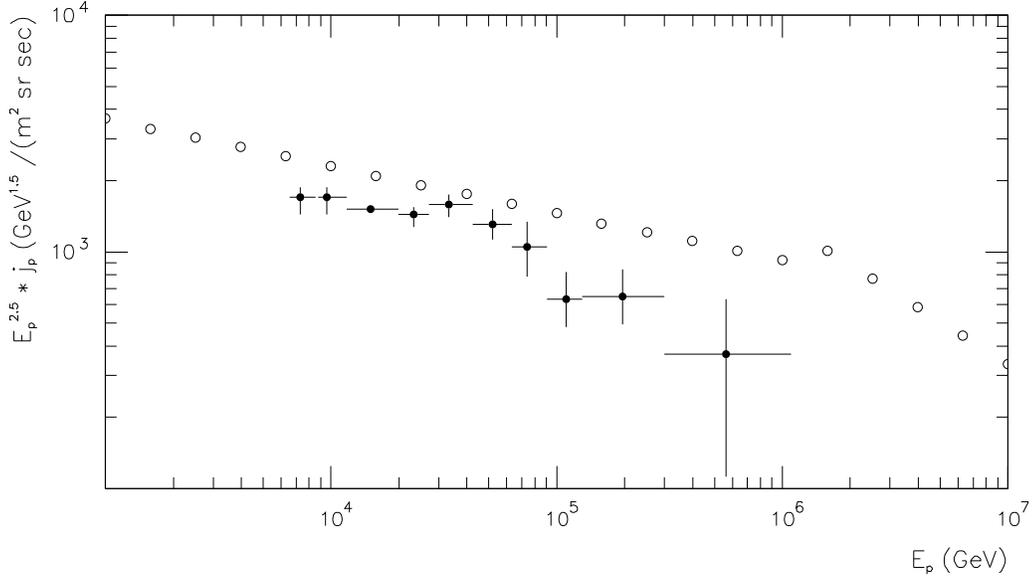,height=8.0cm,width=14.0cm}
 }\\
 \end{center}
 \caption[CR proton differential energy spectra at
 10$^{\qqt 3}$~--~10$^{\qqt 6}$~GeV.]
 {Cosmic ray proton differential energy spectra at
 10$^{\qqt 3}$~--~10$^{\qqt 6}$~GeV as measured by 
 JACEE (black points with error bars)
 (\cite[{\em Asakimori et al., 1995}]{CRspetr:JACEE2}).
 Corresponding differential proton spectrum from
 presented models is shown as open circles.}
 \label{fig:wjacee}
 \end{figure}

 \subsubsection{
 Comparison of proton energy spectra between model and direct
 measurement.}
 In energy range 7~$\cdot$~10$^{\qqt 3}$~--~10$^{\qqt 6}$~GeV/nucleus 
 cosmic ray energy spectra were measured by JACEE.
 The latest results were presented in
 \cite[{\em Asakimori et al., 1995}]{CRspetr:JACEE2}.\\
 Figure~\ref{fig:wjacee} presents a comparison
 between directly measured proton spectrum at the top
 of the atmosphere and the one accepted in the model
 for high energy CR calculations.
 Both presented models in sections~\ref{ApB:nucleon}
 and \ref{ApB:jadro} are the same for protons
 in measurable energy range. The discontinuity
 in differential model spectrum is due
 to sharp change in the slope in the integral
 representation.\\
 The measured proton spectrum has lower
 intensity than the model prediction
 in energy range 7~$\cdot$~10$^{\qqt 3}$~--~10$^{\qqt 6}$~GeV.
 For other elements (not shown here) there are also differences
 in intensities, except relatively good agreement for helium.\\
 It is not easy to accept JACEE spectrum in high energy
 cosmic ray calculations 
 (e.g. single muon spectrum which
 is measured in a few experiments, the physical processes
 involved are relatively well understood, 
 does not agree
 with assumption of JACEE spectrum).\\
 
 This discrepancy requires more experimental effort
 and quite well illustrates problems encountered while 
 comparing results from different experimental techniques
 in cosmic ray physics.

 \newpage
 \vspace{0.5cm}
 \noindent
 {\Large \bf
 Aknowledgements.}\\
 
 I would like to thank my teachers of cosmic ray physics,\\
 Professor Jerzy Wdowczyk 
 and 
 Professor Sir Arnold~W.~Wolfendale, F.R.S.\\
 who taught me the \lq revolutionary' approach,\\
 and those, from whom I have learned the \lq conservative' approach:\\
 Dr.~John L. Osborne 
 and 
 Dr.~Alexander Vladimirovich Voevodsky.\\

 I would like to address special thanks 
 for a very profitable collaboration to\\
 \mbox{Professor Jean No\"{e}l Capdevielle} 
 from Coll\`{e}ge de France.\\
 
 I am very grateful to great number of my friends and colleagues
 who worked and shared their knowledge with me,
 which was essential to my understanding of physics.\\
 In relation to works presented here I would like to
 mention Dr.~Chris~J.~Mayer, Dr.~Ken M.~Richardson and
 Dr.~Xinyu~Chi
 (then at the University of Durham, U.K.),
 Dr.~Valery Borisovich~Petkov from
 Institute of Nuclear Research of
 the Russian Academy of Sciences,
 Dr.~Pierre~Espigat from Coll\`{e}ge de France,
 Dr.~Reda~Attallah 
 and 
 Professor Christian Meynadier from
 University of Perpignan, France.\\

 I owe a special gratitude
 to Professor~Jerzy~Gawin, 
 since without his support and gentle pressure on me this work would
 not have been written.\\

 I am greatly indebted to
 Dr.~Dorota Sobczy\'{n}ska and
 Dr.~Tadeusz~Wibig
 for careful reading of the manuscript,
 their valuable advice and comments,
 and to
 Mr.~Andrzej Wo\'{z}niak for technical assistance.\\

 \noindent
 The paper was prepared using LaTeX 2.09 and
 CERN Program Library PAW.


\newpage
 \begin{thebibliography}{140}
 \setlength{\itemsep}{-2pt}
 \normalsize 

 \bibitem{GAMMA-1:86}
 B.~Agrinier, V.~V.~Akimov, M.~Avignon, V.~M.~Balebanov, 
 A.~R.~Bazer--Bachi, A.~S.~Belousov, I.~D.~Blokhintsev,
 A.~Bouere, F.~Cardon, V.~Yu.~Chesnokov, E.~I.~Chuikin,
 F.~Cotin, J.~Cretolle, M.~B.~Dobrijan, C.~Doulade, G.~Ducros,
 J.~Durand, D.~Fournier, M.~I.~Fradkin, V.~I.~Fuks, A.~M.~Gal'per,
 I.~A.~Gerasimov, V.~A.~Grigor'ev, M.~Gross, M.~V.~Guzenko,
 C.~Hugot, J.~P.~Joli, L.~F.~Kalinkin, P.~Keirle,
 S.~A.~Kirillov--Ugriumov, S.~V.~Koldashev, V.~D.~Kozlov,
 L.~V.~Kurnosova, J.~M.~Lavigne, A.~Leconte, N.~G.~Leikov, J.~P.~Leray,
 P.~Mandrou, P.~Masse, A.~A.~Moiseev, B.~Mougin, J.~Mouli,
 Yu.~I.~Nagornykh, V.~E.~Nesterov, M.~Nobileau, E.~Orsal,
 Yu.~V.~Ozerov, B.~Parlier, J.~A.~Paul, M.~Poivillier,
 V.~E.~Poluektov, A.~V.~Popov, O.~F.~Prilutskii, V.~L.~Prokhin,
 A.~Raviart, V.~G.~Rodin, V.~A.~Rud'ko, M.~F.~Runtso, M.~A.~Rusakovich,
 A.~V.~Serov, G.~Serra, F.~Soroca, S.~R.~Tabaldyev, N.~P.~Topchiev,
 G.~Vedrenne, S.~A.~Voronov, Yu.~T.~Yurkin, V.~M.~Zemskov, V.~G.~Zverev,
 {\em Soviet Astronomy} (1986)
 {\bf 30}, 508--513

 \bibitem{HEGRA:1995}
 F.~Aharonian, A.~G.~Akhperjanian, F.~Arqueros, S.~Bradbury,
 A.~S.~Beglarian, J.~Cortina, A.~Daum, T.~Deckers, E.~Faleiro,
 E.~Feigl, J.~Fernandez, V.~Fonseca, B.~Funk, J.~Gebauer,
 J.~C.~Gonzalez, V.~Haustein, G.~Heinzelmann, V.~Henke,
 G.~Hermann, M.~Hess, A.~Heusler, I.~Holl, W.~Hofmann,
 R.~Kankanian, A.~Karle, O.~Kirstein, C.~K\"{o}hler,
 A.~Konopelko, H.~Krawczynski, F.~Krennrich, H.~Kornmayer,
 A.~Lindner, E.~Lorenz, N.~Magnussen, L.~C.~Martinez, S.~Martinez,
 V.~Matheis, M.~Merk, H.~Meyer, R.~Mirzoyan, A.~Moralejo,
 H.~M\"{o}ller, N.~M\"{u}ller, M.~Panter, L.~Padilla, D.~Petry,
 R.~Plaga, A.~Plyasheshnikov, J.~Prahl, C.~Prosch, 
 G.~Rauterberg, W.~Rhode, M.~R\'{o}\.{z}a\'{n}ska, V.~Sahakian,
 J.~A.~Sanchez, M.~Samorski, D.~Schmele, R.~N.~Sooth, W.~Stamm,
 M.~Ulrich, H.~J.~V\"{o}lk, S.~Westerhoff, B.~Wiebel--Sooth,
 M.~Willmer, C.~Wiedner, H.~Wirth,
 {\em 24-th International Cosmic Ray Conference, Roma, Italy} (1995)
 {\bf 1}, 474--477
 
 \bibitem{GAMMA-1:Moskwa}
 V.~V.~Akimov, V.~M.~Balebanov, A.~S.~Belousov, I.~D.~Blokhintsev,
 I.~A.~Gerasimov, M.~B.~Dobrijan, L.~F.~Kalinkin, V.~D.~Kozlov,
 N.~G.~Leikov, Y.~I.~Nagornih, V.~E.~Nesterov, V.~E.~Poluektov,
 O.~F.~Prilutsky, V.~L.~Prohin, V.~G.~Rodin, S.~R.~Tabaldiev,
 S.~A.~Voronov, A.~M.~Galper, V.~A.~Grigoriev, V.~G.~Zverev,
 V.~M.~Zemskov, S.~A.~Kirillov--Ugriumov,
 Yu.~V.~Ozerov, A.~V.~Popov, M.~F.~Runtso, Yu.~T.~Yurkin,
 L.~V.~Kurnosova, M.~A.~Rusakovich, N.~P.~Topchiev,
 M.~I.~Fradkin, E.~I.~Chuikin,
 M.~Gross, J.~P.~Leray, P.~Masse, B.~Parlier, F.~Soroca,
 A.~R.~Bazer--Bachi, J.~M.~Lavigne,
 {\em 20-th International Cosmic Ray Conference, Moscow, USSR} (1987)
 {\bf 2}, 320--323

 \bibitem{js:GAMMA-1}
 V.~V.~Akimov, V.~M.~Balebanov, A.~S.~Belousov, I.~D.~Blokhintsev,
 M.~N.~Boyarskiv, V.~A.~Volzhenskaya,   E.~A.~Gabrilova,
 L.~F.~Kalinkin, I.~M.~Kuzenkov, V.~D.~Kozlov, N.~G.~Leikov,
 V.~E.~Nesterov, V.~G.~Rodin, A.~A.~Sukhanov, A.~A.~Tikhonov,
 S.~A.~Voronov, A.~M.~Galper, V.~M.~Zemskov,
 S.~A.~Kirillov--Ugriumov, Yu.~V.~Ozerov, A.~V.~Popov, 
 V.~A.~Rud'ko, M.~F.~Runtso, Yu.~T.~Yurkin,
 L.~V.~Kurnosova, M.~A.~Rusakovich, N.~P.~Topchiev,
 M.~I.~Fradkin, 
 P.~N.~Lebedev, 
 E.~I.~Chuikin,
 I.~A.~Gerasimov, V.~E.~Poluektov, A.~V.~Serov, V.~Yu.~Tugaenko,
 D.~Przybycie\'{n}, K.~Kossacki, J.~Szabelski,
 M.~Gross, E.~Baruch, I.~Grenier, T.~Montmerle, A.~R.~Bazer--Bachi,
 J.~M.~Lavigne, J.~F.~Olive,
 {\em 21-st International Cosmic Ray Conference, Adelaide, Australia} (1990)
 {\bf 1}, 233--236

 \bibitem{GAMMA-1:Vela}
 V.~V.~Akimov, V.~G.~Afanassyev, A.~S.~Belaoussov, I.~D.~Blokhintsev,
 L.~F.~Kalinkin, N.~G.~Leikov, V.~E.~Nesterov, V.~A.~Volsenskaya,
 A.~M.~Galper, S.~A.~Kirillov--Ugriumov, B.~I.~Lutchkov, Yu.~V.~Ozerov,
 V.~A.~Rudko, M.~F.~Runtso, V.~M.~Zemskov,
 M.~I.~Fradkin, L.~V.~Kurnosova, M.~A.~Russakovich, N.~P.~Topchiev,
 V.~Yu.~Tugaenko,
 E.~I.~Chuikin,
 M.~Gross, I.~Grenier, E.~Barouch, P.Wallyn,
 A.~R.~Bazer--Bachi, J.~-M.~Lavigne, J.~-F.~Olive,
 J.~Juchniewicz,
 {\em 22-nd International Cosmic Ray Conference, Dublin, Ireland} (1991)
 {\bf 1}, 153--156

 \bibitem{Baksan:EAS}
 E.~H.~Alekseyev, V.~V.~Alekseenko, V.~N.~Bakatanov,
 S.~N.~Boziev, A.~B.~Chernyaev, A.~E.~Chudakov, V.~I.~Guriencov,
 A.~Dudarewicz,
 S.~N.~Karpov, Yu.~N.~Konovalov, N.~F.~Klimenko, G.~D.~Korotky,
 V.~A.~Kozyarivsky, Yu.~V.~Malovichko, D.~D.~Martchuk,
 N.~A.~Metlinsky, V.~B.~Petkov, V.~Ja.~Poddubnyj,
 V.~I.~Razumnyj V.~V.~Rulev, O.~I.~Savun, A.~M.~Semenov,
 A.~M.~Sidorenko, V.~V.~Sklyarov, L.~I.~Slatvitskaya,
 V.~I.~Stepanov, A.~A.~Tarasov, A.~F.~Titenkov, A.~L.~Tsyabuk,
 A.~V.~Voevodsky, V.~I.~Volchenko, A.~F.~Yanin,
 {\em 23-rd International Cosmic Ray Conference, Calgary, Canada} (1993)
 {\bf 2}, 474--476
 
 \bibitem{Andyrchii}
 E.~H.~Alekseyev, A.~V.~Voevodsky, V.~I.~Volchenko, V.~I.~Guriencov,
 S.~N.~Karpov, G.~D.~Korotky, N.~A.~Metlinsky, R.~R.~Minnibaev,
 V.~B.~Petkov, V.~Ja.~Poddubnyj, A.~M.~Semenov, A.~A.~Tarasov,
 A.~B.~Chernyaev, A.~E.~Chudakov, A.~F.~Yanin,
 {\em preprint Russian Academy of Science, Institute for
 Nuclear Research} no. 853/94 (1994) 

 \bibitem{BASA:Gal}
 V. V. Alexeenko, Yu. M. Andreyev, A.~E.~ Chudakov, Ya.~S.~Elensky,
 L.~J.~Graham, A.~S.~Lidvansky, J.~L.~Osborne, V.~V.~Sklyarov,
 V.~A.~Tizengauzen, A.~W.~Wolfendale,
 {\em 23-rd International Cosmic Ray Conference, Calgary, Canada} (1993)
 {\bf 1}, 483--486

 \bibitem{CYGNUS:1995}
 G.~E.~Allen, D.~Berley, S.~Biller, R.~L.~Burman, M.~Cavalli--Sforza,
 C.~Y.~Chang, M.~L.~Chen, P.~Chumney, D.~Coyne, C.~L.~Dion,
 D.~Dorfan, R.~W.~Ellsworth, J.~A.~Goodman, T.~J.~Haines, C.~M.~Hoffman,
 L.~Kelley, S.~Klein, D.~M.~Schmidt, R.~Schnee, A.~Shoup, C.~Sinnis,
 M.~J.~Stark, D.~A.~Williams, J.--P.~Wu, T.~Yang, G.~B.~Yodh,
 {\em 24-th International Cosmic Ray Conference, Roma, Italy} (1995)
 {\bf 2}, 401--404

 \bibitem{MACRO-I}
 M.~Ambrosio, R.~Antolini, G.~Auriemma, R.~Baker, A.~Baldini,
 G.~C.~Barbarino, B.~C.~Barish, G.~Battistoni, R.~Belotti, C.~Bemporad,
 P.~Bernardini, H.~Bilokon, V.~Bisi, C.~Bloise, T.~Bosio, C.~Bower, 
 S.~Bussino,
 F.~Cafagna, M.~Calicchio, D.~Campana, M.~Carboni, M.~Castellano, 
 S.~Cecchini,
 F.~Cei, V.~Chiarella, A.~Corona, S.~Coutu, G.~De~Cataldo, H.~Dekhissi,
 C.~De~Marzo, I.~De~Mitri, M.~De~Vincenzi, A.~Di~Credico, O.~Erriquez, 
 R.~Fantini,
 C.~Favuzzi, C.~Forti, P.~Fusco, G.~Giacomelli, G.~Giannini, N.~Giglietto,
 M.~Goretti, M.~Grassi, A.~Grillo, F.~Guardino, P.~Guarnaccia, C.~Gustavino,
 A.~Habig, K.~Hanson, A.~Hawthhorne, R.~Heinz, J.~T.~Hong, E.~Iarocci,
 E.~Katsavounidis, E.~Kearns, S.~Kyriazopoulou, E.~Lamanna, C.~Lane, 
 D.~S.~Levin,
 P.~Lipari, N.~P.~Longley, M.~J.~Longo, G.~Mancarella, G.~Mandrioli,
 A.~Margiotta--Neri, A.~Marini, D.~Martello, A.~Marzani--Chiesa, 
 M.~N.~Mazziotta,
 G.~Michael, S.~Mikheyev, L.~Miller, P.~Monacelli, T.~Montaruli, M.~Monteno,
 S.~Mufson, J.~Musser, D.~Nicol\'{o}, R.~Nolty, C.~Okada, C.~Orth, 
 G.~Osteria,
 O.~Palamara, S.~Parlati, V.~Patera, L.~Partizii, R.~Pazzi, C.~W.~Peck, 
 S.~Petera,
 P.~Pistilli, V.~Popa, A.~Rain\'{o}, J.~Reynoldson, M.~Ricciardi, F.~Ronga,
 U.~Rubizzo, A.~Sanzgiri, F.~Sartogo, C.~Satriano, L.~Satta, E.~Scapparone,
 K.~Scholberg, A.~Sciubba, P.~Serra--Lugaresi, M.~Severi, M.~Sitta, 
 P.~Spinelli,
 M.~Spinetti, M.~Spurio, R.~Steinberg, J.~L.~Stone, L.~R.~Sulak, A.~Surdo,
 G.~Tarl\'{e}, 
 V.~Togo, V.~Valente, C.~W.~Walter, R.~Webb 
 (The MACRO Collaboration),
 {\em Laboratori Nazionali del Gran Sasso
 preprint INFN/AE--96/28} (1996)

 \bibitem{MACRO-II}
 M.~Ambrosio et al.
 (The MACRO Collaboration),
 {\em Laboratori Nazionali del Gran Sasso
 preprint INFN/AE--96/28} (1996)

 \bibitem{Tibet-II}
 M.~Amenomori, Z.~Cao, B.~Z.~Dai, L.~K.~Ding, Y.~X.~Feng, Z.~Y.~Feng,
 K.~Hibino, N.~Hotta, Q.~Huang, A.~X.~Huo, H.~Y.~Jia, G.~Z.~Jiang,
 S.~Q.~Jiao, F.~Kajino, K.~Kasahara, Labaciren, S.~M.~Liu, D.~M.~Mei,
 L.~Meng, X.~R.~Meng, Mimaciren, K.~Mizutani, J.~Mu, H.~Nanjo,
 M.~Nishizawa, A.~Oguro, M.~Ohnishi, I.~Ohta, T.~Ouchi, J.~R.~Ren,
 To.~Saito, M.~Sakata, Z.~Z.~Shi, M.~Shibata, T.~Shirai, H.~Sugimoto,
 X.~X.~Sun, K.~Taira, Y.~H.~Tan, N.~Tateyama, S.~Torii, H.~Wang,
 C.~Z.~Wen, Y.~Yamamoto, G.~C.~Yu, P.~Yuan, T.~Yuda, C.~S.~Zhang,
 H.~M.~Zhang, L.~Zhang, Zhasang, Zhaxiciren, W.~D.~Zhou,
 {\em 24-th International Cosmic Ray Conference, Roma, Italy} (1995)
 {\bf 3}, 528--531

 \bibitem{CRspetr:JACEE1}
 K. Asakimori, T.~H.~Burnet, M.~L.~Cherry, M.~J.~Christi,
 S.~Dake, J.~H.~Derrickson, W.~F.~Fountain,
 M.~Fuki, J.~C.~Gregory, T.~Hayashi, R.~Holynski, J.~Iwai, A.Iyono,
 W.~V.~Jones, A.~Jurak, O.~Miyamura, K.~H.~Moon, H.~Oda, T.~Ogata,
 T.~A.~Parnell, F.~E.~Roberts, S.~Starusz, Y.~Takahashi,
 T.~Tomiaga, J.~W.~Watts, J.~P.~Wefel, B.~Wilczynska, H.~Wilczynski,
 R.~J.~Wilkes, W.~Wolter, B.~Wosiek (The JACEE Collaboration),
 {\em 23-rd International Cosmic Ray Conference, Calgary, Canada} (1993)
 {\bf 2}, 21--29
 
 \bibitem{CRspetr:JACEE2}
 K. Asakimori, T.~H.~Burnet, M.~L.~Cherry, K.~Chevli, M.~J.~Christi,
 S.~Dake, J.H.Derrickson, W.F.Fountain,
 M.~Fuki, J.~C.~Gregory, T.~Hayashi, R.~Holynski, J.~Iwai, A.Iyono,
 J.~Johnson,
 W.~V.~Jones, M.~Kobayashi, J.~J.~Lord, 
 O.~Miyamura, K.~H.~Moon, H.~Oda, T.~Ogata,
 E.~D.~Olson,
 T.~A.~Parnell, F.~E.~Roberts, K.~Sengupta, T.~Shiina,
 S.~Starusz, T.~Sugitate, Y.~Takahashi,
 T.~Tomiaga, J.~W.~Watts, J.~P.~Wefel, B.~Wilczynska, H.~Wilczynski,
 R.~J.~Wilkes, W.~Wolter, H.~Yokomi, E.~L.~Zager (The JACEE Collaboration),
 {\em 24-th International Cosmic Ray Conference, Roma, Italy} (1995)
 {\bf 2}, 707--709, 728--731, 752--755,
 figures are in the related preprint \#~9509091 from 
 {\em astro-ph@xxx.lanl.gov}

 \bibitem{JS:ozone}
 R.~Attallah, J.~N.~Capdevielle, J.~Gawin, C.~Meynadier,
 B.~Szabelska, J.~Szabelski, A.~Wasilewski,
 {\em The Andrzej So{\l }tan Institute for Nuclear Studies 
 preprint SINS--10/VII} ISSN~1232--5309, (1995)
 
 \bibitem{JS:muRoma}
 R.~Attallah, J.~N.~Capdevielle, C.~Meynadier,
 B.~Szabelska, J.~Szabelski,
 {\em 24-th International Cosmic Ray Conference, Roma, Italy} (1995)
 {\bf 1}, 573--576, 
 
 \bibitem{JS:muIPJ}
 R.~Attallah, J.~N.~Capdevielle, C.~Meynadier,
 B.~Szabelska, J.~Szabelski,
 {\em The Andrzej So{\l }tan Institute for Nuclear Studies 
 preprint SINS--13/VII} ISSN~1232--5309, (1995)
 
 \bibitem{JS:AbrEvap}
 R.~Attallah, J.~N.~Capdevielle, C.~Meynadier,
 B.~Szabelska, J.~Szabelski,
 {\em Journal of Physics G: Nuclear Particle Physics} (1996) {\bf 22},
 1496--1506

 \bibitem{THEMISTOCLE}
 P.~Baillon,
 L.~Behr, S.~Danagoulian, B.~Dudelzak, P.~Espigat, P.~Eschstruth,
 B.~Fabre, G.~Fontaine, R.~George, C.~Ghesqui\`{e}re,
 F.~Kovacs, C.~Meynadier, Y.~Pons, R.~Riskalla, M.~Rivoal, P.~Roy,
 P.~Schune, T.~Socroun, A.~M.~Touchard and J.~Vrana,
 {\em Astroparticle Physics} (1993) {\bf 1} 341--355
 
 \bibitem{Geminga:Nature}
 C.~D.~Bailyn,
 {\em Nature} (1992) {\bf 357}, 191

 \bibitem{UHE:Fly's_eyeI}
 R. M. Baltrusaitis, G. L. Cassiday, J. W. Elbert,
 P.~R.~Gerhardy, E.~C.~Loh, Y.~Mizumoto, P.~Sokolsky,
 D.~Steck,
 {\em Physical Review} (1985) {\bf D~31}, 2192--2198

 \bibitem{UHE:SUGAR_2}
 C.~J.~Bell, A.~D.~Bray, S.~A.~David, B.~V.~Denehy,
 L.~Goorevich, L.~Horton, J.~G.~Loy, C.~B.~A.~McCusker, P.~Nielsen,
 A.~K.~Outhred, L.~S.~Peak, J.~Ulrichs, L.~S.~Wilson, M.~M.~Winn
 {\em Journal of Physics A: Math., Nucl. Gen.} 
 (1974) {\bf 7}, 990--1009

 \bibitem{CygX-3:COSB}
 K. Bennett, G. Bignami, W.~Hermsen, 
 H.~A.~Mayer-Hasselwander, J.~A.~Paul, L.~Scarsi,
 {\em Astronomy and Astrophysics} (1977)
 {\bf 59}, 273--274

 \bibitem{COMPTEL:pulsary}
 K. Bennett, M. Busetta, R. Buccheri, A. Carraminana,
 A. Connors, R.~Diehl, W.~Hermsen, L.~Kuiper,
 G.~Lichti, J.~Ryan, V.~Schonfelder, A.~Strong,
 {\em 23-rd International Cosmic Ray Conference, Calgary, Canada} (1993)
 {\bf 1}, 172--175

 \bibitem{Gaisser:gamy}
 V. S. Berezinsky, T. K. Gaisser, F. Halzen, Todor~Stanev,
 {\em Astroparticle Physics} (1993) {\bf 1}, 281--287

 \bibitem{CGRO:Geminga}
 D. L. Bertsch, K. T. S. Brazier, C. E. Fichtel,
 R. C. Hartman, S.~D~Hunter, G.~Kanbach, D.~A.~Kniffen,
 P.~W.~Kwok, Y.~C.~Lin, J.~R.~Mattox,
 H.~A.~Mayer-Hasselwander, C.~v.~Montigny, P.~F.~Michelson,
 P.~L.~Nolan, K.~Pinkau, H.~Rothermel, E.~J.~Schneid,
 M~Sommer, P.~Sreekumar, D.~J.~Thompson,
 {\em Nature} (1992) {\bf 357}, 306--307

 \bibitem{COSB:Geminga}
 G. F. Bignami, P. A. Caraveo,
 {\em Nature} (1992) {\bf 357}, 287

 \bibitem{UHE:Fly's_eyeIV}
 D.~J.~Bird, S.~C.~Corbat\'{o}, H.~Y.~Dai, B.~R.~Dawson,
 J.~W.~Elbert, T.~K.~Gaisser, K.~D.~Green, M.~A.~Huang,
 D.~B.~Kieda, S.~Ko, C.~G.~Larsen, E.~C.~Loh, M.~Luo,
 M.~H.~Salamon, J.~D.~Smith, P.~Sokolsky, P.~Sommers,
 T.~Stanev, J.~K.~K.~Tang, S.~B.~Thomas, S.~Tilav,
 {\em Physical Review Letters} (1993) {\bf 71}, 3401--3404
 
 \bibitem{BhatAWW:gradient}
 C.~L.~Bhat, M.~R.~Issa, B.~P.~Houtson,
 C.~J.~Mayer, A.~W.~Wolfendale,
 {\em Nature} (1985) {\bf 314} 511--515
 
 \bibitem{Erice:W-R}
 J.~B.~Blake and D.~S.~.P.~ Dearborn
 {\em \lq \lq Genesis and Propagation of Cosmic Rays",
 M.~M.~Shapiro and J.~P.~Wefel (eds.), NATO ASI Series {\bf C220}}
 D.~Reidel Publ. Co. (1998), 153--162

 \bibitem{Bloemen:Phd}
 J.~B.~G.~M.~Bloemen, 
 {\em PhD Thesis, 
 Leiden, Netherlands} 
 (1985)
 
 \bibitem{COSB:Bloemen}
 J.~B.~G.~M.~Bloemen, 
 {\em Annual Reveiw of Astronomy and Astrophysysics} 
 (1989) {\bf 27}, 469--516
 
 \bibitem{CasaMia-noCyg}
 A.~Borione, M.~C.~Chantell, C.~E.~Covault, J.~W.~Cronin,
 B.~E.~Fick, J.~W.~Fowler, L.~F.~Fortsen, K.~G.~Gibbs,
 K.~D.~Green, B.~J.~Newport, R.~A.~Ong, S.~Oser, L.~J.~Rosenberg,
 M.~A.~Catanese, M.~A.~K.~Glasmacher, J.~Matthews, D.~F.~Nitz,
 D.~Sinclair, J.~C.~van~der~Velde, D.~B.~Kieda,
 {\em Physical Review} (1997) {\bf 55}, 1714--1731

 \bibitem{UHE:HP_2}
 A.~J.~Bower, G. Brooke, D. Pearce, J.~C.~Perrett, A.~A.~Watson,
 {\em Journal of Physics G: Nuclear Physics} (1983) {\bf 9},
 1569--1576

 \bibitem{Durham:Cyg_rev}
 K. T. S. Brazier, A. Carrami\~{n}ana,
 P.~M.~Chadwick, N.~A.~Dipper, E.~W.~Lincoln,
 T.~J.~L.~McComb, K.~J.~Orford, S.~M.~Rayner, K.~E.~Turver,
 {\em Proc. 23-rd ESLAB Symposium on Two-Topics in X-Ray Astronomy,
 Bologna, Italy} (1989)
 ESA SP-296, 315--319

 \bibitem{Durham:teles}
 K. T. S. Brazier, A. Carrami\~{n}ana,
 P.~M.~Chadwick, T.~R.~Currell, N.~A.~Dipper, E.~W.~Lincoln,
 V.~G.~Mannings,
 T.~J.~L.~McComb, K.~J.~Orford, S.~M.~Rayner, K.~E.~Turver,
 {\em Experimental Astronomy} (1989) {\bf 1} 77--99

 \bibitem{jlo:nontherm}
 A. Broadbent, C. G. T. Haslam, J. L. Osborne,
 {\em Mon. Not. R. astr. Soc.} (1989) {\bf 237}, 381--410

 \bibitem{JACEE:widmo}
 T. H. Burnet, S. Dake, J. H. Derrickson, W. F. Fountain,
 M.~Fuki, J.~C.~Gregory, T.~Hayashi, R.~Holynski, J.~Iwai,
 W.~V.~Jones, A.~Jurak, J.~J.~Lord, O.~Miyamura, H.~Oda, T.~Ogata,
 T.~A.~Parnell, F.~E.~Roberts, S.~Starusz, T.~Tabuki, Y.~Takahashi,
 T.~Tomiaga, J.~W.~Watts, J.~P.~Wefel, B.~Wilczynska, H.~Wilczynski,
 R.~J.~Wilkes, W.~Wolter, B.~Wosiek (The JACEE Collaboration),
 {\em Astrophysical Journal} (1990) {\bf 349}, L25--L28

 \bibitem{JS:muCher}
 H.~Cabot, C.~Meynadier, D.~Sobczy\'{n}ska, B.~Szabelska,
 J.~Szabelski, T.~Wibig,
 {\em The Andrzej So{\l }tan Institute for Nuclear Studies 
 preprint SINS--18/VII} (1997) ISSN~1232--5309
 
 \bibitem{KfK4998}
 J. N. Capdevielle, P. Gabriel, H.~J.~Gils, P.~Grieder,
 D.~Heck, J.~Knapp, H.~J.~Mayer, J.~Oehlschl\"{a}ger, H.~Rebel,
 G.~Schatz, T.~Thouw,
 {\em Kernforschungszentrum Karlsruhe preprint KfK~4998} (1992)

 \bibitem{Capd:Model}
 J. N. Capdevielle,
 {\em Journal of Physics G: Nuclear Particle Physics} (1989) {\bf 15},
 909--924

 \bibitem{Capd:AbrEvap}
 J. N. Capdevielle,
 {\em 23-rd International Cosmic Ray Conference, Calgary, Canada} (1993)
 {\bf 4}, 52--54

 \bibitem{Geminga:parallax}
 P.~A.~Caraveo, G.~F.~Bignami, R.~Mignani, L.~G.~Taff,
 {\em Astrophysical Journal} (1996) {\bf 461} L91--L94

 \bibitem{UHE:Fly's_eyeII}
 G. L. Cassiday,
 {\em Annual Review of Nuclear and Particle Science}
 (1985) {\bf 35}, 321--349

 \bibitem{js:uheJap}
 X. Chi, J. Szabelski, M. N. Vahia, A.~W.~Wolfendale,
 {\em Proc. of the ICRR International Symposium
 \lq \lq Astrophysical Aspects of the Most Energetic Cosmic Rays",
 M.~Nagano and F.~Takahara (eds.), World Scientific} (1991), 140--145

 \bibitem{Akeno_1}
 N. Chiba, K. Hashimoto, N. Hayashida, K.~Honda, M.~Honda,
 N.~Inoue, F.~Kakimoto, K.~Kamata, S.~Kawaguchi, N.~Kawasumi,
 Y.~Matsubara, K.~Murakami, M.~Nagano, S.~Ogio, H.~Ohoka, To.~Saito,
 Y.~Sakuma, I.~Tsushima, M.~Teshima, T.~Umezawa, S.~Yoshida, H.~Yoshii,
 {\em 22-nd International Cosmic Ray Conference, Dublin, Ireland} (1991)
 {\bf 2}, 700--703
 
 \bibitem{AECh:Kyoto}
 A.~E.~Chudakov,
 {\em 16-th International Cosmic Ray Conference, Kyoto, Japan} (1979)
 {\bf 10}, 192--197
 
 \bibitem{CASA-MIA}
 J.~W.~Cronin, B.~E.~Fick, K.~G.~Gibbs, H.~A.~Krimm, N.~C.~Mascarenhas,
 T.~A.~McKay, D.~M\"{u}ller, B.~J.~Newport, R.~A.~Ong, L.~J.~Rosenberg,
 {\em Physical Review} (1992) {\bf D~45} 4385--4391

 \bibitem{UHE:HPAcat}
 G.~Cunningham, D.~M.~Edge D.~England, A.~C.~Evans,
 J.~D.~Hollows, S.~J.~Hopper, J.~Lapikens, B.~Liversedge,
 J.~Lloyd-Evans, P.~Odgen, M.~Patel, D.~Pearse, 
 A.~M.~T.~Pollock, R.~J.~O.~Reid, R.~M.~Tennent, R.~Walker,
 A.~A.~Watson, J.~G.~Wilson, A.~M.~Wray,
 {\em Catalogue of the Highest Energy Cosmic Rays, Haverah Park}
 (1980) {\bf No.~1}, World Data Center C2 for Cosmic Rays,
 Institute of Physical and Chemical Research, 
 Itabashi, Tokyo, Japan, pp.~61--97

 \bibitem{UHE:HP_I}
 G. Cunningham, J. Lloyd-Evans, A. M. T. Pollock, R.~J.~E.~Reid,
 A.~A.~Watson,
 {\em Astrophysical Journal} (1980) {\bf 236}, L71--L75

 \bibitem{Celeste}
 D.~Dumora, B.~Giebels, J.~Procureur, J.~Qu\'{e}bert, D.~A.~Smith,
 R.~Attallah, B.~Fabre, C.~Meynadier,
 A.~Cordier, P.~Eschstruth, B.~Merkel, Ph.~Roy,
 P.~Fleury, E.~Par\'{e}, J.~Vrana,
 P.~Espigat,
 E.~Ass\'{e}o, 
 I.~Grenier,
 G.~Henri, A.~Marcowith, G.~Pelletier,
 H.~Sol, L.~Vincente, G.~Remy,
 M.~Jires, F.~M\"{u}nz, L.~Rob,
 M.~Hrabovsky, M.~Palatka, P.~Schovanek,
 S.~P.~Ahlen, A.~Martin, J.~Rohlf,
 M.~H.~Salomon, C.~Jui, D.~Kieda,
   {\em Cerenkov Low Energy Sampling \& Timing Experiment
   CELESTE -- Experimental Proposal} 1996

 \bibitem{TDzik:PhD}
 T. J. Dzikowski,
 {\em PhD Thesis, University of {\L }\'{o}d\'{z}} (1985), p.~105
 (in Polish)

 \bibitem{Krab_RS}
 T. Dzikowski, B. Grochalska, J. Gawin, J.~Wdowczyk,
 {\em Philosophical Transactions Royal Society of London} (1981)
 {\bf A~301}, 641--644

 \bibitem{Krab_JP}
 T. Dzikowski, J. Gawin, B. Grochalska, J.~Wdowczyk,
 {\em Journal of Physics G: Nuclear Physics} (1983) {\bf 9},
 459--465
 
 \bibitem{Krab_Kosice}
 T. Dzikowski, J. Gawin, B. Grochalska, J. Korejwo, J.~Wdowczyk,
 {\em Acta Universitatis Lodziensis, folia Physica} (1984)
 {\bf 7}, 5--16

 \bibitem{UHE:YKScat}
 N.~N.~Efimov, T.~A.~Egorov, D.~D.~Krasilnikov, M.~I.~Pravdin,
 I.~Ye.~Sleptsov,
 {\em Catalogue of the Highest Energy Cosmic Rays, Yakutsk}
 (1988) {\bf No.~3}, World Data Center C2 for Cosmic Rays,
 Institute of Physical and Chemical Research, Wako, Saitama, Japan

 \bibitem{HEAO-3:nuclei}
 J.~J.~Engelmann, P.~Ferrando, A.~Soutoul, P.~Goret,
 E.~Juliusson, L.~Koch-Miramond, N.~Lund, P.~Masse,
 B.~Peters, N.~Petrou, I.~L.~Rasmussen,
 {\em Astronomy and  Astorphysics} (1990) {\bf 233}, 96--111

 \bibitem{Feynman:scal}
 R.~P.~Feynman, 
 {\em Physical Review Letters} (1969) {\bf 23}, 1415--1417
 
 \bibitem{SAS-II}
 C.E.Fichtel, R.C.Hartman, D.A.Kniffen, D.J.Thompson,
 G.F.Bignami, H.B. \"{O}gelman, M.E.\"{O}zel, T.T\"{u}mer,
 {\em Astrophysical Journal} (1975) {\bf 198}, 163--182

 \bibitem{SAS-II:tables}
 C.E.Fichtel, R.C.Hartman, D.A.Kniffen, D.J.Thompson,
 H.B. \"{O}gelman, T.T\"{u}mer, M.E.\"{O}zel,
 {\em preprint NASA GSFC Technical Memorandum 79650} (1978) 
 
 \bibitem{CR:Gaisser}
 T.~K.~Gaisser,
 {\em Cosmic Rays and Particle Physics},
 Cambridge University Press, 1990 (ISBN~0~521~32667~2)

 \bibitem{Lodz:mupoorI}
 J. Gawin, J. Hibner, A. Zawadzki,
 {\em International Cosmic Ray Conference, Jaipur, India} (1963) 
 {\bf 4}, 180--186

 \bibitem{Lodz:mupoorII}
 J. Gawin, J. Hibner, J. Wdowczyk, A. Zawadzki, R. Maze,
 {\em International Cosmic Ray Conference, London, England} (1965) 
 {\bf 2}, 639--641

 \bibitem{bs:con_acc}
 M.~Giler, J.~L.~Osborne, V.~S.~Ptuskin, B.~Szabelska,
 J.~Wdowczyk, A.~W.~Wolfendale,
 {\em Astronomy and Astrophysics} (1989) {\bf 217}, 311--318

 \bibitem{Baksan:tel1}
 V.~I.~Gurencov, 
 {\em preprint Institute for Nuclear Research,
 Academy of Science of the USSR}, P-0379, (1984)

 \bibitem{Rosat:Geminga}
 J.~P.~Halpern, S.~S.~Holt,
 {\em Nature} (1992) {\bf 357}, 222--224
 
 \bibitem{Hermsen:PhD}
 W.~Hermsen,
 {\em PhD Thesis, 
 Leiden, Netherlands} 
 (1980)

 \bibitem{COMPTEL:VelaGem}
 W.~Hermsen, K. Bennett, R. Buccheri, M. Busetta,
 A. Carrami\~{n}ana, A.~Connors, R.~Diehl, I.~A.~Grenier, L.~Kuiper,
 G.~Lichti, J.~Ryan, V.~Sch\"{o}nfelder, A.~W.~Strong,
 {\em 23-rd International Cosmic Ray Conference, Calgary} (1993)
 {\bf 1}, 176--179

 \bibitem{UHE:Hillas}
 A. M. Hillas,
 {\em Annual Review of Astronomy and Astrophysics} 
 (1984) {\bf 22}, 435--444

 \bibitem{CRspetr:Sokol}
 I.~P.~Ivanenko, V.~Ya.~Shestoperov, L.~O.~Chikova,I.~M.~Fateeva,
 L.~A.~Khein, D.~M.~Podoroznyi, I.~D.~Rapoport, G.~A.~Samsonov,
 V.~A.~Sobinyakov, A.~N.~Turundaevskyi, I.~V.~Yashin,
 {\em 23-rd International Cosmic Ray Conference, Calgary, Canada} (1993)
 {\bf 2}, 17--20

 \bibitem{EGRET:project}
 G.~Kanbach, D.~L.~Bertsch, A.~Favele, C.~E.~Fichtel,
 R.~C.~Hartman, R.~Hofstadter, E.~B.~Hughes, S.~D~Hunter, 
 B.~W.~Hughlock,  D.~A.~Kniffen, Y.~C.~Lin,
 H.~A.~Mayer-Hasselwander, P.~L.~Nolan, K.~Pinkau,
 H.~Rothermel, E.~J.~Schneid, M~Sommer, D.~J.~Thompson,
 {\em Space Science Reviews} (1988) {\bf 49} 69--84
 
 \bibitem{Soudan2}
 S.~M.~Kasahara, W.~W.~M.~Allison, G.~J.~Alner, D.~S.~Ayres, W.~L.~Barret,
 C.~R.~Bode,
 P.~M.~Border, C.~B.~Brooks, J.~H.~Cobb, D.~J.~A.~Cockerill, R.~J.~Cotton,
 H.~Courant,
 D.~M.~DeMuth, B.~Even, T.~H.~Fields, H.~R.~Gallagher, M.~C.~Goodman,
 R.~W.~Gran, R.~N.~Gray, K.~Johns,
 T.~Kafka, W.~Lesson, P.~J.~Litchfield, N.~P.~Longley,
 M.~J.~Lowe, W.~A.~Mann M.~L.~Marshak, E.~N.~May, R.~H.~Milburn, 
 W.~H.~Miller,
 L.~Mualem, A.~Napier, W.~Oliver, G.~F.~Pearce, E.~A.~Peterson, L.~E.~Price,
 D.~M.~Roback, K.~Ruddick, D.~J.~Schmid, J.~Schneps, M.~H.~Schub, 
 R.~V.~Seidlein,
 M.~A.~Shupe, N.~Sundaralingam, J.~L.~Thron, H.~J.~Trost, J.~L.~Uretsky,
 V.~Vassiliev, G.~Villaume, S.~P.~Wakely, D.~Wall, S.~J.~Werkema, N.~West,
 {\em Physical Review} (1997) {\bf D~55} 5282--5294

 \bibitem{CORSIKAv450}
 J.~Knapp, D.~Heck,  
 {\em Forschungszentrum Karlsruhe} (1995) private communication

 \bibitem{kocz:1990}
 G. E. Kocharov et al.,
 {\em 21-st International Cosmic Ray Conference, Adelaide} (1990)
 {\bf 7}, 120--123

 \bibitem{UHE:Fly's_eyeIII}
 J. Linsley,
 {\em Scientific American} (1978) {\bf 239}, 48--58

 \bibitem{UHE:VRAcat}
 J. Linsley,
 {\em Catalogue of the Highest Energy Cosmic Rays, Volcano Ranch}
 (1980) {\bf No.~1}, World Data Center C2 for Cosmic Rays,
 Institute of Physical and Chemical Research, 
 Itabashi, Tokyo, Japan, pp.~3--59

 \bibitem{CygX3:muony}
 M.~L.~Marshak, J.~Bartelt, H.~Courant, K.~Heller, T.~Joyce,
 E.~A.~Peterson, K.~Ruddick, M.~Shupe, D.~S.~Ayres, J.~Dawson,
 T.~Fields, E.~N.~May, L.~E.~Price, K.~Sivaprasad,
 {\em Physical Review Letters} (1985) {\bf 54}, 2079--2082

 \bibitem{js:CRgrad}
 C. J. Mayer, K. M. Richardson, M. J. Rogers, J. Szabelski,
 A.~W.~Wolfendale,
 {\em Astronomy and Astrophysics} (1987) {\bf 180}, 73--78

 \bibitem{CygX3:SciAm}
 P. K. MacKeown, T. C. Weekes,
 {\em Scientific American} (1985) {\bf 253}, 40--49

 \bibitem{VHE:sources}
 D.E.Nagle, T.K.Geisser, R.J.Protheroe,
 {\em Annual Review of Nuclear and Particle Science}
 (1988) {\bf 38} 609--657

 \bibitem{Punch:CAT}
 M.~Punch and CAT Collaboration
 {\em the Proceedings of the Padova Workshop on 
 TeV Gamma~--~Ray Astrophysics \lq \lq Towards a Major
 Atmospheric Cherenkov Detector--IV", 
 ed.~M.~Cresti}
 (1995),
  pp.~356--362

 \bibitem{js:arm-interarm}
 M. J. Rogers, M. Sadzi\'{n}ska, J. Szabelski,
 D.~J.~van~der~Walt, A.~W.~Wolfendale,
 {\em Journal of Physics G.: Nuclear Physics} 
 (1988) {\bf 14}, 1147--1156

 \bibitem{SPASE}
 J.~van~Stekelenborg, T.~K.~Gaisser, J.~C.~Perrett, J.~P.~Petrakis,
 T.~S.~Stanev, J.~Beaman, A.~M.~Hillas, P.~A.~Johnson, J.~Lloyd--Evans,
 N.~J.~T.~Smith, A.~A.~Watson,
 {\em Physical Review} (1993) {\bf D~48} 4504--4517

 \bibitem{COS-B:database}
 A.~W.~Strong, J.~B.~G.~M.~Bloemen, F.~Lebrun, W.~Hermsen, 
 H.~A.~Mayer--Hasselwander, R.~Buccheri,
 {\em Astronomy and Astrophysics Supplement Series}
 (1987) {\bf 67}, 283--296

 \bibitem{AWS:gradient}
 A.~W.~Strong, J.~R.~Mattox,
 {\em Astronomy and Astrophysics} (1996) {\bf 308}, L21--L24

 \bibitem{COSB:2CGcat}
 B. N. Swanenburg, K. Bennett, G. F. Bignami, R. Buccheri,
 P. Caraveo, W.~Hermsen, G.~Kanbach, G.~G.~Lichti, J.~L~Masnou,
 H.~A.~Mayer-Hasselwander, J.~A.~Paul, B.~Sacco, L.~Scarsi,
 R.~D.~Wills,
 {\em Astrophysical Journal} (1981) {\bf 243}, L69--L73

 \bibitem{Spacelab:nulc}
 S.~P.~Swordy, D.~M\"{u}ller, P.~Meyer, J.~L'Heureux,
 J.~M.~Grunsfeld,
 {\em Astrophysical Journal} (1990) {\bf 349}, 625--633

 \bibitem{js:koczarov}
 B. Szabelska, J. Szabelski, A. W. Wolfendale,
 {\em Journal of Physics G.: Nuclear Physics} 
 (1991) {\bf 17}, 545--553

 \bibitem{js:antypI}
 J. Szabelski, J. Wdowczyk, A. W. Wolfendale,
 {\em Nature} (1980) {\bf 285}, 386--388

 \bibitem{js:UHEold1}
 J. Szabelski, J. Wdowczyk, A. W. Wolfendale,
 {\em Journal of Physics G.: Nuclear Physics} 
 (1986) {\bf 12}, 1433--1442

 \bibitem{CGRO:PSR1706}
 D.~J.~Thompson, Z. Arzoumanian,
 D. L. Bertsch, K. T. S. Brazier, N.~D'Amico, C.~E.~Fichtel,
 J.~M.~Fierro, R.~C.~Hartman, S.~D~Hunter, 
 S.~Johnston, G.~Kanbach, V.~M.~Kaspi, D.~A.~Kniffen,
 P.~W.~Kwok, Y.~C.~Lin, A.~G.~Lyne, R.~N.~Manchester, J.~R.~Mattox,
 H.~A.~Mayer-Hasselwander, P.~F.~Michelson, C.~v.~Montigny,
 H.~I.~Nel, D.~Nice, P.~L.~Nolan, K.~Pinkau, H.~Rothermel, 
 E.~J.~Schneid, M~Sommer, P.~Sreekumar, J.~H.~Taylor,
 {\em Nature} (1992) {\bf 359}, 615--616
 
 \bibitem{EGRET:IstCatalog}
 D.~J.~Thompson, D.~L.~Bertsch, B.~L.~Dingus, J.~A.~Esposito,
 A.~Etienne, C.~E.~Fichtel, D.~P.~Friedlander, R.~C.~Hartman,
 S.~D.~Hunter, D.~J.~Kendig, J.~R.~Mattox, L.~M.~McDonald,
 C.~von~Montigny, R.~Mukherjee, P.~V.~Ramanamurthy, P.~Sreekumar,
 J.~M.~Fierro, Y.~C.~Lin, P.~F.~Michelson, P.~L.~Nolan,
 S.~K.~Shriver, T.~D.~Willis, G.~Kanbach, H.~A.~Mayer-Hasselwander,
 M.~Merch, H.~-D.~Radecke,  
 D.~A.~Kniffen, E.~J.~Schneid,
 {\em Astrophysical Journal Supplement Series}
 (1995) {\bf 101}, 259--286,
 {\em NASA preprint 662-94-04} (1994)
 
 \bibitem{EGRET:calib}
 D.~J.~Thompson,
 D.~L.~Bertsch, C.~E.~Fichtel,
 R.~C.~Hartman, R.~Hofstadter, E.~B.~Hughes, S.~D~Hunter, 
 B.~W.~Hughlock, G.~Kanbach, D.~A.~Kniffen,
 Y.~C.~Lin, J.~R.~Mattox,
 H.~A.~Mayer-Hasselwander, C.~von~Montigny, P.~L.~Nolan,
 H.~I.~Nel, K.~Pinkau, H.~Rothermel, 
 E.~J.~Schneid, M~Sommer, P.~Sreekumar, D.~Tieger, A.~H.~Walker,
 {\em preprint NASA LHEA 92--026} (1992)

 \bibitem{EGRET:IIndCatalog}
 D.~J.~Thompson, D.~L.~Bertsch, B.~L.~Dingus, J.~A.~Esposito,
 A.~Etienne, C.~E.~Fichtel, D.~P.~Friedlander, R.~C.~Hartman,
 S.~D.~Hunter, D.~J.~Kendig, J.~R.~Mattox, L.~M.~McDonald,
 C.~von~Montigny, R.~Mukherjee, P.~V.~Ramanamurhty, P.~Sreekumar, 
 J.~M.~Fierro, 
 B.~B.~Jones, %tylko supplement
 Y.~C.~Lin, P.~F.~Michelson, P.~L.~Nolan,
 S.~K.~Shriver, %bez supplement
 W.~Tompkins, T.~D.~Willis, 
 G.~Kanbach, H.~A.~Mayer-Hasselwander, M.~Merk, 
 H.--D.~Radecke, %bez supplement 
 M.~Pohl, %tylko supplement
 D.~A.~Kniffen,
 E.~J.~Schneid,
 {\em anonymous ftp from gamma.gsfc.nasa.gov,
 subdirectory pub/second\_catlog} (1997)

 \bibitem{Crab:Whipple_91}
 G.~Vacanti, M.~F.~Cawley, E.~Colombo, D.~J.~Fegan, A.~M.~Hillas,
 P.~W.~Kwok, M.~J.~Lang, R.~C.~Lamb, D.~A.~Lewis, D.~J.~Macomb,
 K.~S.~O'Flaherty, P.~T.~Reynolds, T.~C.~Weekes,
 {\em Astrophysical Journal} (1991) {\bf 377}, 467--479

 \bibitem{JS:Moskwa94} 
 A.~V.~Voevodsky, V. B. Petkov, A.~M.~Semenov, A. Tsyabuk,
 A.~E.~Chudakov, J.~Szabelski,
 {\em Izviestia Akademii Nauk, Seria fiziceskaya} 
 (1994) {\bf 58}, no~9, 127--129
 (in Russian)

 \bibitem{WW:scalbr}
 J.~Wdowczyk, A.~W.~Wolfendale,
 {\em Il Nuovo Cimento} (1979) {\bf 54A}, 444--450
 
 \bibitem{pr:Webber}
 W.~R.~Webber, R.~L.~Golden, S.~A.~Stephens,
 {\em 20-th International Cosmic Ray Conference, Moscow} (1987)
 {\bf 1}, 325--328
 
 \bibitem{Crab:Whipple_89}
 T.~C.~Weekes, M.~F.~Cawley, D.~J.~Fegan, K.~G.~Gibbs, A.~M.~Hillas,
 P.~W.~Kwok, R.~C.~Lamb, D.~A.~Lewis, D.~Macomb, N.~A.~Porter,
 P.~T.~Reynolds,  G.~Vacanti,
 {\em Astrophysical Journal} (1989) {\bf 342}, 379--395

 \bibitem{UHE:SUGAR_1}
 M. M. Winn, J. Ulrichs, L. S. Peak, C. B. A. McCusker, L.~Horton,
 {\em Journal of Physics G: Nuclear Physics} (1986) {\bf 12},
 653--674

 \bibitem{UHE:SUGcat}
 M. M. Winn, J. Ulrichs, L. S. Peak, C. B. A. McCusker, L.~Horton,
 {\em Catalogue of the Highest Energy Cosmic Rays, SUGAR}
 (1986) {\bf No.~2}, World Data Center C2 for Cosmic Rays,
 Institute of Physical and Chemical Research, 
 Itabashi, Tokyo, Japan

 \end{thebibliography}
 \end{document}